\documentclass[bibyear]{aa} 

\usepackage{graphicx}
\usepackage{txfonts}
\usepackage{longtable}
\begin{document}

   \title{European VLBI Network imaging of 6.7~GHz methanol masers}


   \author{A. Bartkiewicz
          \inst{1},
          M. Szymczak
          \inst{1},
          \and
          H.J. van Langevelde
          \inst{2,}\inst{3}
          }

   \institute{Centre for Astronomy, Faculty of Physics, Astronomy and Informatics, Nicolaus Copernicus University, Grudziadzka 5, 87-100 Torun, Poland,
\email{[annan;msz]@astro.umk.pl}
         \and
             Joint Institute for VLBI ERIC (JIVE), Postbus 2, 7990 AA Dwingeloo, The
Netherlands, \email{langevelde@jive.eu}
         \and
             Sterrewacht Leiden, Leiden University, Postbus 9513, 2300 RA Leiden, The Netherlands
             }

   \date{Received 12 October 2015; accepted 04 January 2016}

 
  \abstract
   {Methanol masers at 6.7~GHz are well known tracers of high-mass star-forming regions (HMSFRs).
 However, their origin is still not clearly understood.}
   {We aimed to determine the morphology and velocity structure for a large sample of the maser emission with generally lower peak flux densities than those in previous surveys.}
   {Using the European VLBI Network (EVN) we imaged the remaining sources from a sample of sources that were selected from the unbiased survey using the Torun 32~m dish. In this paper we report the results for 17 targets. Together they form a database of a total of 63 source images with high sensitivity (3$\sigma_{rms}$=15--30~mJy~beam$^{-1}$), milliarcsecond angular resolution (6--10~mas) and very good spectral resolution (0.09~km~s$^{-1}$ or 0.18~km~s$^{-1}$) for detailed studies.}
   {We studied in detail the properties of the maser clouds and calculated the mean and median values of the projected size (17.4$\pm$1.2~au and 5.5~au, respectively) as well as the FWHM of the line (0.373$\pm$0.011~km~s$^{-1}$ and 0.315~km~s$^{-1}$ for the mean and median values, respectively), testing whether it was consistent with Gaussian profile. We also found maser clouds with velocity gradients (71\%) that ranged from 0.005~km~s$^{-1}$~au$^{-1}$ to 0.210~km~s$^{-1}$~au$^{-1}$. We tested the kinematic models to explain the observed structures of the 6.7~GHz emission. There were targets where the morphology supported the scenario of a rotating and expanding disk or a bipolar outflow. Comparing the interferometric and single-dish spectra we found that,  typically, 50--70\% of the flux was missing. This phenomena is not strongly related to the distance of the source.}
   {The EVN imaging reveals that in the complete sample of 63 sources the ring-like morphology appeared in 17\% of sources, arcs were seen in a further 8\%, and the structures were complex in 46\% cases. The ultra-compact (UC) H~{\small II} regions coincide in position in the sky for 13\% of the sources. They are related both to extremely high and low luminosity masers from the sample.}


\titlerunning{EVN imaging of 6.7~GHz methanol masers}
\authorrunning{A. Bartkiewicz et al.}

   \maketitle
%

\section{Introduction}
The 6.7~GHz methanol maser line (Menten \cite{m91}) is closely associated with massive young stellar objects (MYSOs) (e.g. Breen et al.~\cite{b13} and references therein) and is used for the identification of these regions. However the question 'where does the 6.7~GHz methanol maser arise?' is still not answered. The detailed spatial structure of methanol maser emission derived using the very long baseline interferometry (VLBI) provided a few clues to their origin. For example, linear structures with velocity gradients were interpreted as parts of circumstellar disks seen edge-on (Minier et al.~\cite{m00}). However, they could also have been created when a planar shock propagated through a rotating dense mole cular clump (Dodson et al.~\cite{d04}) or in outflows (De Buizer \& Minier~\cite{dm05}). Various structures have also been detected: simple -- containing a few maser spots -- curved, complex -- with no regularity in the velocity distribution -- and paired (Norris et al.~\cite{n93}; Philips et al.~\cite{p98}; Walsh et al.~\cite{w98}; Minier et al.~\cite{m00}). Proper motion studies have resulted in valuable information, however they have only been undertaken for a few targets. Sanna et al.~(\cite{s10a}) report that in G16.59$-$0.05 the 6.7~GHz methanol masers originate in a disk or toroid, which is rotating around a MYSO. In G23.01$-$0.41 they detect slow motions of radial expansion and rotation related to this maser line (Sanna et al.~\cite{s10}). Goddi et al.~(\cite{g11}) measure an infall of a molecular envelope onto an intermediate- to high-mass protostar using 6.7~GHz methanol maser spots. In IRAS~20126$+$4104 Moscadelli et al.~(\cite{m11}) identify that one group of methanol masers was  rotating in the disk and a second one was moving away perpendicularly from the disk, probably tracing the disk material was being marginally entrained by the jet. 

To put more constraints on the methanol maser's origin, we performed imaging of the 6.7~GHz methanol maser line at a milliarcsecond (mas) scale with a few mJy sensitivity using the European VLBI Network (EVN\footnote{The European VLBI Network is a joint facility of independent European, African, Asian, and North American radio astronomy institutes. Scientific results from data presented in this publication are derived from the following EVN project code: ES060.}) for a sample of sources that were discovered in a blind survey of the Galactic plane $20\degr<l<40\degr$ and $|b|<0\fdg5$ using the 32~m Torun dish (Szymczak et al.~\cite{sz02}). The main aims were to obtain accurate coordinates at a level of a few mas and to determine the morphology for a large sample of sources for which the peak flux densities, measured with the 32~m dish, were lower compared to previous VLBI studies, the median being 18~Jy. In Bartkiewicz et al.~(\cite{b09}) we present 31 images of the 6.7~GHz methanol maser emission. The diversity of morphologies was observed as being similar to previous observations. However, a new class of objects, the ring-like structures, was reported. Similar categories were observed using the East-Asian VLBI Network (EAVN) by Fujisawa et al.~(\cite{f14}). The ring-like (or elliptical) morphology was also detected in the nearby high-mass star-forming region (HMSFR) Cep~A (Torstensson et al.~\cite{t11}). Next, we imaged an   additional 15 methanol maser sources that were selected on the basis of the complexity of their spectra as registered using the Torun dish. The results were presented in Bartkiewicz et al.~(\cite{b14}), the positions and morpholgies were determined and we searched their mid-infrared counterparts. In this paper, we report the remaining data set of methanol masers that were discovered by Szymczak et al.~(\cite{sz02}). With this unique database of more than 60 sources imaged with the mas resolution and a few mJy sensitivity, we summarise the results.

\section{Observations}
\subsection{Target selection}
The sample from this work was derived from the untargeted observations of the 6.7~GHz methanol maser line 
of the Galactic plane area 20\degr$<$$l$$<$40\degr and $|b|$$<$0\fdg52 (Szymczak et al.~\cite{sz02}). We selected 18 target sources whose positions were not measured with the radio interferometers by 
Walsh et al. (\cite{w98}) and Bartkiewicz et al. (\cite{b09}). Their coordinates have been refined with MERLIN
(Wong-McSweeney \cite{wong08}) to achieve the astrometric accuracy required for follow-up VLBI studies.
The MERLIN data for the sources G20.24$+$00.07 and G35.00$+$00.34 suggest that they are double, therefore
two pointing positions were scheduled for each of them. A total of 20 pointing positions were observed
(Table~\ref{table1}).

\subsection{EVN observations}
The EVN observations 
were carried out at 6668.519~MHz on 2009 May 31 (Run 1), June 1 (Run 2), and on 2010 March 12 (Run 3) 
and 13 (Run 4) for 10~hr each. The following antennas were used: Cambridge,  
Jodrell Bank, Effelsberg, Medicina, Onsala, Noto, Torun, Westerbork, and
Yebes. The Onsala antenna was not used in Run 1, while the Cambridge and Jodrell Bank antennae were 
not used in Run 3. A phase-referencing scheme was applied 
 to determine the absolute positions of the targets at the level of a few mas.
The reference sources and their angular separations from the targets are listed in Table~\ref{table1}.  
We used a cycle time between the maser and phase-calibrator of 195~s$+$105~s. This yielded
a total integration time for each individual source of $\sim$50~min. 
The bandwidth was set to 2\,MHz yielding 90\,km\,s$^{-1}$ velocity coverage. In Table~\ref{table1} we list 
the local standard of rest (LSR) velocity range for each source.
Data were correlated at JIVE with 1~s integration time and 1024 spectral channels. The field of view over which the response to point source was degraded by less than $\sim$10\% was 37\arcsec, and the spectral resolution was 0.089\,km\,s$^{-1}$.

\begin{table*}
\centering
\caption{Details of EVN observations.}
\label{table1}
\begin{tabular}{@{}lllccccc@{}}
\hline
Source$^{a}$ & \multicolumn{2}{c}{Pointing positions (J2000)} & Velocity & Phase-calibrator & Separation
& Observing & Synthesized beam\\
 & & & coverage & & & run & \\
Gll.ll$\pm$bb.bb& RA (h m s) & Dec ($^{\rm o}$ ' '') & $({\rm km~s}^{-1})$ &  & (\fdg) & &
(mas$\times$mas;PA(\degr))\\
\hline
G20.24$+$00.07 & 18 27 44.572 & $-$11 14 54.18 & 7:97 & J1825$-$0737 & 3.66 & 1 & 9.0$\times$5.4;$-$6 \\
               & 18 27 44.961 & $-$11 14 47.91 & 7:97 & J1825$-$0737 & 3.66 & 1 & 9.0$\times$5.4;$-$6 \\
G21.56$-$00.03 & 18 30 36.067 & $-$10 07 10.88 & 72:162 & J1825$-$0737 & 2.78 & 2 & 7.8$\times$5.3;$-$10 \\
G22.44$-$00.17 & 18 32 43.827 & $-$09 24 32.85 & 7:97 & J1825$-$0737 & 2.50 & 1 & 7.7$\times$5.2;$-$12\\
G23.01$-$00.41 & 18 34 40.297 & $-$09 00 38.26 & 7:97 & J1825$-$0737 & 2.63 & 1 & 7.0$\times$5.0;$-$5 \\
G24.33$+$00.14 & 18 35 08.145 & $-$07 35 01.79 & 72:162 & J1825$-$0737 & 2.36 & 2 & 5.7$\times$4.7; 2 \\
G24.49$-$00.04 & 18 36 05.909 & $-$07 31 25.13 & 72:162 & J1825$-$0737 & 2.60 & 2 & 7.8$\times$5.3;$-$12 \\
G24.79$+$00.08 & 18 36 12.574 & $-$07 12 10.91 & 72:162 & J1825$-$0737 & 2.66 & 2 & 7.4$\times$5.2;$-$14 \\
G24.85$+$00.09 & 18 36 18.398 & $-$07 08 50.95 & 72:162 & J1825$-$0737 & 2.69 & 2 & 6.9$\times$5.2; 5 \\
G28.01$-$00.43 & 18 43 57.969 & $-$04 34 24.09 & $-$25:65 & J1834$-$0301 & 2.85 & 3 & 7.8$\times$5.3;$-$18 \\
G29.98$-$00.05 & 18 46 12.962 & $-$02 39 01.38 & 35:125 & J1834$-$0301 & 3.01 & 4 & 8.7$\times$5.4;$-$9 \\
G30.20$-$00.17 & 18 47 03.061 & $-$02 30 34.01 & 35:125 & J1834$-$0301 & 3.24 & 4 & 8.3$\times$5.4;$-$8 \\
G30.21$-$00.17 & 18 47 03.093 & $-$02 30 04.91 & 35:125 & J1834$-$0301 & 3.24 & 4 & 8.5$\times$5.3;$-$8 \\ 
G32.75$-$00.08 & 18 51 21.995 & $-$00 12 02.82 & $-$25:65 & J1907$+$0127 & 4.29 & 3 & 6.4$\times$4.7;$-$31 \\
G34.25$+$00.14 & 18 53 21.442 & $+$01 14 26.02 & 35:125 & J1907$+$0127 & 3.47 & 4 & 9.4$\times$4.3; 26 \\
G34.26$+$00.15 & 18 53 18.633 & $+$01 14 57.40 & 35:125 & J1907$+$0127 & 3.48 & 4 & 9.3$\times$6.4; 11 \\
G34.39$+$00.22 & 18 53 19.070 & $+$01 24 05.93 & 35:125 & J1907$+$0127 & 3.47 & 4 & 17.6$\times$7.1; 30 \\
G35.00$+$00.34 & 18 54 00.195 & $+$01 59 37.40 & $-$25:65 & J1907$+$0127 & 3.34 & 3 & 5.7$\times$4.7;$-$19 \\
               & 18 54 00.727 & $+$02 01 09.91 & $-$25:65 & J1907$+$0127 & 3.34 & 3 & 5.7$\times$4.7;$-$19 \\
G38.26$-$00.07 & 19 01 26.251 & $+$04 42 19.96 & $-$25:65 & J1907$+$0127 & 3.56 & 3 & 5.3$\times$4.0;$-$50 \\
\hline
\end{tabular}
\tablefoot{$^a$Names are the Galactic coordinates derived from the MERLIN results (Wong-McSweeney \cite{wong08}).}\\
\end{table*}

\subsection{Data reduction and imaging}
The data calibration and reduction were carried out with the Astronomical Image Processing System (AIPS), 
developed by the National Radio Astronomy Observatory (NRAO). We used standard procedures for spectral line observations 
and the Effelsberg antenna was set as reference. The amplitude was calibrated through measurements of the system 
temperature at each telescope and application of the antenna gain curves. The parallactic angle corrections were subsequently 
added to the data. The source 3C345 was used as a delay, rate, and bandpass calibrator. The phase-calibrators
J1825$-$0737, J1834$-$0301 and J1907$+$0127 were imaged and flux densities of 265, 180 and 160~mJy were obtained, respectively. 
During Run 3, the flux densities of J1834$-$0301 and J1907$+$0127 were found to be lower: 150 and 133~mJy, respectively. 
The maser data were corrected for the effects of the Earth's rotation and
its motions within the solar system and towards the LSR.

To find the positions of the emission for each target, we created a dirty map of size 8\arcsec$\times$8\arcsec\, centred at the pointing position of the brightest feature (Table~\ref{table1}) seen in the vector--averaged spectrum. 
In the case of the clear appearance of a maser spot, we produced a smaller map (1\arcsec$\times$1\arcsec) centred at its position.
When the emission was not seen in the large dirty map, we imaged the region nearby consistently, shifting 
the map centre in right ascension and declination in the range from $-$18\arcsec\, to 18\arcsec. This allowed us to find emission towards most of targets. Then we ran a self-calibration 
procedure using the clean components of the compact and bright maser spot map. 
The first map was applied as a model to keep the unchanged absolute position of the dominant component.
Finally, naturally-weighted maps of spectral channels were produced over the velocity range where
the emission was seen in the scalar-averaged spectrum. A pixel separation of 1\,mas in both coordinates was
used for the imaging.  The resulting synthesized beams are listed in Table~\ref{table1}. The rms noise level
(1$\sigma_{\rm rms}$) in emission line-free channels was typically a few mJy for each source. To obtain the spectrum of each source, we recovered the flux density from an image cube using the task ISPEC integrated over the masing area.

To fit two-dimensional Gaussian models and to obtain the positions of the methanol maser spots 
in all channel maps, we used the AIPS task JMFIT. The formal fitting errors were typically less than 0.1\,mas.
The absolute position accuracy of maser spots was about a few mas (Bartkiewicz et al.~\cite{b09}).
In the case of two objects G24.33$+$00.14 and G32.744$-$00.076 the absolute position uncertainty was larger because the phase-referencing failed (no emission was found on large maps successfully shifted in the field of view) and the FRING task was used. 
For the first source we list the coordinates obtained using the EVLA (Hu et al.~{\it in prep.}) 
and for the second source we list the coordinates and uncertainties estimated from the EVN dirty image. They are consistent with results presented by Breen et al.~(\cite{breen15}).

No emission was found in the object G30.21$-$00.17 at the coordinates from Wong-McSweeney (\cite{wong08}). Nothing was also reported in that direction by Breen et al.~(\cite{breen15}). However, based on this data set, we imaged the emission G30.224$-$00.180 at the coordinates reported by Breen et al.~(\cite{breen15}) for this source. No emission was also found towards the position given by the first entry for G35.00+00.34 in Table~\ref{table1}. Breen et al.~(\cite{breen15}) also did not report any source there. The non-detection of these two sources is likely because of misidentification in Wong-McSweeney (\cite{wong08}) and not because of their variability.
 We also did not obtain proper images for two sources, G21.56$-$00.03 and G38.26$-$00.07, although the emission (which was a few times weaker compared to single-dish data) in the spectra was seen and the input coordinates are consistent within 2" of those in Breen et al.~(\cite{breen15}). Among possible reasons for this failure 
may be a poor uv-coverage, too faint emission or/and too distant (2\fdg78 and 3\fdg56) phase-calibrators. 

\section{Results}
A total of 17 targets were successfully imaged.
Table~\ref{results} contains the revised source name (the Galactic coordinate of the peak flux density), the coordinates, the velocity of the brightest spot
($V_{\rm p}$), its intensity ($S_{\rm p}$), the velocity width of emission ($\Delta V$), the distance estimates
and the morphology class. The distribution of the maser emission and the spectrum for each source are shown in Figs.~\ref{distrib} and \ref{distribapp}. When the emission appeared in at least three consecutive spectral channels, coinciding in position within half of the synthesized beam, we defined it as a maser cloud and verified its properties. These clouds with Gaussian velocity profiles are summarised in Table~\ref{clouds} and presented in Fig.~\ref{gauss}. For each target, the following parameters are listed: relative position of the brightest maser spot ($\Delta$RA, $\Delta$Dec), its velocity (V$_{\rm p}$), and flux density (S$_{\rm p}$), fitted velocity (V$_{\rm fit}$), line full width at half maximum (FWHM) and flux density (S$_{\rm fit}$). In addition, the projected length (L$_{\rm proj}$) and, if detected, the velocity gradient (V$_{\rm grad}$) are given.

We calculated the kinematic distances using the code of Reid et al.~(\cite{r09}), assuming that 
the systemic velocity of each source is equal to the peak velocity of its 6.7~GHz maser emission. 
For a few objects the distance ambiguity has been resolved or the direct estimation of the distance through the trigonometric parallax was available in the literature (Table~\ref{results}), whereas for the other sources we adopted the near kinematic distances.

We have also classified the obtained maser emission according to the criteria presented in Bartkiewicz et al.~(\cite{b09}). 
This classification was done by eye and based on the projected-on-the-sky distribution of maser spots, which is summarised in Table~\ref{results}.

\subsection{Individual sources}
\begin{figure*}
\centering
\includegraphics[scale=0.57]{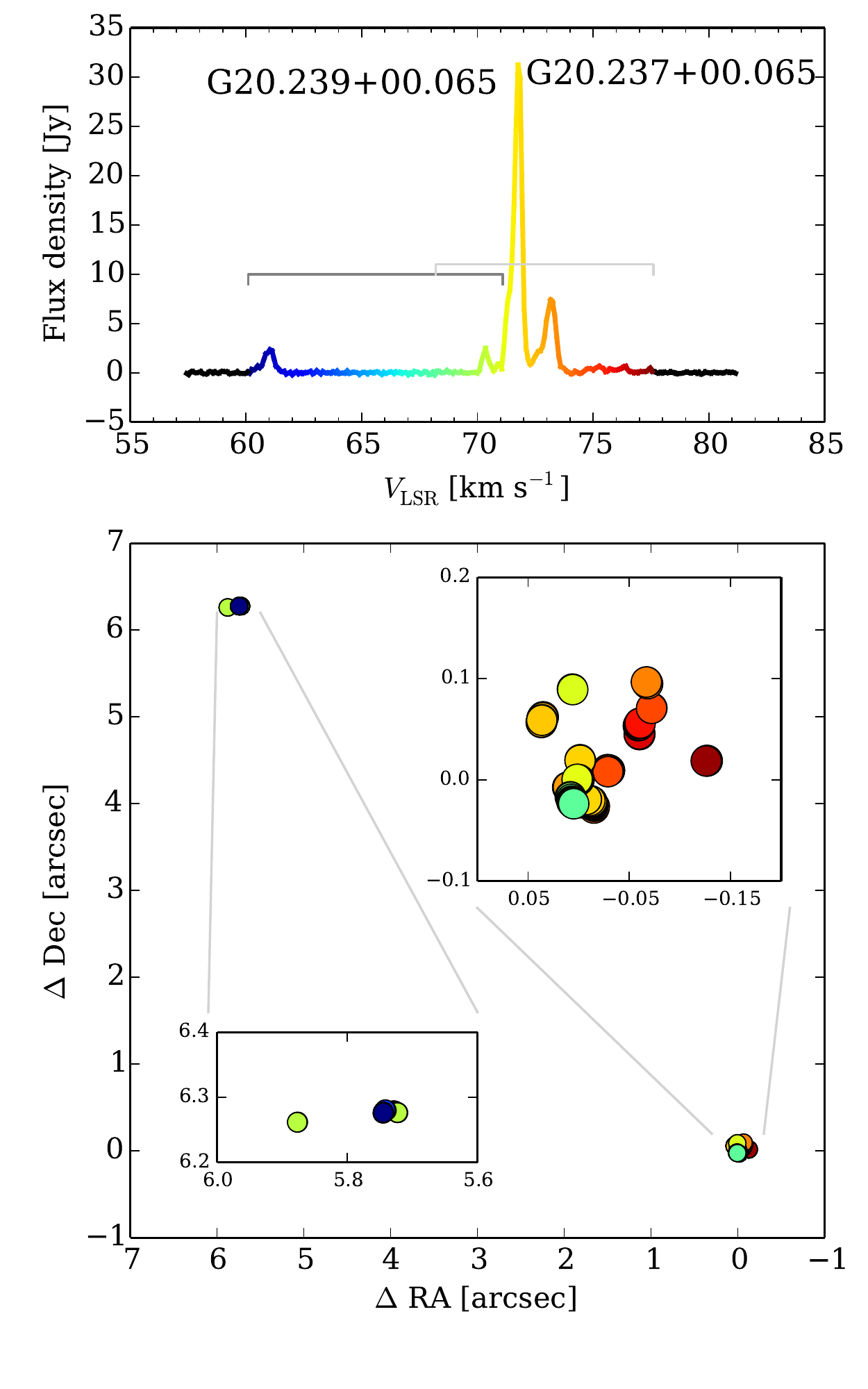}
\includegraphics[scale=0.57]{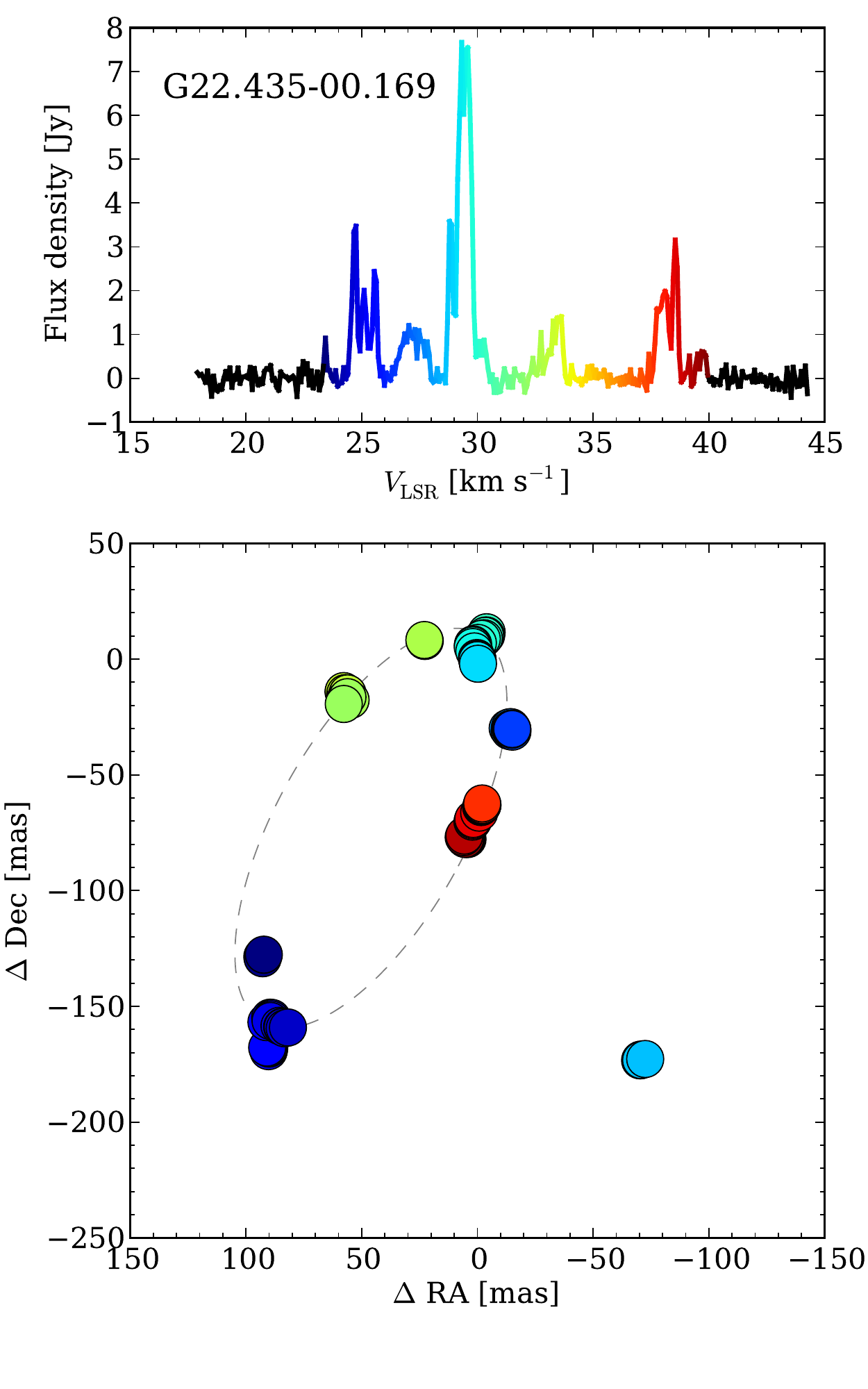}
\includegraphics[scale=0.57]{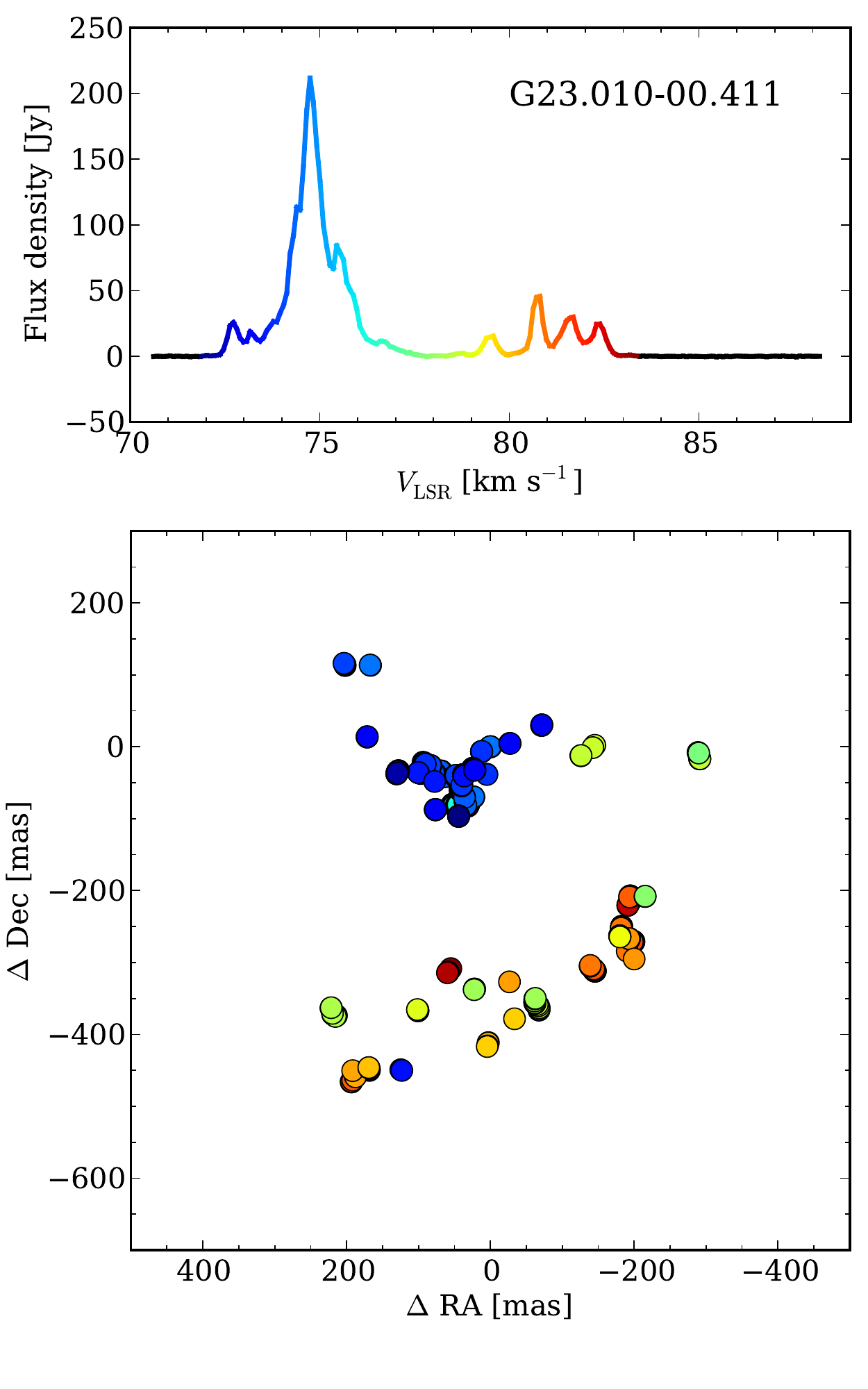}
\includegraphics[scale=0.57]{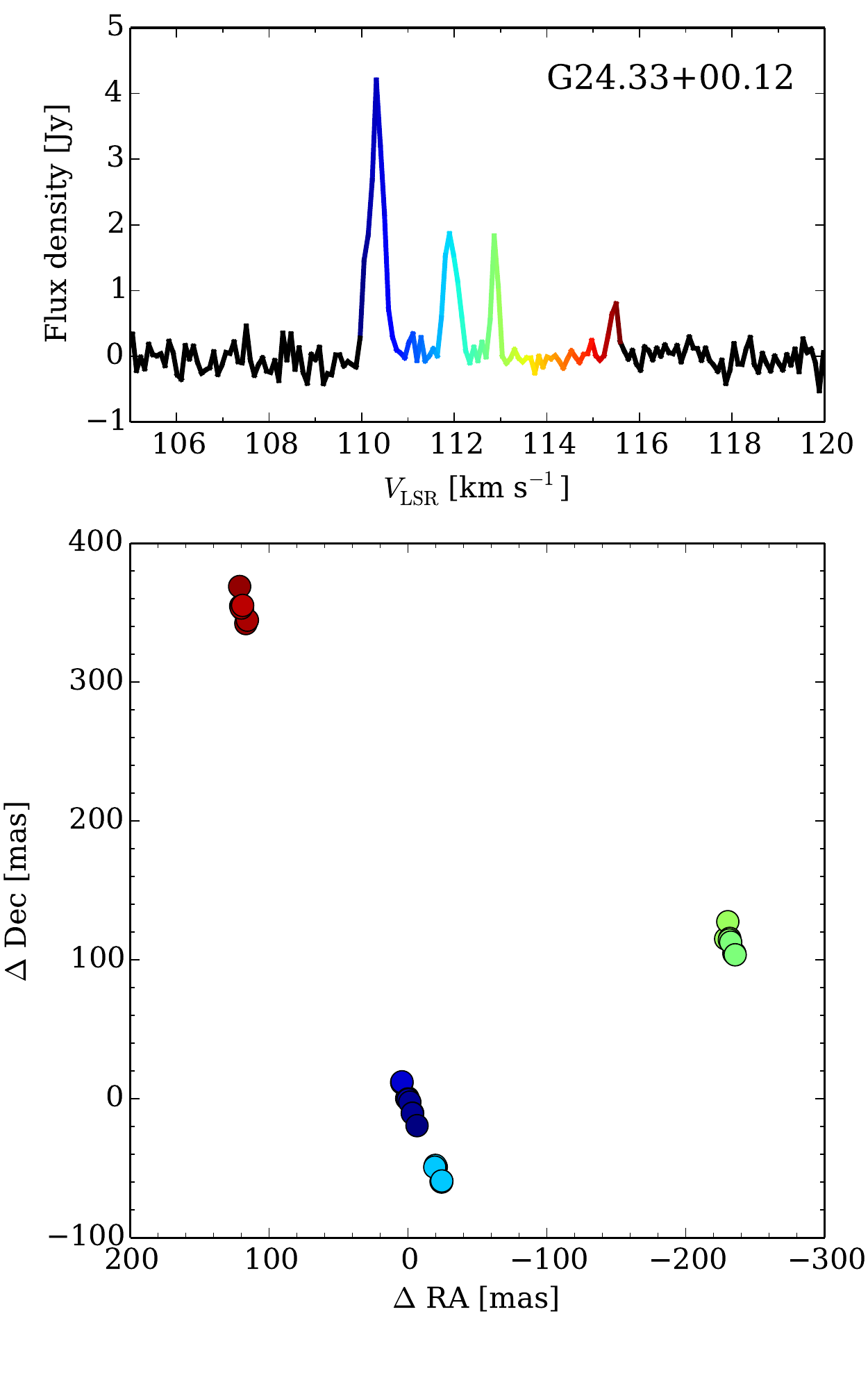}
\caption{Spectra and maps of 6.7~GHz methanol masers detected using the EVN.
    The names are the Galactic coordinates of the brightest spots listed
    in Table~\ref{results} (the (0,0) locations).The colors of circles relate to the LSR velocities 
as shown in the spectra. Note, that in the first target the scale in distribution is presented in arcsec, 
while the rest of sources are described by a mas scale. The grey ellipses mark the best-fit ellipses for the sources with the ring-like morphology. The plots for the remaining targets are presented in the online material (Appendix~A).} 
\label{distrib}  
\end{figure*}

{\it G20.237$+$00.065 and G20.239$+$00.065} \hspace{0.25cm}
There are two distinct maser sources separated by 8\farcs6 that corresponds to 38100~au (0.18pc) for the near kinematic distance of 4.4~kpc (Fig.~\ref{distrib}). The source G20.237$+$00.065 consists of 102 maser spots in the LSR velocity range from 68.2 to 77.6~km~s$^{-1}$ 
and the source G20.239$+$00.065 consists of 37 maser spots in the two velocity ranges: 60.1--63.3~km~s$^{-1}$  and 70.1--71.1~km~s$^{-1}$. The maser spots of source G20.237$+$00.065 show a complex distribution over
$\sim$880~au$\times$880~au without any regularity in velocity. Source G20.239$+$00.065 is extended in 
the east--west direction over an area of $\sim$220~au$\times$880~au.  
The coordinates from the EVN maps agree within $\sim$70\,mas with those determined from the East-Asian VLBI 
Network observation (Fujisawa et al.~\cite{f14}) but the morphologies of both maser sources appear
more complex owing to the higher sensitivity and better spectral resolution of EVN. 
Caswell~(\cite{c09}), using the Australian Telescope Compact Array (ATCA), first recognized the site as two
maser sources in the velocity range from 60 to 78~km~s$^{-1}$. We note that his value of the declination 
of source 20.239$+$00.065 differs by 1\farcs1 from ours, that is within the ATCA uncertainty.  
Source G20.237$+$00.065 coincides with the OH maser emission at 1665, 1720, and 6035~MHz (Caswell~\cite{c03,c04}). No OH emission was found towards G20.239$+$00.065 (Caswell~\cite{c09}). 

{\it G22.435$-$00.169} \hspace{0.25cm}
We detected 91 maser spots from 23.4 to 40.0~km~s$^{-1}$ with the brightest spot of 2.44~Jy~beam$^{-1}$ 
at 29.3~km~s$^{-1}$. The morphology is ring-like with a few spots lying 50--100~mas offset to the south-west direction (Fig.~\ref{distrib}). Using the GNU Octave script developed by Fitzgibbon et
al.~(\cite{f99}), the ellipse fitting to the majority of spots of G22.435$-$00.169, excluding a few blueshifted spots, 
gives values 192 and 82~mas for the major and minor axes with the position angle of the major axis 
of $-$61\degr. The relative coordinates of the ellipse centre are 46~mas and $-$73~mas (Fig.~\ref{distrib}). 
For the assumed distance of 2.2~kpc the ellipse size corresponds to 422~au$\times$180~au.
The coordinates and emission range agree well with those observed with the ATCA (Caswell et al.~\cite{c09}). The 6035~MHz OH masers coincide within $\sim$1\arcsec\, with the methanol emission (Caswell~\cite{c01}). 

{\it G23.010$-$00.411} \hspace{0.25cm}
We found 361 methanol maser spots in the velocity range from 71.8 to 83.4~km~s$^{-1}$ (Fig.~\ref{distrib}). 
The brightest spot of 72.6~Jy~beam$^{-1}$ appeared at 74.7~km~s$^{-1}$. The blueshifted part of the emission 
is generally located in the northern elongated structure, while the redshifted one is located in the southern structure.
A total extent of maser emission is 0\farcs6$\times$0\farcs6 and the morphology closely matches that seen in 
the previous EVN observations (Sanna et al.~\cite{s10}).

{\it G24.33$+$00.14} \hspace{0.25cm}
The source consists of three groups of maser spots separated by $\sim$0\farcs2--0\farcs4 (Fig.~\ref{distrib}).
The most redshifted emission, which is composed of eight spots, originated from the northern group, the blueshifted emission
of 21 spots was located to the south, and the emission at the intermediate velocities came from eight spots of 
the western group. The emission was so weak (S$_{\rm p}$=1.62Jy~beam$^{-1}$) that the map was obtained using
additional phase calibration with FRING. Table~\ref{results} gives the coordinates taken from
the EVLA observations in 2012 (Hu et al.~{\it in prep}). Comparison of the two data sets revealed that
the spectrum from EVN is three times weaker than that from EVLA; the redshifted emission near 115~km~s$^{-1}$ 
was not seen in the EVLA maps, whereas 
the extreme redshifted emission near 119~km~s$^{-1}$ was only detected with the EVLA. This suggests that the source 
experienced strong variability. Indeed, the single-dish monitoring revealed an outburst in 2011 (Szymczak et al. {\it in prep.}). 

{\it G24.494$-$00.038} \hspace{0.25cm} 
The source has a complex morphology where 73 methanol maser spots in the velocity range from 107.9 to 116.2~km~s$^{-1}$ 
are distributed over a region of 0\farcs4$\times$0\farcs4 (Fig.~\ref{distribapp}). It corresponds to 2320~au$\times$2320~au for the assumed near kinematic distance of 5.8~kpc. 

{\it G24.790$+$00.083} \hspace{0.25cm}
This source contains two groups of masers, each with a ring-like distribution (Fig.~\ref{distribapp}).
We found 98 spots located in the south-east and 73 spots in the north-west, separated by $\sim$1\farcs5 that corresponds to 11550~au for the 7.7~kpc distance. 
Moscadelli et al.~(\cite{m07}) explored this region with methanol and water masers, and compared it  
with available thermal continuum and molecular line tracers. Our methanol groups correspond nicely 
to their masers, which are associated with G24~A1 and G24~A2 mm subcores. Because we had four times better resolution 
in velocity, we detected more methanol spots within the same velocity range 106.3--116.8~km~s$^{-1}$. 
We note that the overall structure persisted for six years. 

\begin{table*}
 \centering
  \caption{Results of EVN observations}
  \begin{tabular}{lllcccrcc}
  \hline
   Source$^a$ & \multicolumn{2}{c}{Position of the brightest spot (J2000)} & V$_{\rm p}$ & $\Delta$V & S$_{\rm p}$ & D$_{\rm near}$/D$_{\rm far}$ & D$_{\rm adopted}$ & Class$^b$\\
   Gll.lll$\pm$bb.bbb & RA (h m s) & Dec (\degr ' ") & (km~s$^{-1}$) & (km~s$^{-1}$) & (Jy~beam$^{-1}$) & (kpc) & (kpc) &\\
 \hline
G20.237$+$00.065 & 18 27 44.56429 & $-$11 14 54.0938 & 71.8  & 9.4  & 22.98 & 4.4/11.2 & 4.4$^d$& C\\
G20.239$+$00.065 & 18 27 44.95441 & $-$11 14 47.8165 & 61.0  & 11.0 & 2.41  & 3.9/11.7 & & L\\
G22.435$-$00.169 & 18 32 43.81865 & $-$09 24 32.7838 & 29.3  & 16.6 & 2.44  & 2.2/13.1 & & R\\
G23.010$-$00.411 & 18 34 40.28623 & $-$09 00 38.1037 & 74.7  & 11.6 & 72.6  & 4.3/11.0 & 4.59$^e$& C\\
G24.33$+$00.14$^{c1}$& 18 35 08.119  & $-$07 35 04.13   & 110.3 & 5.6  & 1.62  & 5.7/9.5  & 9.5$^d$&C\\
G24.494$-$00.038 & 18 36 05.72977 & $-$07 31 19.2012 & 115.1 & 8.3  & 2.85  & 5.8/9.3 & & C\\
G24.790$+$00.083 & 18 36 12.56216 & $-$07 12 10.8211 & 113.4 & 10.5 & 55.83 & 5.8/9.4 & 7.7$^f$ & (P)R,R\\ 
G24.850$+$00.087 & 18 36 18.38754 & $-$07 08 50.8051 & 110.2 & 7.9  & 5.9   & 5.7/9.5 & 8.0$^g$ & C \\
G28.011$-$00.426 & 18 43 57.96865 & $-$04 34 24.0938 & 16.1  & 12.7 & 1.26  & 1.1/13.4 & & C\\
G29.978$-$00.048 & 18 46 12.96122 & $-$02 39 01.3798 & 103.5 & 8.9  & 31.2  & 5.6/8.8  & 8.8$^d$ & C\\
G30.198$-$00.169 & 18 47 03.07004 & $-$02 30 36.2597 & 108.2 & 10.4 & 14.7  & 5.9/8.5 & & C\\
G30.224$-$00.180 & 18 47 08.29781 & $-$02 29 29.3199 & 113.4 &  3.0 & 2.58  & 6.6/7.9 & & C \\
G32.744$-$00.076$^{c2}$ & 18 51 21.8671& $-$00 12 05.3  & 33.4  & 11.1 & 10.5  & 2.3/11.8 & 11.8$^d$ & (P)A,L\\
G34.245$+$00.134 & 18 53 21.45175 & $+$01 13 46.0317 & 55.0  &  8.2 & 1.4   & 3.3/10.4 & 3.3$^h$&C\\
G34.258$+$00.153 & 18 53 18.64476 & $+$01 15 00.4115 & 57.6  & 1.8  & 5.9   & 3.4/10.3 & 3.4$^h$&C\\
G34.396$+$00.222 & 18 53 19.09345 & $+$01 24 13.862  & 55.6  & 7.8  & 5.2   & 3.3/10.4 & 3.6$^i$&C\\
G35.025$+$00.350 & 18 54 00.65696 & $+$02 01 19.3260 & 43.9  & 5.5  & 5.3   & 2.7/10.9 & 2.32$^j$&L\\
\hline
\end{tabular}
\tablefoot{$^a$Names are the Galactic coordinates derived from the EVN results. $^b$ Class of morphology as described 
in Bartkiewicz et al.~(\cite{b09}): \\L -- linear, R -- ring, C -- complex, A -- arched, P -- pair. \\
Coordinates with less accuracy (see Sect.~2.3): $^{c1}$ from the BeSSeL survey (EVLA) by Hu et al. ({\it in prep.}) 
and $^{c2}$ estimated from uncleaned EVN images with uncertainties of $\Delta$RA=0\fs0005 and $\Delta$Dec=1\farcs7.\\
Distances derived from trigonometric parallaxes by $^e$ Brunthaler et al.~(\cite{bru09}), $^j$ Wu et al.~(\cite{w14}). 
Kinematic distance ambiguity resolved from: \\$^d$ Green \& McClure-Griffiths (\cite{green11}), $^f$ Moscadelli et al.~(\cite{m07}), 
$^g$ Hill et al.~(\cite{h05}), $^h$ Kuchar \& Bania~(\cite{k94}), $^i$ He et al.~(\cite{he12}). In the remaining\\
targets the near kinematic distances were assumed.}
\label{results}
\end{table*}

{\it G24.850$+$00.087} \hspace{0.25cm}
In this source 73 maser spots are spread over a region of 110~mas$\times$90~mas that corresponds 
to 880~au$\times$720~au for a distance of 8~kpc. The maser emission was detected in the velocity range 
from 107.7 to 115.6~km~s$^{-1}$. The blueshifted emission, i.e., at velocities lower than $\sim$111~km~s$^{-1}$), 
is concentrated in the north-east, while the redshifted emission is found in the south-west (Fig.~\ref{distribapp}). Becker et al.~(\cite{becker94}) detected a 5~GHz radio continuum souce lying approximately 2\farcs9 westward from the masers. The H~{\small II} region is also listed in the JVLA sample (Hu et al.~{\it in prep.}).

{\it G28.011$-$00.426} \hspace{0.25cm}
This source contains 61 maser spots in the velocity range from 15.8 to 28.5~km~s$^{-1}$ (Fig.~\ref{distribapp}). 
The blueshifted emission (15.8--18.0~km~s$^{-1}$) is spread over a region of 100~mas$\times$100~mas, 
while the redshifted spots (23.3--28.5~km~s$^{-1}$) are concentrated in the eastern group of size 20~mas$\times$20~mas. 

{\it G29.978$-$00.048} \hspace{0.25cm}
157 spots from the LSR velocity range from 96.7 to 105.6~km~s$^{-1}$ were detected (Fig.~\ref{distribapp}). 
The emission of complex morphology is distributed over a region of 150~mas$\times$150~mas that corresponds 
to 1320~au$\times$1320~au for the far kinematic distance of 8.8\,kpc. We note that the blueshifted emission
is located in the northern elongated structure, whereas the redshifted emission forms the extended structure located in the south.

{\it G30.198$-$00.169} \hspace{0.25cm}
The emission of a total of 107 spots occurs in the velocity range of 100.6 $-$ 111.0~km~s$^{-1}$.
The redshifted and brightest emission comes from a small region located in the north-east (Fig.~\ref{distribapp}). Here, 51 spots of blueshifted and weak ($\sim$0.1--1~Jy~beam$^{-1}$) emission are distributed over 250$\times$100~mas area in the south-west. 
Our coordinates differ by about 3" in declination to those obtained by Xu et al.~(\cite{xu09}) using ATCA, but are consistent with Breen et al.~(\cite{breen15}). 

{\it G30.224$-$00.180} \hspace{0.25cm}
We detected 43 maser spots covering a velocity range from 111.0~km~s$^{-1}$ to 114.0~km~s$^{-1}$ (Fig.~\ref{distribapp}). This source, the most compact one in the angular size among the sample presented in this paper, is distributed over a region of 50~mas$\times$20~mas, which corresponds to 330~au$\times$132~au. The emission was found according to Breen et al.~(\cite{breen15}) coordinates, using data for G30.21$-$00.17 with a significant shift (78" in RA and 35\farcs6 in Dec) (Sect.~2.3).

{\it G32.744$-$00.076} \hspace{0.25cm} 
Two major groups of maser spots are separated by $\sim$1" from each other (Fig.~\ref{distribapp}). This angular separation corresponds to 11800~au for the far kinematic distance of 11.8\,kpc, as derived by H~{\small I} self-absorption method (Green \& McClure-Griffiths~\cite{green11}). In this case, these two groups of masers are unlikely to be physically related. For the near kinematic distance of 2.3~kpc, the separation would be 2300~au. The redshifted emission in the velocity range of 36.2$-$39.4~km~s$^{-1}$ emerges from the northern elongated structure of $\sim$160~mas length. The blueshifted emission  
(28.3$-$34.5~km~s$^{-1}$) in the south forms a complex structure of size $\sim$200$\times$300\,mas.
The OH 1665~MHz emission shows a similar morphology and velocity range (Argon et al.~\cite{a00}).
Caswell~(\cite{c01}) reports the OH 6035~MHz emission and there may also be a water maser emission 
in the region (Caswell et al.~\cite{c83}). However, an uncertainty of the absolute position in declination is large and we are not able to verify the detailed association.

{\it G34.245$+$00.134} \hspace{0.25cm} 
The source contains 86 spots in the velocity range from 54.5 to 62.7~km~s$^{-1}$, distributed over a region of 
150~mas$\times$150~mas (Fig.~\ref{distribapp}). The blue- ($<$58.5~km~s$^{-1}$) and redshifted ($>$58.5~km~s$^{-1}$) emissions
are located in the west and east, respectively. Caswell et al. (\cite{c95}) reported weak 12.2~GHz 
methanol emission at the velocity of the blueshifted 6.7~GHz emission. 
The methanol masers coincide with a submm source that was identified with a deeply embedded proto-B star
(Hunter et al.~\cite{h98}). The EVN images from 1999 showed a similar complex structure, but with a much 
stronger blueshifted emission (Yi et al.~\cite{yi02}).

{\it G34.258$+$00.153} \hspace{0.25cm}
This source lies 1\farcm4 to the north-west from G34.245$+$00.134 and the emission from the narrow range from
56.5 to 58.3~km~s$^{-1}$ is 2--3 times stronger. Six of the most blueshifted spots lie to the east, 
while the remaining 19 spots are grouped to the west (Fig.~\ref{distribapp}). Both groups are separated by $\sim$150~mas that
corresponds to 510~au at a distance of 3.4~kpc. They are also in agreement with earlier EVN images (Yi et al.~\cite{yi02}).
This site is known as the prototypical cometary ultra-compact (UC) H~{\small II} region (e.g.~Garay et al.~\cite{g86}). 
The methanol masers lie close to the west side of the B component that meets the criteria of
HC H~{\small II} region (Garay et al.~\cite{g86}; Avalos et al.~\cite{a09}; Sewi{\l}o et al.~\cite{s11}).
Caswell et al.~(\cite{c95}) detected the 12.2~GHz methanol maser emission and absorption in
the same velocity range as that of the 6.7\,GHz maser. A weak (0.6~Jy) OH emission found at 6035~MHz 
(Caswell~\cite{c01}) coincides within 0\farcs5 with the 6.7\,GHz maser. 
Water masers were also detected at velocities from 50 to 65~km~s$^{-1}$ but they are offset by $\sim$1" 
from the methanol masers to the south (Hofner et al.~\cite{h96}). OH 1665~MHz spots are closer;  
they coincide with the peak or lie on the east side of the HC H~{\small II} 
component B lying $\sim$0\farcs2-1" from the blueshifted methanol maser spots Zheng et al.~(\cite{z00}).

{\it G34.396$+$00.222} \hspace{0.25cm}
Methanol emission appears in 69 spots between 55.3 to 63.1~km~s$^{-1}$ (Fig.~\ref{distribapp}). 
There are two groups of masers that are separated by $\sim$130~mas, which corresponds to 470~au at the distance of 3.6\,kpc. 
The maser lies at the
southern edge of a mm source that may contain massive B0.5 protostar (Rathborne et al.~\cite{r05}). 
The source belongs to the complex region which contains more than 20 cluster members (e.g., Shepherd et al.~\cite{sh07}). 

{\it G35.025$+$00.350}  \hspace{0.25cm}
The source contains 79 maser spots within the velocity range from 41.3 to 46.8~km~s$^{-1}$ (Fig.~\ref{distribapp}).
The methanol emission comes from a linear structure of a size of $\sim$450~mas (1040~au) aligned along the position angle 
of $-$30\degr \,(north to east). No velocity gradient is seen over the whole 
maser region. Pandian et al.~(\cite{p11}) reported a similar structure within the same velocities 
but their MERLIN spectrum is 2-3 times brighter. Their coordinates agree to within 0\farcs1 with ours. 
Wu et al.~(\cite{w14}) imaged the 22~GHz water masers in the same region and found two clusters:  
a blueshifted one at SE with bright spots near the velocity of 45~km~s$^{-1}$ and a redshifted 
cluster near 56~km~s$^{-1}$ in the north-west. The proper motions of water masers suggest a bipolar outflow 
with a size of 650~au and a speed of 25~km~s$^{-1}$. The methanol spots are spread over the region 
between these two water maser clusters. The OH 1665~MHz maser in the velocity range 40.4--51.4~km~s$^{-1}$
(Argon et al.~\cite{a00}) coincides within $\sim$1\farcs3 in declination with the methanol maser.
The OH 6035~MHz maser in this region is strong and a single feature near 45.4~km~s$^{-1}$ (Caswell~\cite{c01})
coincides within 0\farcs1 with the methanol maser.

\section{Discussion}
In this section we discuss and summarise the results of imaging the large sample of 6.7~GHz methanol maser lines, their structures, and milliarcsecond details. We test the kinematic models available in the literature and search for radio continuum counterparts. Using single-dish data, we can also study the missing flux of the interferometer maps.

\subsection{VLBI morphology}
Here, we classify the maser emission that was obtained according to the morphologies presented in Bartkiewicz et al.~(\cite{b09}): single, linear, curved, complex, double, ring-like 
(Table~\ref{results}). We agree with Pandian et al.~(\cite{p11}) that, in many cases, the VLBI 
resolves a significant fraction of the maser emission and it may happen that some sources may have been  classified incorrectly because of this. 
However, as we discuss in Sect.~4.6, weak and diffuse maser emission that is filtered out is likely from halos of maser spots in  majority of cases and the overall emission extent is also seen on the VLBI images.

In conclusion, 12 sources are classified as complex (e.g.~G20.237$+$00.065). 
In these targets we do not note any regularities in spatial distribution nor overall velocity gradients. 
Two sources, G24.790$+$00.083 and G32.744$-$00.076, can be classified as pairs (angular separation of $\sim$1") 
although the velocity separation between two groups of masers are less than 10~km~s$^{-1}$ 
as originally defined in this class of morphology by Phillips et al.~(\cite{p98}). Taking substructures of every source into consideration, 
we note that in G24.790$+$00.083 two groups of methanol maser show a ring-like morphology, 
while in G32.744$-$00.076 we see the arc and linear distributions in the south and north, respectively. The remaining sources either show a ring-like emission (G22.435$-$00.169) or a linear one (G20.237$+$00.065 and G35.025$+$00.350). 

Taking the complete sample of 63 targets, we can summarise that by using the EVN we derived the following morphologies: simple in one source, linear in 13 sources, ring-like in 11, arched in five, pair in four cases, and the remaining 29 showed a complex morphology. 

In general it is difficult to relate the 6.7~GHz maser morphology at the mas scale to the specific physical phenomena and pinpoint the position of the MYSO. Even the ring-like sources do not bring a direct implication of where MYSOs are (De Buizer et al.~\cite{d12}). Proper motion studies are a way of verifying what is going on in these HMSFRs combined with high-angular studies at other wavelengths like infra-red or radio continuum (e.g. Moscadelli et al.~\cite{m07}; Sanna et al.~\cite{s10}). 

\subsection{Maser clouds}
In total we extracted 201 maser clouds with Gaussian velocity profiles towards all 17 targets (Table~\ref{clouds}). Among them, 33 showed double (e.g., Clouds {\it 8} and {\it 12} in G20.237$+$00.065) and four (e.g., Cloud {\it 2} in G32.744$-$00.076) showed triple Gaussian profile complexity. In total, we obtained 242 Gaussian velocity profile components. Their projected lengths range from 1~au to 117.7~au with a mean value of 17.4$\pm$1.2~au and a median of 5.5~au. The longest maser cloud comes from the target G24.790$+$00.083. The FWHM ranged from 0.06~km~s$^{-1}$ to 1.20~km~s$^{-1}$ with an averaged value of 0.373$\pm$0.011~km~s$^{-1}$, and a median of 0.315~km~s$^{-1}$. 
171 of the maser clouds (i.e. 71\%) showed a velocity gradient with a magnitude between 0.005 and 0.210~km~s$^{-1}$~au$^{-1}$ with a median of 0.039~km~s$^{-1}$~au$^{-1}$. Two clouds showed higher values of 
0.434~km~s$^{-1}$~au$^{-1}$ (Cloud {\it 1} in G30.198$-$00.169) and 0.529 (Cloud {\it 6} in G35.025$+$00.350), respectively. All these values correspond nicely with the previous sample that was presented earlier in Bartkiewicz et al.~(\cite{b14}).

\begin{figure}
\centering
\includegraphics[width=10cm]{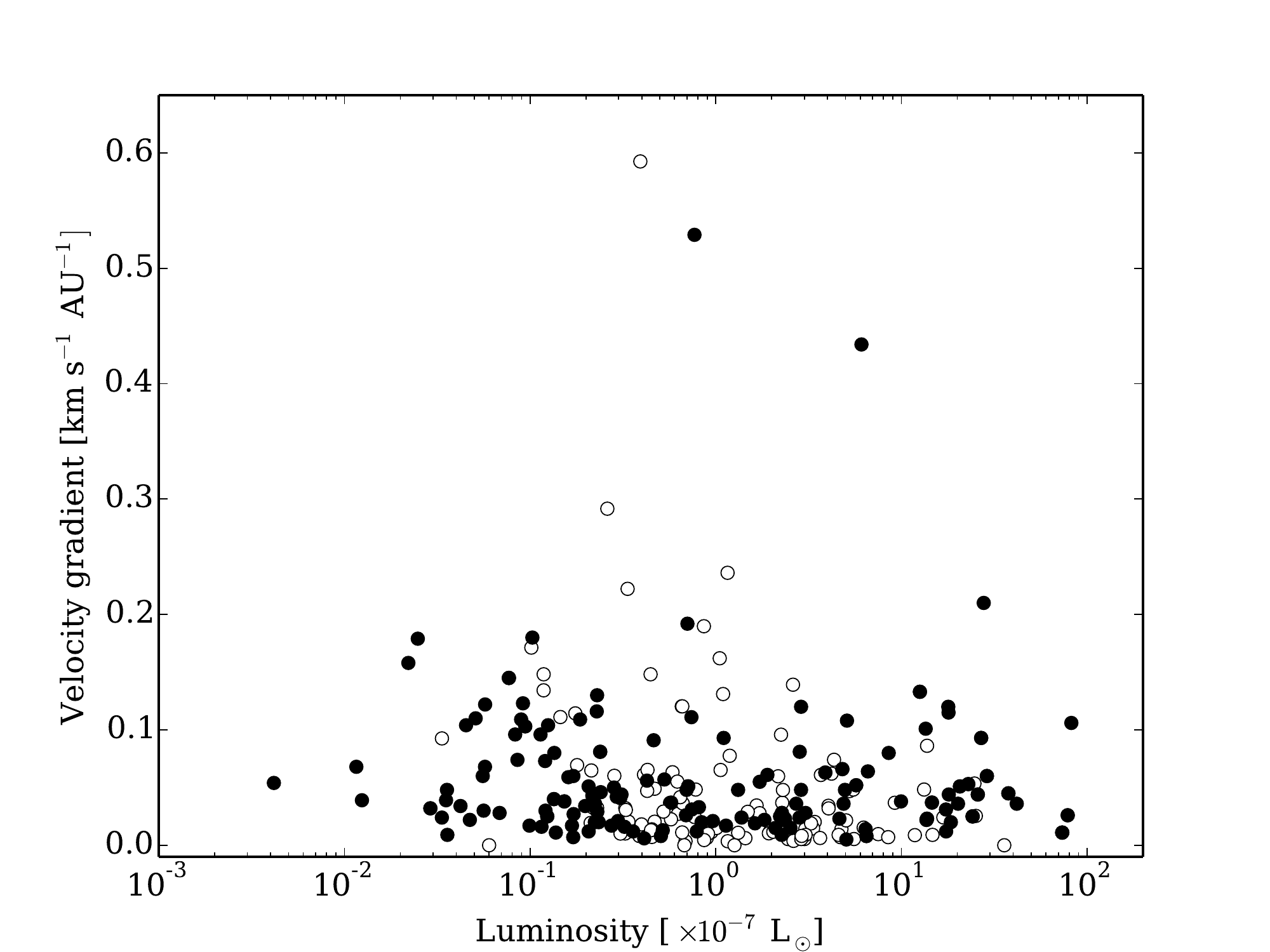}
\caption{The luminosity of a single maser cloud vs. its velocity gradient. The black dots represent data from this work, while the open circles refer to the data from Bartkiewicz et al.~(\cite{b14}).}   
\label{grad_lum}
\end{figure}

In Fig.~\ref{grad_lum} we summarise the relationship between the luminosity of a single methanol maser cloud and its velocity gradient. The luminosity was calculated according to the formula: $L_{6.7GHz}[L_{\sun}] = 6.9129 \times 10^{-9} D^{2}[\rm kpc] S_{int}[\rm Jy~km~s^{-1}]$. The values range is (0.012 -- 73.4)$\times 10^{-7} L_\odot$ with a mean of (7.9$\pm1.2) \times 10^{-7} L_\odot$. In addition, we plotted the data from Bartkiewicz et al.~(\cite{b14}) (Fig.~\ref{grad_lum}). They are all consistent: we do not notice any trends; low and high luminous clouds have similar gradients.

\subsubsection{'Amplification-bounded' maser}
We note two interesting cases among all 201 maser clouds with the Gaussian velocity profiles:  Clouds {\it 1} in G22.435$-$00.169 and {\it 5} in G24.33$+$00.12 (Fig.~\ref{gauss}) in which the double Gaussian fittings give two lines that have significantly different FWHMs with ratios of the wider to narrower line of 5 and 7, respectively. The peak velocities of the narrow and wide features are slightly different. In Fig.~\ref{cloud1}, we present the four brightest spots from the wider and narrower velocity profile components of G22.435$-$00.169. The subtraction of different spots shows that the 'emission excess' (at 39.9~km~s$^{-1}$) is located in the same position as the emission at remaining velocities (Fig.~\ref{cloud1}). However, the signal-to-noise ratio at the relevant image is only 4, in which case we cannot be conclusive. 
In G24.33$+$00.12, however, the subtracted images show the signal-to-noise ratio above 10 and it is clear that both components come from the same region within the beamsize (Fig.~\ref{cloud2}). 

We verified a relationship between an observed component's size and intensity using the method described in Richards et al.~(\cite{r11}) to distinguish between 'amplification-bounded' and 'matter-bounded' masers. These terms were defined by Elitzur et al.~(\cite{e92}) concerning the maser geometry and a beaming type. From any given direction a spherical maser appears identical to a cylindrical maser aligned along the line of sight. However, in the spherical maser the observed size is significantly smaller than the physical size as a result of the beaming, i.e. the strong the amplification, the smaller the observed area. This is referred to as an 'amplification-bounded' maser. In the cylindrical maser, the projected physical size is equal to the observed one and the beaming is controlled by the matter distribution, i.e. 'matter-bounded' maser. 
We plot a deconvolved maser spot size vs. natural logarithm of its peak flux density, with both values derived using the JMFIT procedure in AIPS (Fig.~\ref{cloud2}). The estimated slope is: $\alpha=-1.4\pm0.6$ and 
this suggests the existence of 'amplification-bounded' maser (Richards et al.~\cite{r11}). The gain decreases with frequency shift away from the line centre, the appearance of a spherical maser varies across the line profile; the observed area is smallest at line centre, increasing towards the line wings. Moreover, a slope steeper than $-$0.5 suggests a partially saturated maser. We note that the emission in G24.33$+$00.12 is resolved out up to 50\%, comparing to that detected using the single-dish (see Sect.~4.6) and consequently the larger, fainter components are most severely affected. However, both the apparent spot size and flux density are reduced, so the 'net effect' is just to increase the uncertainty of the faintest components. 

To the best of our knowledge these are the first cases of very narrow ($<$0.1~km~s$^{-1}$) features at 6.7\,GHz which might be interpreted as 
'amplification-bounded' masers. However, we note that the weakness of this interpretation is due to limited data in the fitting procedure; in fact the narrow Gaussian component is based on only one point (Figs~\ref{cloud1} and \ref{cloud2}). Future VLBI studies with higher spectral resolution may help to examine this phenomenon. Moreover, the analysis of a large sample of maser clouds for the all 63 sources may bring more cases for verification, as presented above.

\subsection{Kinematics models}
Below, we test two kinematic models that exist in the literature and that can be applied to one epoch data of the presented sources in this work as a first approach; 
a rotating and expanding disk (Uscanga et al.~\cite{u08}) and
a bipolar outflow (Moscadelli et al.~\cite{mos00}, \cite{m05}). A detailed description of these models and fitting procedures is also given in Bartkiewicz et al.~(\cite{b09}).

The ring-like methanol maser emission in G22.435$-$00.169 
suggests it is physically associated with a rotating torus around a central object, probably a MYSO. Torstensson et al.~(\cite{t11}) report a similar elliptical structure for  
6.7~GHz methanol masers in the nearby HMSFR Cep~A. They conclude that methanol maser  emission probably occured between the infalling gas 
and the accretion disk since the model of a rotating and expanding disk did not give a reasonable fit. When we apply the model of Uscanga et al.~(\cite{u08}) to maser spots in G22.435$-$00.169 (excluding the blueshifted group 
of spots at the LSR velocities from 28.6 to 29.0~km~s$^{-1}$) we also do not find a convincing solution. We tested the model for a range of the rotation and radial (an infall or an expansion) velocities from $-$20 to 20~km~s$^{-1}$ with
a step of 0.2~km~s$^{-1}$. The inconsistency with a rotating and expanding disk is clearly seen in the spot distribution (Fig.~\ref{distrib}), where the most blue- and redshifted spots are not located at the extreme positions, as would be expected for an inclined disk. We also note that without any additional information we cannot determine the sign of the inclination angle of the disk and the direction of the rotation as well as the expansion and infall are ambigous.

\begin{figure}
\centering
\includegraphics[scale=0.60]{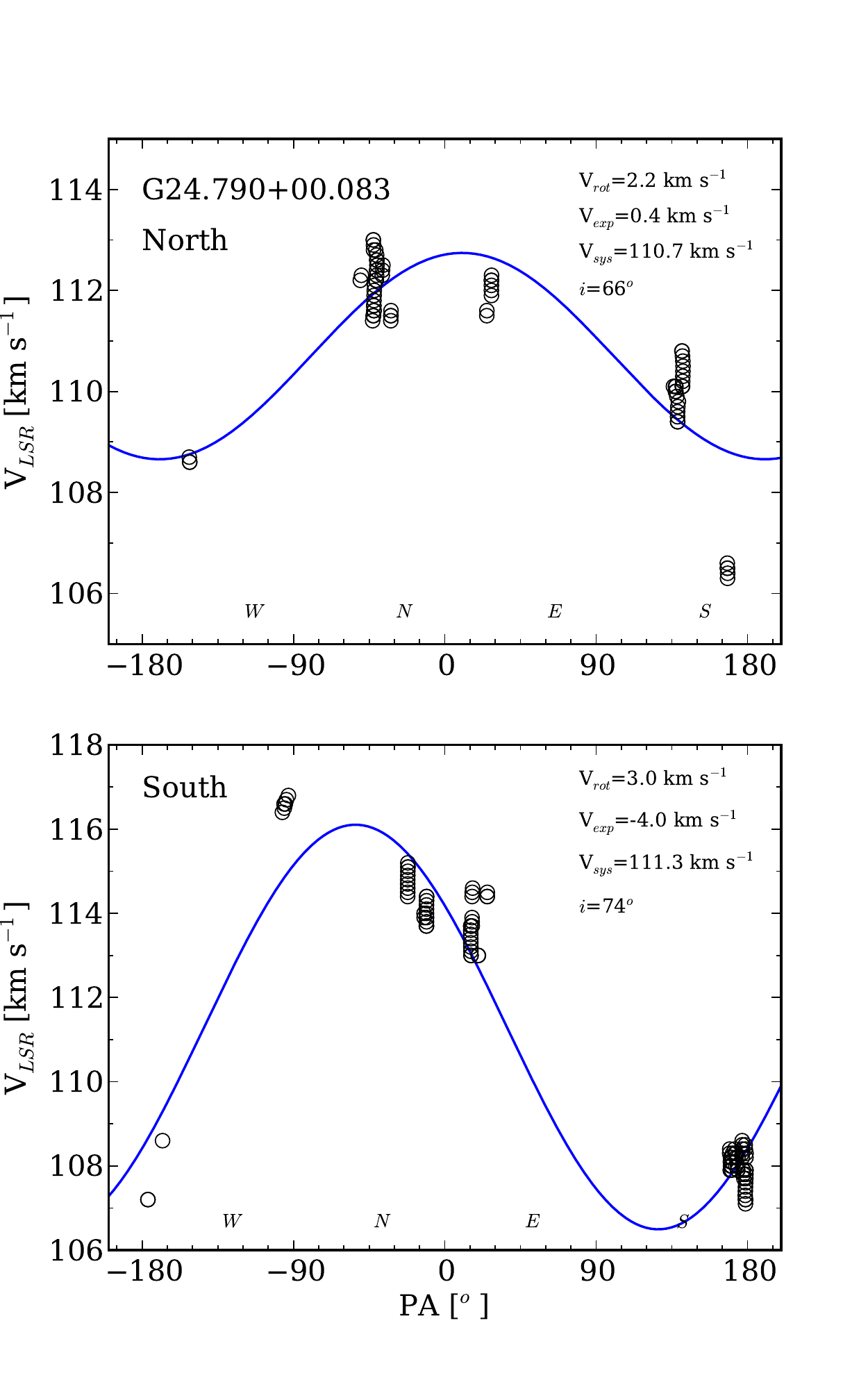}
\caption{Velocity of the maser spots in G24.790$+$00.083 north (top) and south (bottom) versus
 azimuth angle measured from the major axis (north to east). Circles represent the data, while 
the lines represent the best kinematical models of rotating and expanding disks with infintesimal 
thickness, as described in Sect.~4.3. The main parameters of the fit are listed in the figures. 
The N, E, S, and W directions are also indicated for clarity.}   
\label{disk}
\end{figure}

The rings in G24.790$+$00.083 are described in Moscadelli et al.~(\cite{m07}). The 6.7~GHz methanol emission coincides with the mm subcores A1 and A2. The variations of LSR velocities across the masers agree with the velocity gradient that was observed using CH$_3$CN (12-11) line, with redshifted ($\sim$112~km~s$^{-1}$) to the NE and the blueshifted (110.5~km~s$^{-1}$) to the SW. Moscadelli et al.~(\cite{m07}) point out that the velocity characteristics of methanol masers (with a three times larger spread of LSR velocity across a three times smaller region, compared to the thermal region) was consistent with the conservation of angular momentum and a rotation model of the cores was supported. With our new measurements with their better spectral resolution, we obtain good fits for the rotating and expanding/infalling disks for both maser rings. 
The northern group is fitted well with 2.2~km~s$^{-1}$,  0.4~km~s$^{-1}$, 110.7~km~s$^{-1}$ for 
the velocities of rotation, expansion, and the systemic one, respectively, with an inclination of 
a thin disk of 66\degr~\, ($\chi^2_{\nu}=102$). The southern group is also reproduced nicely with 
the following parameters: 3~km~s$^{-1}$,  $-$4~km~s$^{-1}$, 111.3~km~s$^{-1}$, 
respectively with $i$=74\degr~\,($\chi^2_{\nu}=94.5$). The fits of the model to the data are presented in Fig.~\ref{disk}. 

The source G32.744$-$00.076 also consists of two groups of masers separated by $\sim$1". However,  
the northern group is linear, while the southern one is arched. 
The fit of the rotating and expanding disk to the southern part gives the following parameters: 
velocity of rotation of $-$1.6~km~s$^{-1}$, velocity of expansion of 0.8~km~s$^{-1}$, 
systemic velocity of 31.1~km~s$^{-1}$. However, the best fit still has a high value of $\chi^2_{\nu}$ that equals 323. 
The northern linear structure, assuming that it represents an edge-on disk of a
size of 1830~au (155~mas at 11.8~kpc), with a velocity gradient of 1.7~m~s$^{-1}$~au$^{-1}$ 
would correspond to a mean central mass of 20~M$_\odot$ using the method of Minier et al.~(\cite{m00}). The available images of 
radio continuum in that region show H~{\small II} regions with peaks separated by $\sim$0\farcs8 
and surrounded by 1.6~GHz OH masers (Argon et al.~\cite{a00}). 
The distribution and morphology of 6.7~GHz methanol masers support the existence of at least two centres of star formation. 

\begin{figure*}
\centering
\includegraphics[scale=0.60]{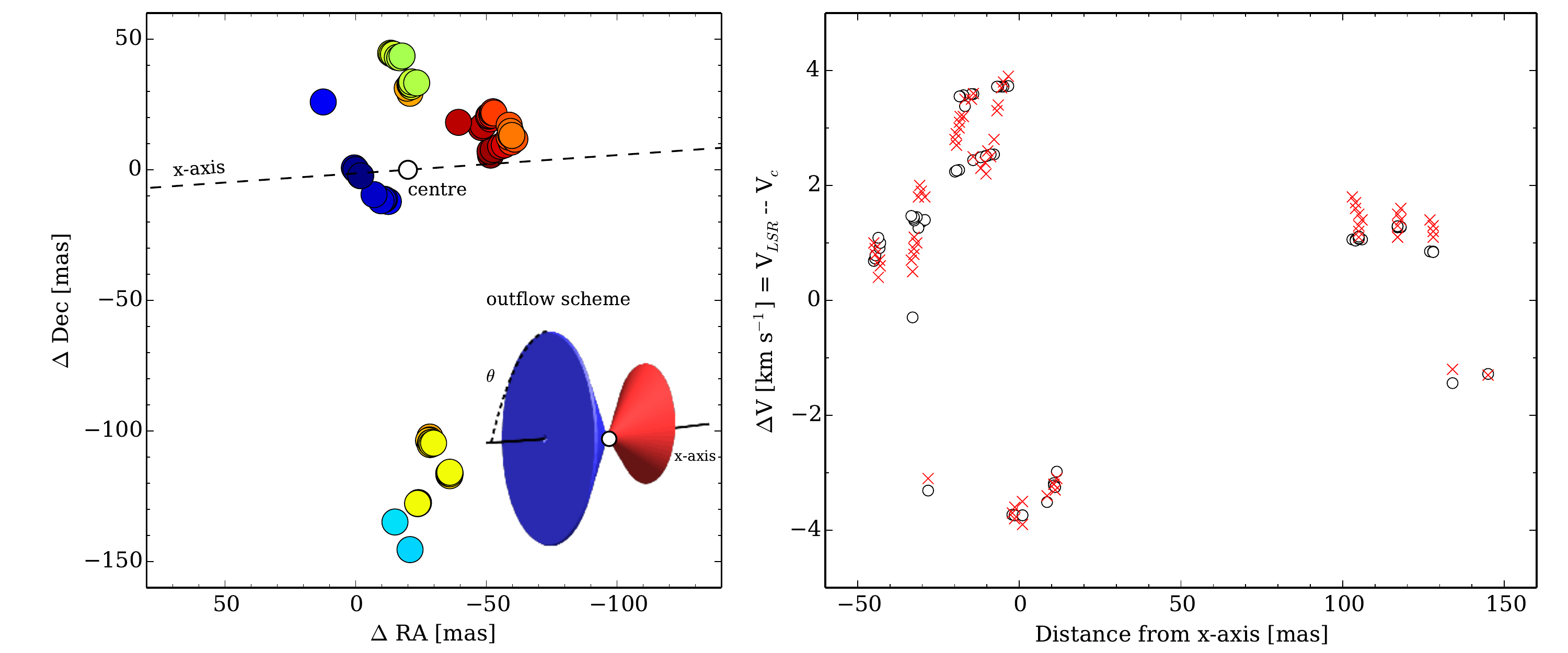}
\caption{Outflow model fitted to 6.7~GHz methanol masers  
     in G34.396$+$00.222 according to the model of Moscadelli et al.~(\cite{m00}). 
     The relevant parameters are listed in Table~\ref{outfits}. The dashed line and the white circle in the left panel trace the outflow axis (x-axis) and the powering source (the cone vertex). The jet is aligned to the observer at the angle of 41$^{\rm o}$ and the opening angle of the outflow is 162$^{\rm o}$. Because of a projection effect the outflow covers the whole area, and is therefore not marked and the scheme for clarity is presented. The right panel presents a comparison of obtained data (an x symbol) vs. model (open circles). V$_{\rm c}$ is the systemic LSR velocity as given in Sect.~4.3.}   
\label{outflow}
\end{figure*}

Below, we summarise the results of fitting the bipolar outflow model by Moscadelli et al.~(\cite{mos00}, \cite{m05}). The systemic velocity (V$_{\rm c}$) was assumed as a central velocity of the 6.7~GHz maser emission. We obtained reasonable fits for nine sources. These are summarised in Table~\ref{outfits}, the vertex of the cone ($\Delta$RA, $\Delta$Dec, both sampled with a step of 10~mas within the range of the whole visible emission, roughly corresponding to the size of the presented images in Figs.~\ref{distrib} and \ref{distribapp}), the position angle of the x-axis that coincides with the projection of the outflow on the plane of the sky (PA) (with a step of 5$^{\rm o}$), the inclination angle between the outflow axis and the line of sight (i.e. the z-axis), $\Psi$ (with a step of 10$^{\rm o}$), the opening angle of the outflow, 2$\theta$ (with a step of 5$^{\rm o}$) and the $\chi^2$ values as in Eq.~(3) of Moscadelli et al.~(\cite{mos00}). An example of the fitted outflow model is presented in Fig.~\ref{outflow}. We note that in all results the opening angle of the fitted outflow is large, 2$\theta$ above 112$^{\rm o}$ and, in fact, the outflow directed to the observer covers the whole area. This could be a characteristic of the outflows that are related to the current methanol maser sample or could indicate the weakness of the model.

Among these nine sources, there are five towards which de Villiers et al.~(\cite{dev14}) imaged the CO outflows. We do not find consistency between the alignment of $^{\rm 13}$CO red- and blueshifted wings (with arcsecond resolution) and fitted projections of the outflows to methanol maser spot distributions. Moreover, they are usually perpendicular to each other. This shows the need for high-angular resolution studies using thermal lines to verify if the mas structure of methanol masers is associated with the large-scale outflows.

\begin{table}
 \centering
  \caption{Parameters derived from fitting the model of bipolar outflow by Moscadelli et al.~(\cite{mos00}).}
  \begin{tabular}{lcllll}
  \hline
   Source & $\Delta$RA, $\Delta$Dec$^\star$ & PA & $\Psi$ & 2$\theta$ & $\chi^2$\\ 
   Gll.lll$\pm$bb.bbb & (mas, mas) &($^{\rm o}$) &($^{\rm o}$) &($^{\rm o}$) & \\
 \hline
G22.435$-$00.169 & 10, $-$90 & 44 & 41 & 112 & 1.75\\
G24.33$+$00.14 & 130, $-$50 & 49 & 31 & 154 & 0.08 \\
G24.494$-$00.038 & 10, $-$10 & $-$41 & 21 & 172 & 1.08 \\
G24.790$+$00.083 & $-$620, 210 & 69 & 51 & 142 & 2.22\\
G24.850$+$00.087 & $-$80, $-$10 & 9 & 141 & 172 & 0.34\\
G28.011$-$00.426 & 80, $-$20 & $-$61 & 121 & 142 & 0.39 \\
G32.744$-$00.076 & $-$160, 850 & $-$66 & 11 & 172 & 1.83\\
G34.258$+$00.153 & $-$10, $-$40 & $-$61 & 51 & 162 & 0.35\\
G34.396$+$00.222 & $-$20, 0 & $-$86 & 41 & 162 & 0.67\\
\hline
\end{tabular}
\tablefoot{$^\star$ Coordinates relative to the brightest spot as listed in Table~\ref{results}. }
\label{outfits}
\end{table}

\subsection{Bipolar morphology}
The G23.010$-$00.411 source was studied in detail by Sanna et al.~(\cite{s10}) using the methanol and water maser lines and continuum emission. Their proper motion studies of 6.7~GHz methanol masers suggested a composition of expansion and rotation around a YSO of about 20~M$_\odot$ and physically associated with the internal parts of a molecular toroid. The disk scenario was also proposed by Polushkin \& Val'tts~(\cite{pol11}) after their analysis of the same 6.7~GHz methanol maser data set, one epoch only. We  note that the characteristic of this source is a clear separation on the sky of the blue- and redshifted emission with a spatial separation of 0\farcs35 (1600~au) and a separation in velocity of ca.5~km~s$^{-1}$ (Fig.~\ref{distrib}). We note that the following targets from our sample show similar behaviour (the spatial and in velocity separations between blue- and redshifted groups are listed in brackets): \\G24.33$+$00.12 (2300~au, 5~km~s$^{-1}$), G24.494$-$00.038 (1200~au, 5~km~s$^{-1}$ ), G24.850$+$00.087 (800~au, 5~km~s$^{-1}$), G28.011$-$00.426 (110~au, 8~km~s$^{-1}$), G29.978$-$00.048 (600~au, 5~km~s$^{-1}$), G30.198$-$00.169 (600~au, 4~km~s$^{-1}$), G32.744$-$00.076 (two groups of masers would be physically related for the near kinematic distance; 2300~au, 5~km~s$^{-1}$), G34.245+$00.134$ (170~au, 4~km~s$^{-1}$), G34.258$+$00.153 (500~au, 1~km~s$^{-1}$). 
One can obtain the mass of the central object with the above parameters, 
assuming a Keplerian rotating disk. For five targets (G24.33$+$00.12, G24.494$-$00.038, G24.850$+$00.087, G30.198$-$00.169, and G32.744$-$00.076) we calculate masses of 6--16M$_\odot$, which supports the toroid scenario and physical relation to the molecular toroids around YSOs. In the remaining sources the derived masses are below 2.7M$_\odot$ and are not convincing for HMSFRs. High-angular resolution data of continuum and/or infra-red emission are needed to verify this hypothesis and to confirm the existence of a YSO. 

\subsection{Radio continuum counterparts}
We search the CORNISH catalogue to find the radio emission from possible associated H~{\small II} regions (Hoare et al.~\cite{h12}, Purcell et al.~\cite{p13}). Near four targets there are sources identified as UC~H~{\small II} regions, near one more there is an object described as a dark H~{\small II} region. The details are summarised in Table~\ref{cornish}. In 12 cases there was no radio continuum emission within 30''. However, we note that for source G023.010$-$00.411 there is no counterpart listed in the catalogue, but the VLA search detected emission at the 1.3~cm and 3.6~cm (Sanna et al.~\cite{s10}).

\begin{table*}
 \centering
  \caption{Radio continuum emission counterparts from the CORNISH Catalogue.}
  \begin{tabular}{llllll}
  \hline
   Source &  CORNISH& Source & Peak & Angular & Separation\\
          & object & type & & size & \\
   Gll.lll$\pm$bb.bbb & Gll.lll$\pm$bb.bbb & & (mJy beam$^{-1}$) & ('') & ('')\\
 \hline
G24.494$-$00.038 & G024.4921$-$00.0386 & UC~H~{\small II} & 19.30 & 4.44 & 4.1\\
G24.790$+$00.083 & G024.7898$+$00.0833 & Dark H~{\small II} & 12.54 & 1.50 & 0.1\\
G24.850$+$00.087 & G024.8497$+$00.0881 & UC~H~{\small II} & 19.93 & 3.67 & 3.1\\
G34.258$+$00.153 & G034.2581$+$00.1533 & UC~H~{\small II} & 35.87 & 1.50 & 0.6\\
G35.025$+$00.350 & G035.0242$+$00.3502 & UC~H~{\small II} & 10.70 & 1.55 & 2.7\\
\hline
\end{tabular}
\label{cornish}
\end{table*}

\begin{figure}
\centering
\includegraphics[width=10cm]{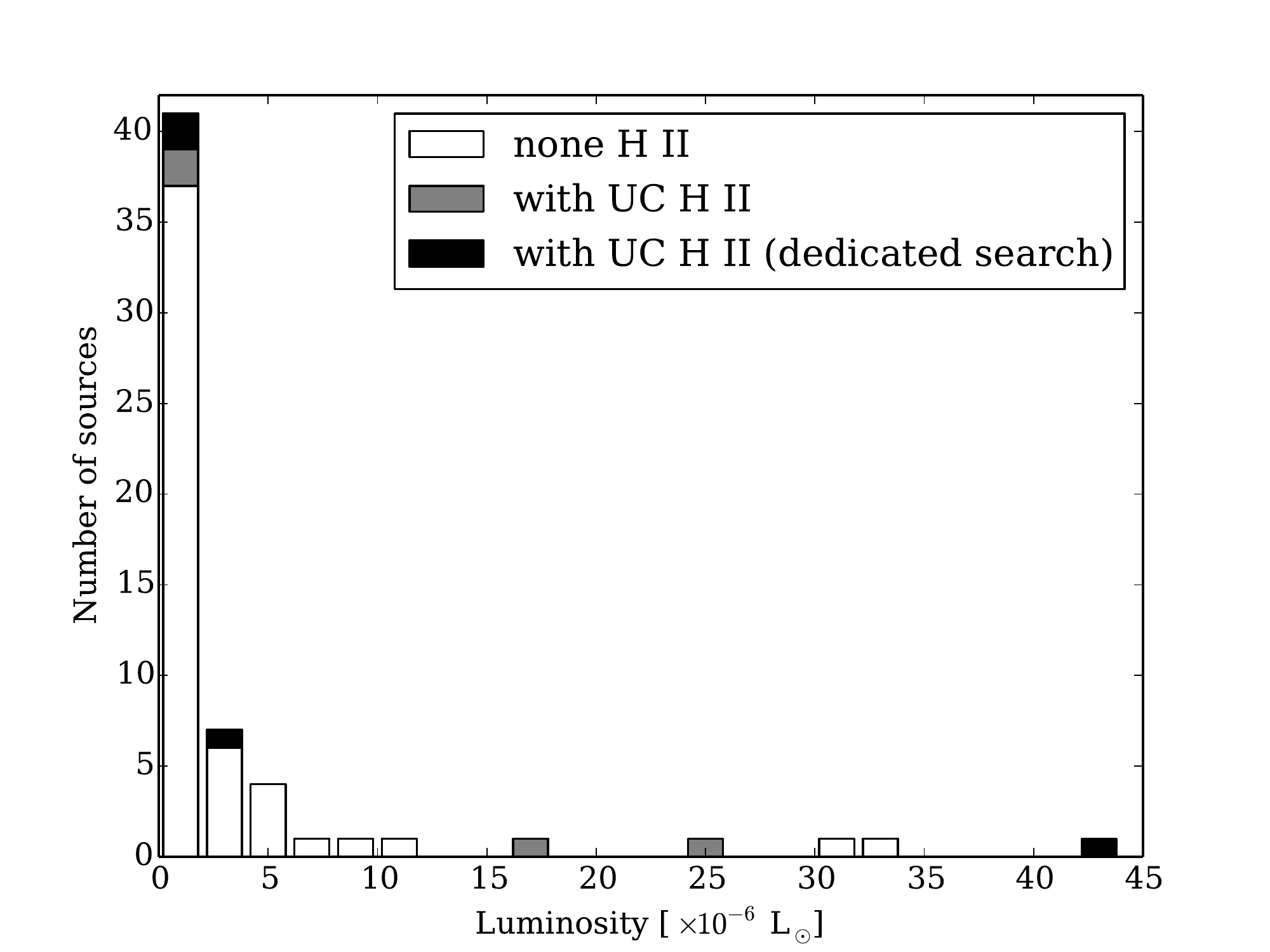}
\caption{Histogram of luminosities of 6.7~GHz methanol masers imaged using EVN. The masers with UC H~{\small II} counterparts are divided in two categories: the dedicated searches for the radio continuum emission were done (Bartkiewicz et al.~\cite{b09}; Sanna et al.~\cite{s10}) and search from the CORNISH catalogue (Hoare et al.~\cite{h12}, Purcell et al.~\cite{p13}).}   
\label{lum_hii}
\end{figure}

In the cases of G24.790$+$00.083 and G34.258$+$00.153, an association between the masers and UC H~{\small II} regions is very likely since the separations are 0\farcs1 and 0\farcs6, which correspond to 770~au and 2000~au, respectively. In a dedicated VLA search, three more masers were related to radio continuum emission among 31 sources from the first part of our sample; these are: 
G24.148$-$00.009, G28.817$+$00.365 and G36.115$+$00.552 (Bartkiewicz et al.~\cite{b09}).\footnote{For G26.598$-$00.024 the methanol masers appeared to be related to a younger object clearly visible at the infra-red wavelengths, while the radio continuum coincided with a more evolved young star (Bartkiewicz et al.~\cite{b10}).} Searching the CORNISH catalogue for H~{\small II} counterparts for  sources from Bartkiewicz et al.~(\cite{b14}) gives another two possible associations: 
G040.425$+$00.700 (separation of 0\farcs4 corresponding to 4600~au) and G043.149$+$00.013 (0\farcs3, 3300~au). In total, that provides eight (13\%) associations or possible associations between 6.7~GHz methanol masers and UC H~{\small II} regions among the sample of 63 sources. In Fig.~\ref{lum_hii} we present a histogram of the number of sources as a function of luminosity of 6.7~GHz methanol emission, as registered using the EVN. Four sources, G23.010$-$00.411, 24.790$+$00.083, 32.744$-$00.076, and  
G40.425$+$00.700 showed luminosity above 16$\times10^{-6}$L$_\odot$. Three of them are associated with radio continuum emission. This seems to agree with a suggestion of Breen et al.~(\cite{breen10,b11}) that "more luminous 6.7~GHz methanol masers are generally associated with a later evolutionary phase  of massive star formation than less luminous 6.7~GHz methanol maser sources". That still leaves five methanol maser sources with likely H~{\small II} counterparts among the two lowest ranges of luminosity, below 4$\times10^{-6}$L$_\odot$ (Fig.~\ref{lum_hii}). However, with this data set we are not able to convincingly test the dependence of the luminosity of the masers and the evolutionary phase of the central object.

\subsection{Missing flux}
We compared the methanol maser profiles presented in this work and the observations that were obtained using the 32~m Torun telescope with the dates as close as possible. The single-dish data are part of a monitoring programme described in Szymczak et al.~(\cite{sz12}). The calculated integral fluxes for both instruments are presented in Table~\ref{intfl}. The maser flux density obtained using the EVN 
(Sect.~2.3)
is typically at a level of 30--50\% of that detected by the single-dish. 
Two sources, G29.978$-$00.048 and G30.198$-$00.169, showed the missing fluxes at a level of only 25\% and 7\%, respectively. Three targets showed missing flux at the level of 81\% and 75\% (G24.33$+$00.14, G34.258$+$00.153,  and G34.396$+$00.222). We do not note any special characteristics of these targets to explain this behaviour. This was also noted in the EAVN studies by Fujisawa et al.~(\cite{f14}). 

In Fig.~\ref{intfl_fig}, we plot the ratios describing the missing fluxes and the distances. We consider the full sample, i.e. including data from Bartkiewicz et al.~(\cite{b09}, \cite{b14}), for which  the single-dish data were available. The interferometric data were recovered from consistently naturally--weighted image cubes in all cases. In total, 52 objects are plotted. The least-square fitting 
resulted in the following coefficients: $a=1.5\pm 1.1$, $b=37\pm 7$. We conclude that the missing flux does not strongly depend on the distance to the source. Comparison of the EVN data and images from shorter baseline interferometers for a few cases suggests that the missing flux in the EVN data was caused by resolving-out of the diffuse and weak maser emission that comes from the halos of maser spots.
 Source G23.010$-$00.411 was imaged using the VLA by Cyganowski et al.~(\cite{cyg09}). Both interferometers  show a similar extent of methanol maser emission, as well as the spot distribution. G35.025$+$00.350 was also imaged using the VLA (Cyganowski et al.~(\cite{cyg09}) and MERLIN (Pandian et al.~\cite{p11}). The same size and morphology, in particular with the MERLIN data, can clearly be seen. However the spectrum is twice as weak as that seen using EVN.

\begin{table}
 \centering
  \caption{Integral fluxes}
  \begin{tabular}{llll}
  \hline
   Source & S$_{\rm int(EVN)}$ &S$_{\rm int(32 m)}$ & Ratio  \\
   Gll.lll$\pm$bb.bbb & (Jy~km~s$^{-1}$) & (Jy~km~s$^{-1}$) &(per cent)\\
 \hline
G20.237$+$00.065$^a$ & 22.954 & 48.118 & 48\\  
G22.435$-$00.169 & 12.237 & 28.485 & 43\\
G23.010$-$00.411 &295.688 &749.472 & 39\\
G24.33$+$00.14   &  2.256 & 12.101 & 19\\
G24.494$-$00.038 &  5.882 & 13.237 & 45\\
G24.790$+$00.083$^b$ & 73.511 &139.927 & 52\\
G28.011$-$00.426 &  2.283 &  6.610 & 34\\
G29.978$-$00.048 & 58.515 & 77.631$^c$ & 75\\
G30.198$-$00.169 & 23.854 & 25.523 & 93\\
G30.224$-$00.180 & 4.094  &10.810 & 38 \\
G32.744$-$00.076 & 34.948 & 93.483 & 37\\
G34.245$+$00.134 &  3.030 &  9.053 & 33\\
G34.258$+$00.153 &  3.250 & 12.842 & 25\\
G34.396$+$00.222 & 4.358  & 17.104 & 25\\
G35.025$+$00.350 &  6.505 & 22.084 & 30\\
\hline
\end{tabular}
\tablefoot{$^a$ The sum spectra of G20.237$+$00.065 and G20.239$+$00.065. $^b$ The sum spectra of G24.790$+$00.083 and G24.850$+$00.087. $^c$ The integral flux taken from the V$_{\rm LSR}$ ranges 96.7--99.5~km~s$^{-1}$ and 101.7--106.8~km~s$^{-1}$. }
\label{intfl}
\end{table}

\begin{figure}
\centering
\includegraphics[width=10cm]{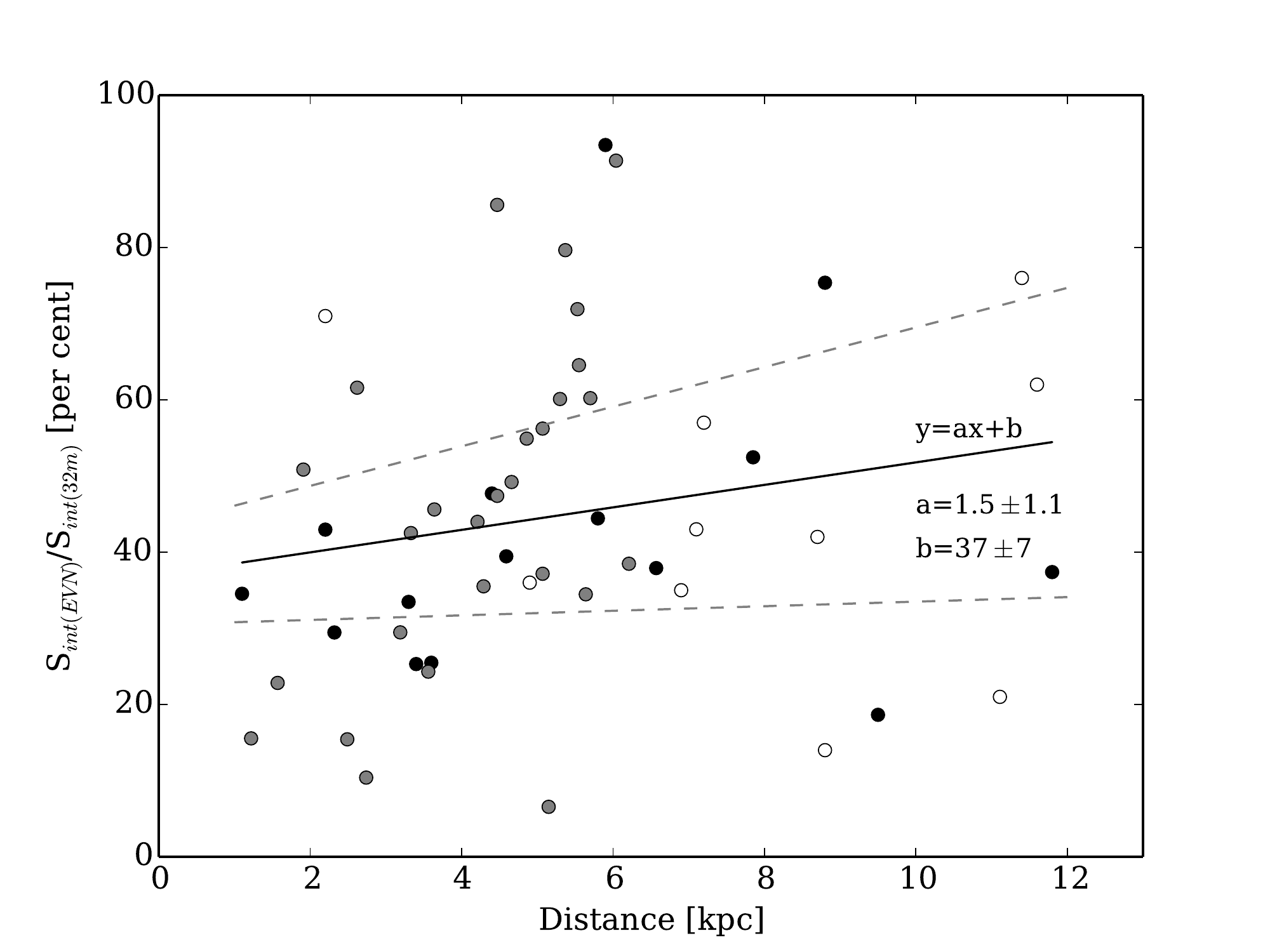}
\caption{Ratios of integral fluxes (obtained using EVN and 32~m dish) vs. distances to targets. The black dots represent data from this work, while the grey and open circles to the data from Bartkiewicz et al.~(\cite{b09}, \cite{b14}), respectively. The least-square fit to all points is presented by the black line and the dashed grey lines limit the region due to the uncertainties of the fit.}   
\label{intfl_fig}
\end{figure}

\section{Conclusions}
We successfully imaged the 6.7~GHz emission towards 17 targets and in 15 cases absolute positions with mas accuracy are given. We studied the characteristics of the velocity profiles of maser clouds. In two clouds out of 201, we found a narrow component imposed onto the wider one. We interpret this as the likely existence of an 'amplification-bounded' maser, but further studies are needed to confirm these type of phenomena. Considering the complete sample of 63 methanol maser sources that were imaged successively using the EVN, we find that one source showed the simple morphology (1.6\%), 13 sources a linear one (21\%), eleven showed ring-like (18\%), five showed arched (8\%), four showed paired (6.4\%) and 29 showed complex morphology (45\%). In the majority of sources, the missing flux is at the level of 30--50\%. We do not find any clear dependence on a missing flux in a function of a distance to the source.

\begin{acknowledgements}
We thank Dr.~hab.~Krzysztof Katarzy\'nski from TCfA for encouragment and help with Python programming and Dr.~Anita Richards from JBCA for the discussion about amplification-bounded masers. We are grateful to Prof.~Gary Fuller from JBCA for sharing the MERLIN results. We also thank the referee, Prof.~Simon Ellingsen, for detailed reading and constructive  comments on this publication. 
AB and MS acknowledge support from the National Science Centre Poland through grant 2011/03/B/ST9/00627. This work has also been supported by the European Community
Framework Programme 7, Advanced Radio Astronomy in Europe, grant agreement
No.~227290. 
\end{acknowledgements}


\Online
\begin{appendix}
\section{Figures}
\begin{figure*}
\centering
\includegraphics[scale=0.6]{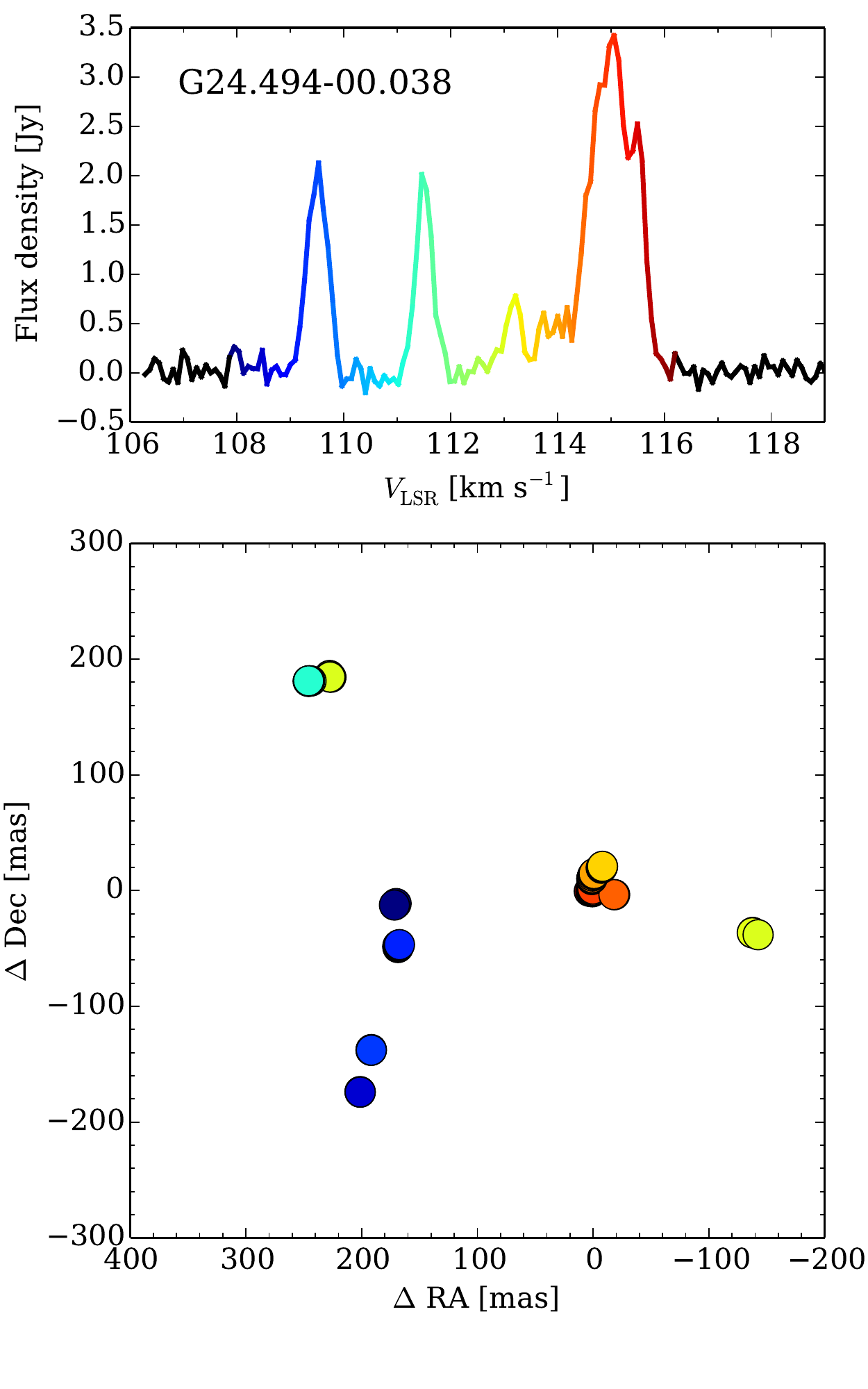}
\includegraphics[scale=0.6]{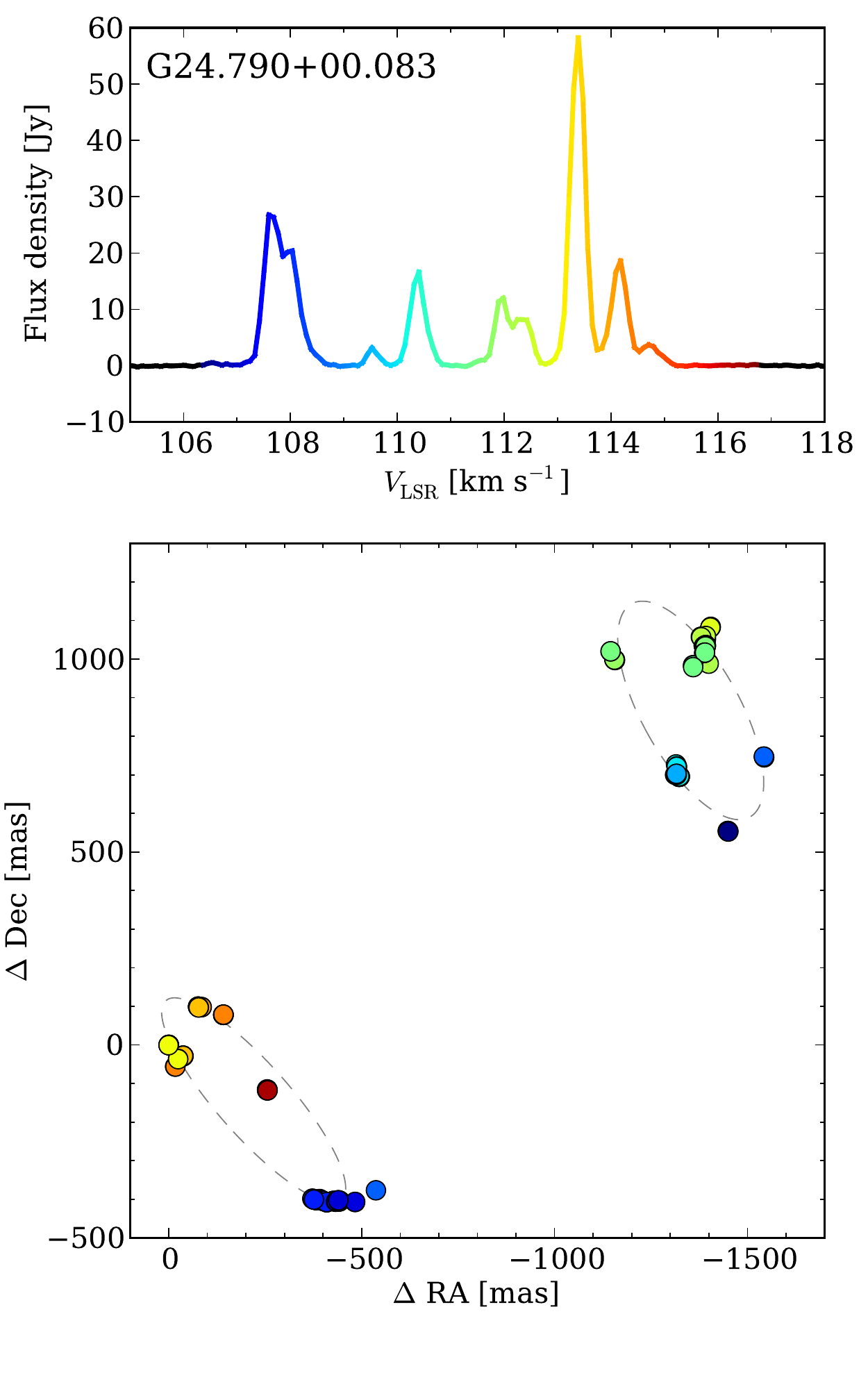}
\includegraphics[scale=0.6]{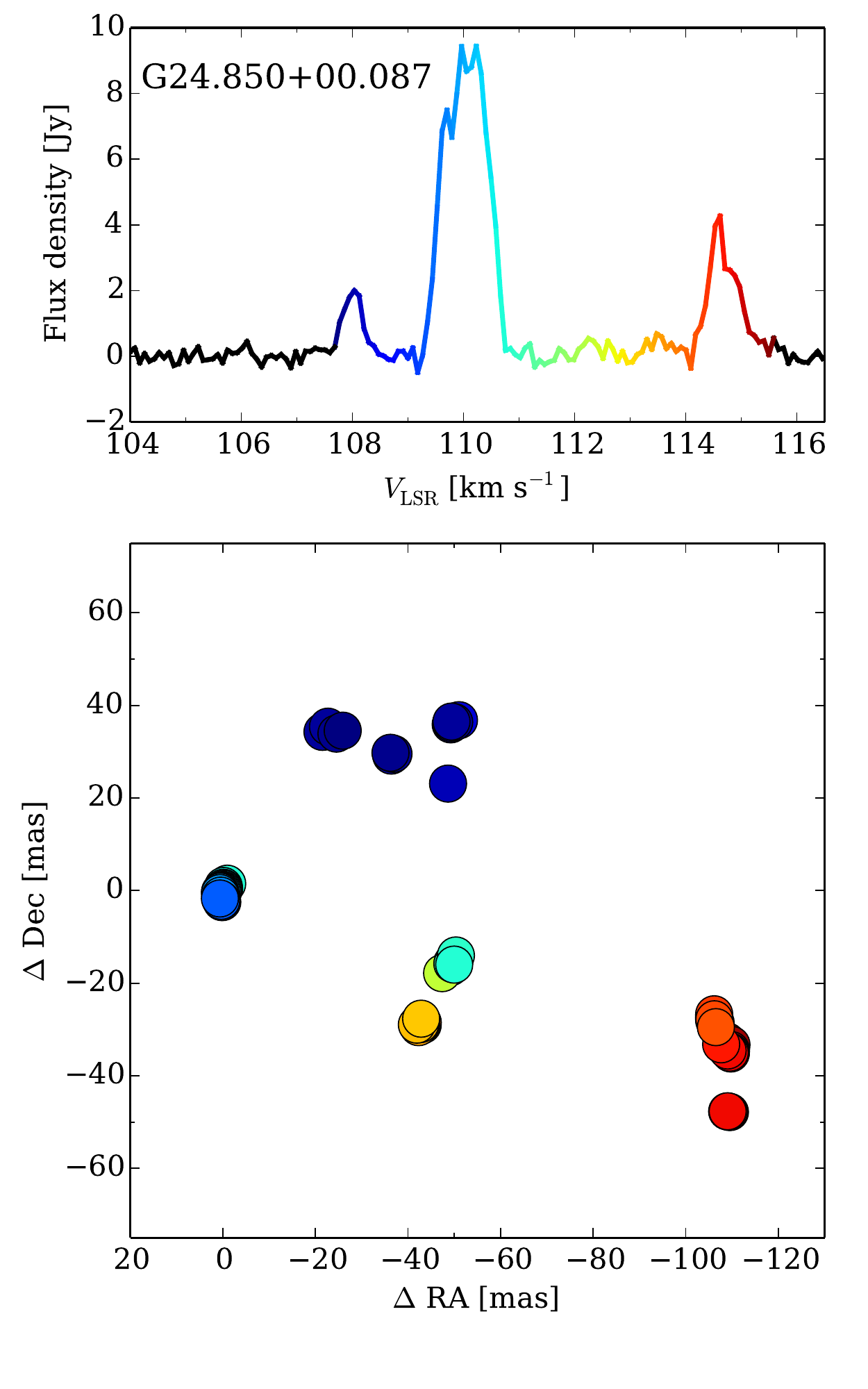}
\includegraphics[scale=0.6]{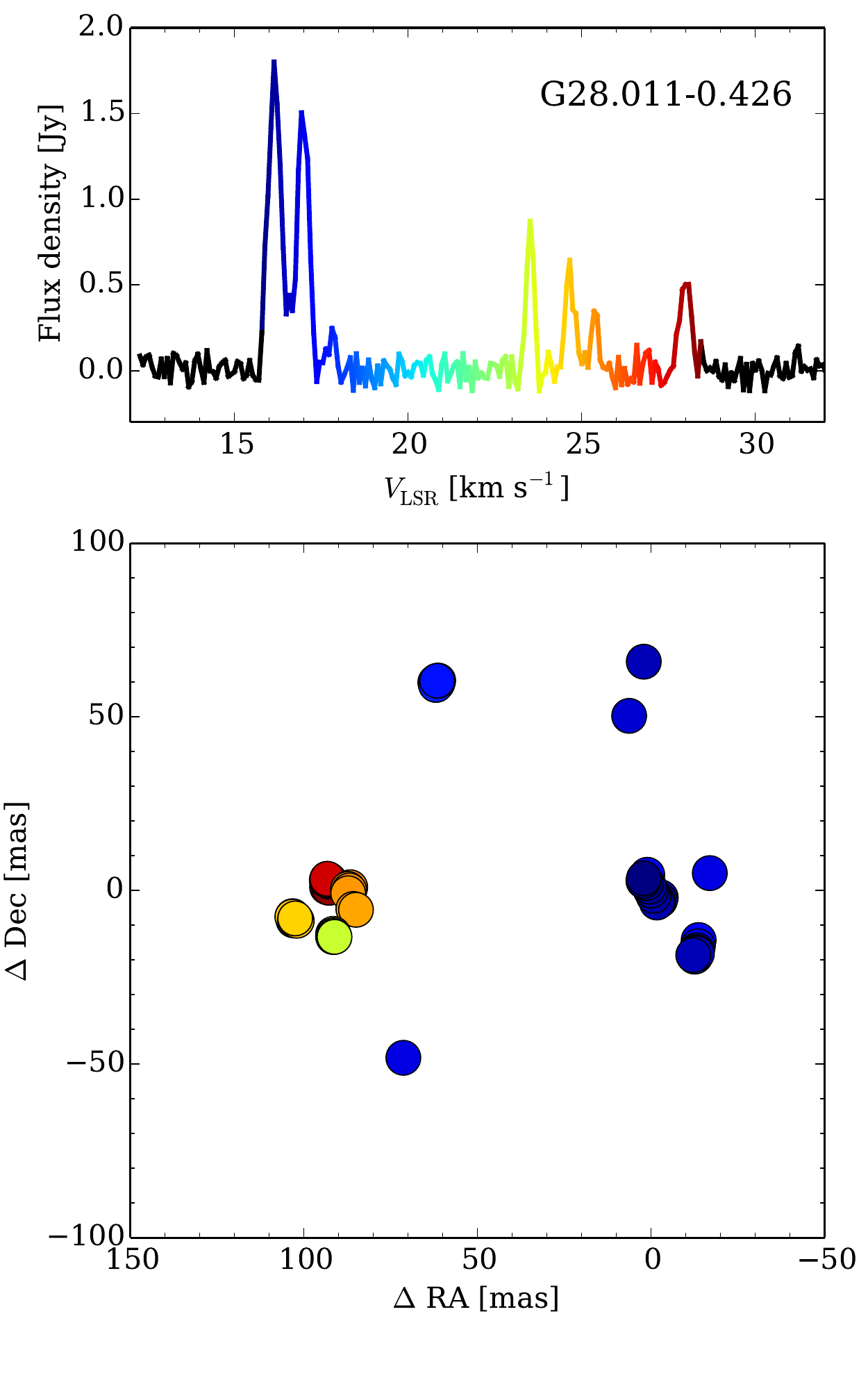}
\caption{Spectra}
\addtocounter{figure}{-1}
\end{figure*}

\begin{figure*}
\centering
\includegraphics[scale=0.6]{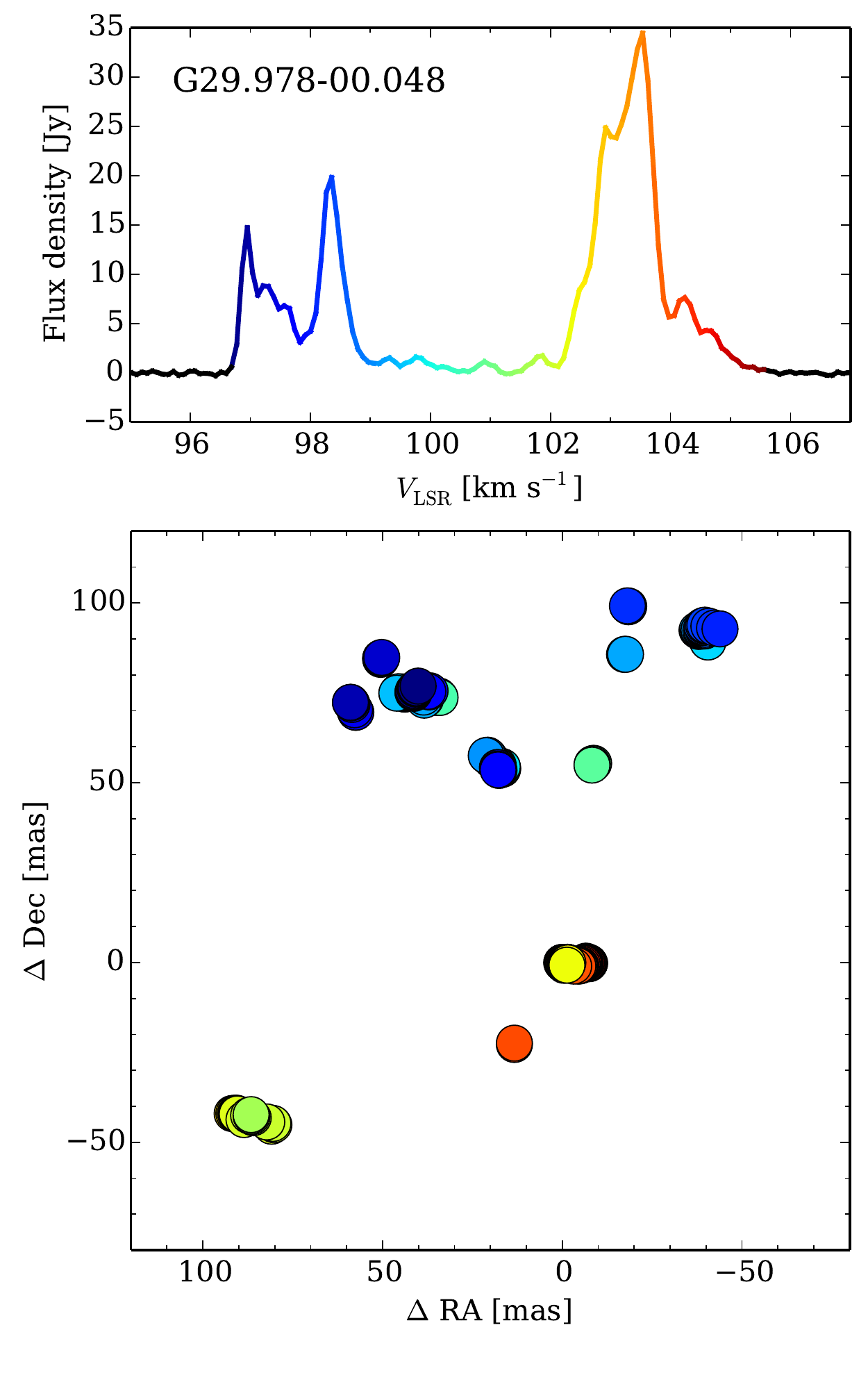}
\includegraphics[scale=0.6]{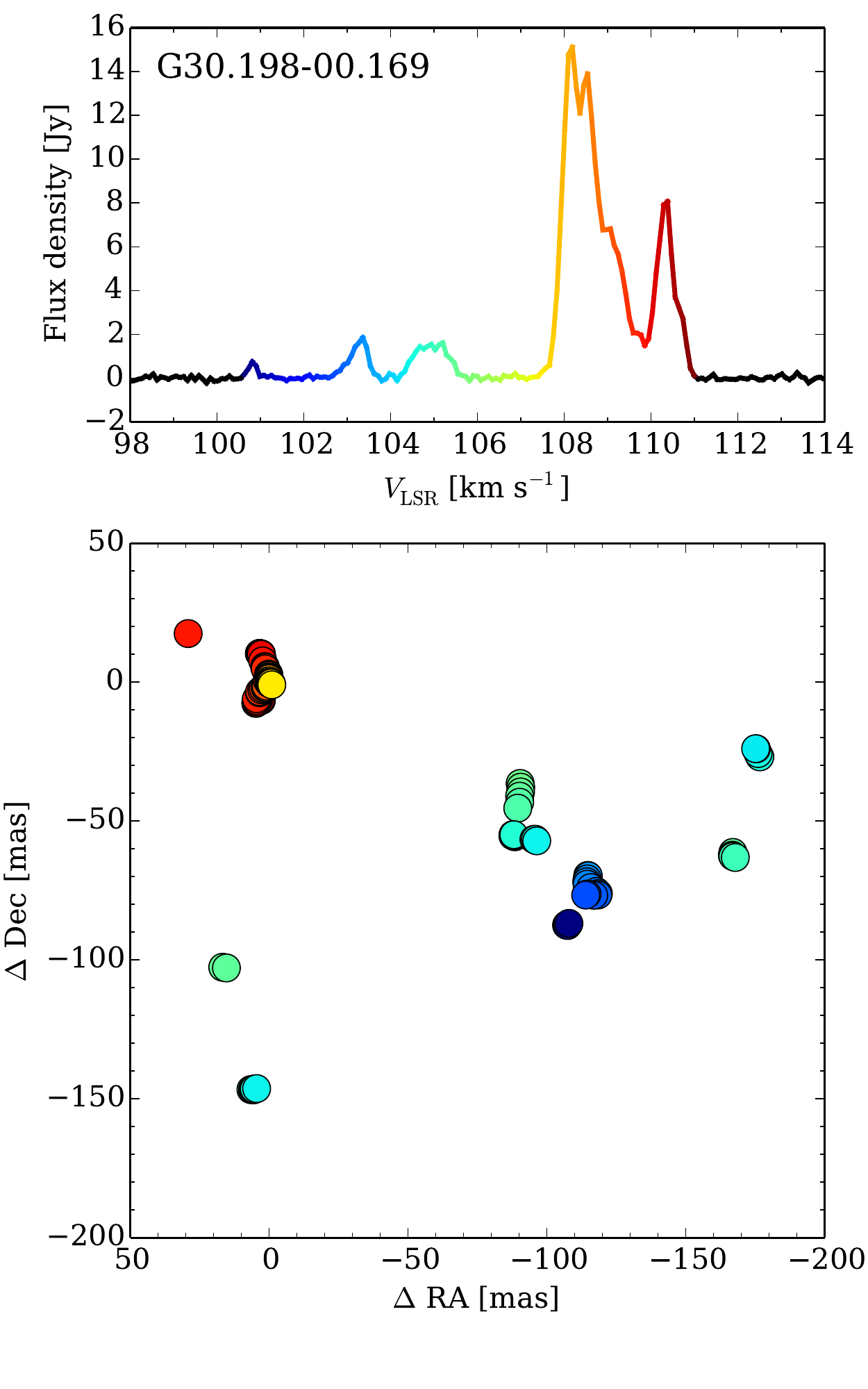}
\includegraphics[scale=0.6]{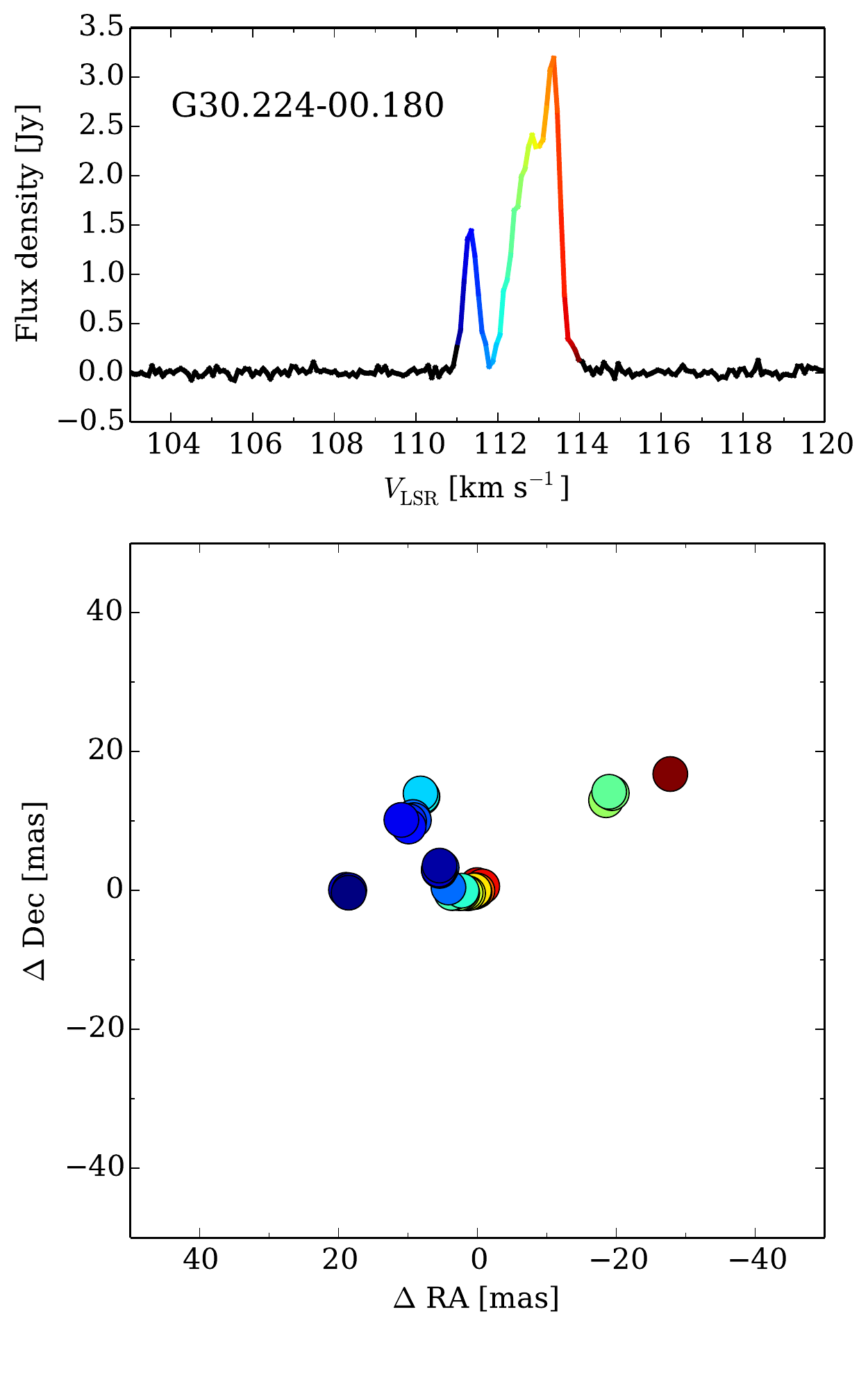}
\includegraphics[scale=0.6]{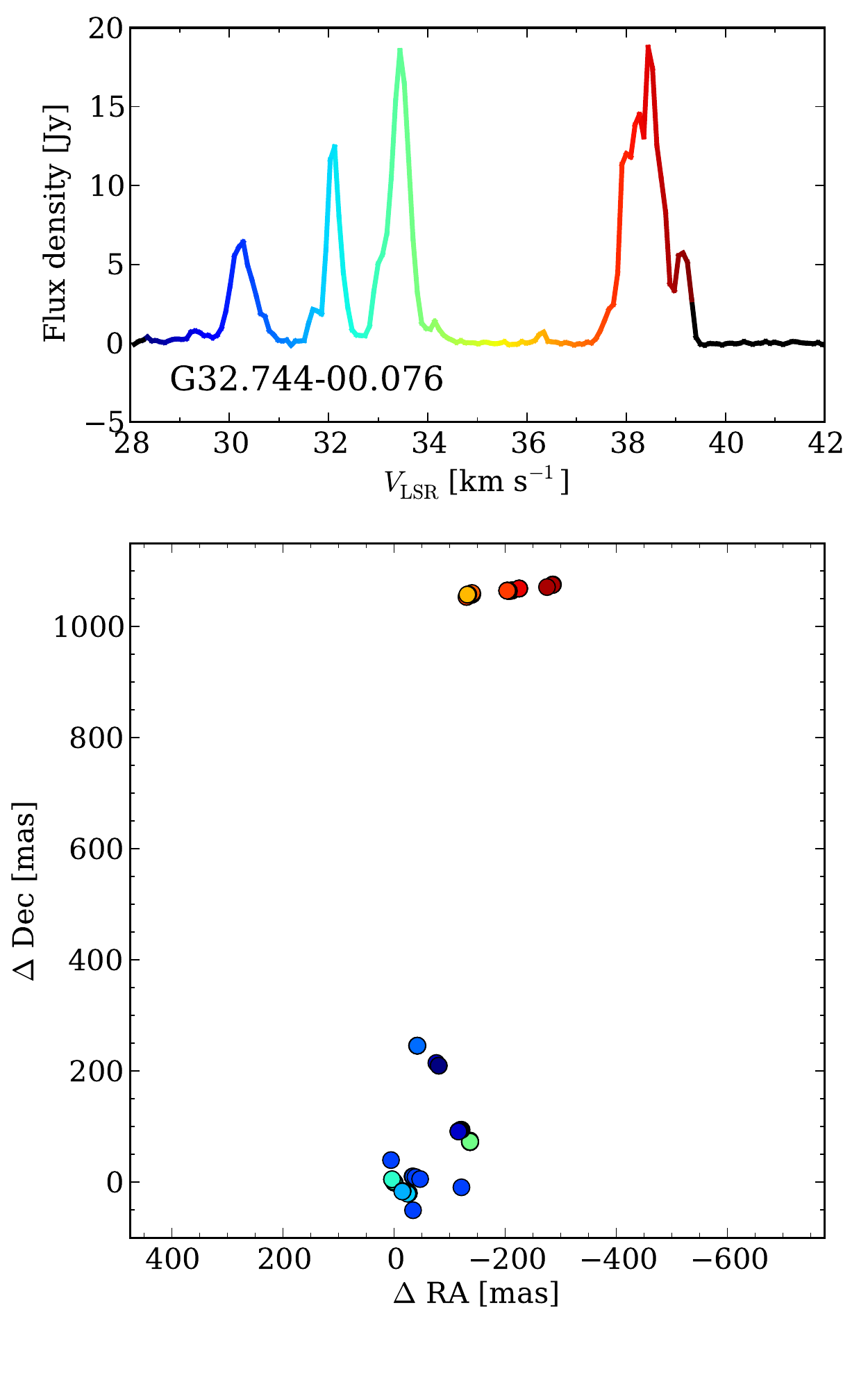}
\caption{continued.}
\addtocounter{figure}{-1}
\end{figure*}

\begin{figure*}
\centering
\includegraphics[scale=0.6]{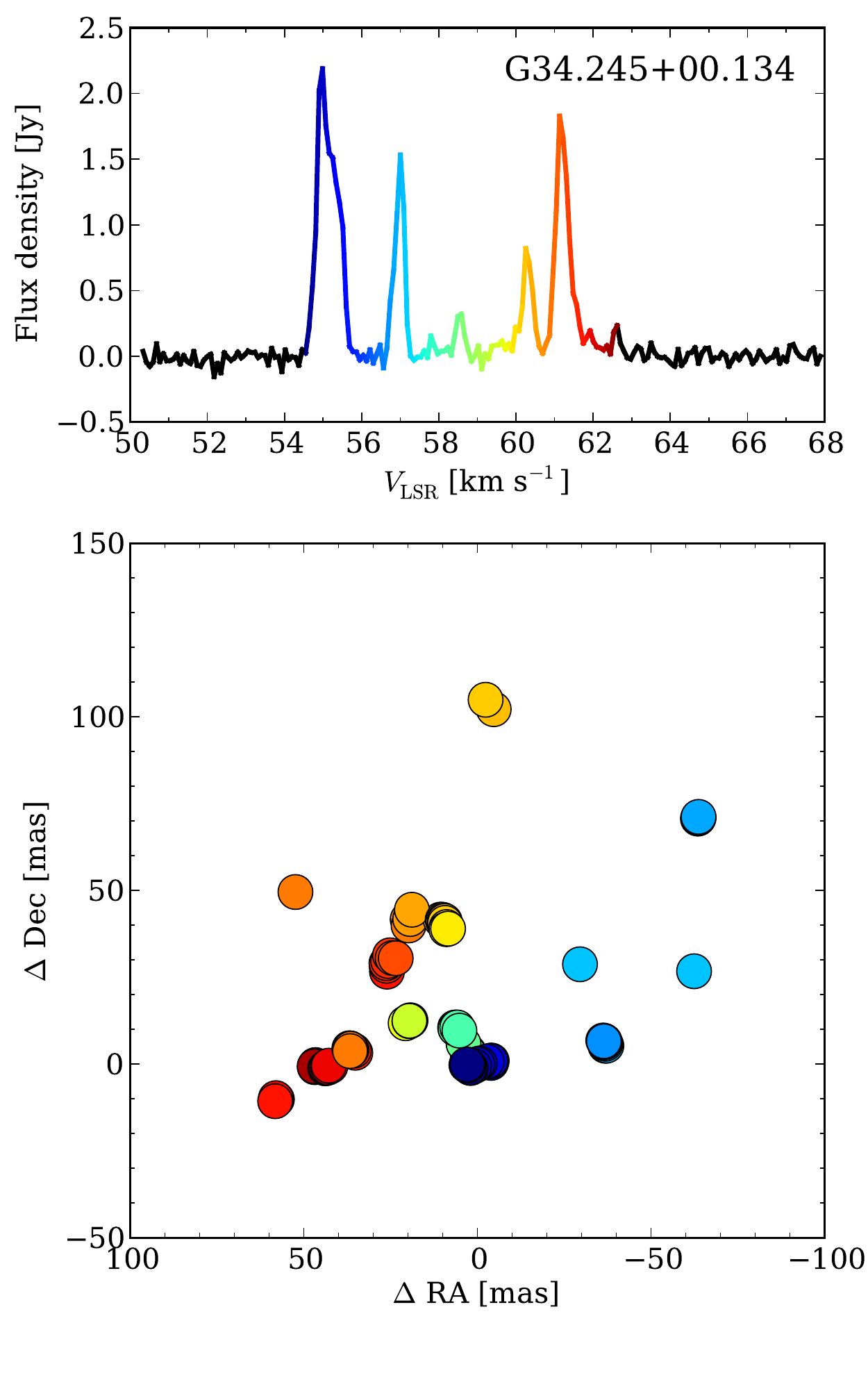}
\includegraphics[scale=0.6]{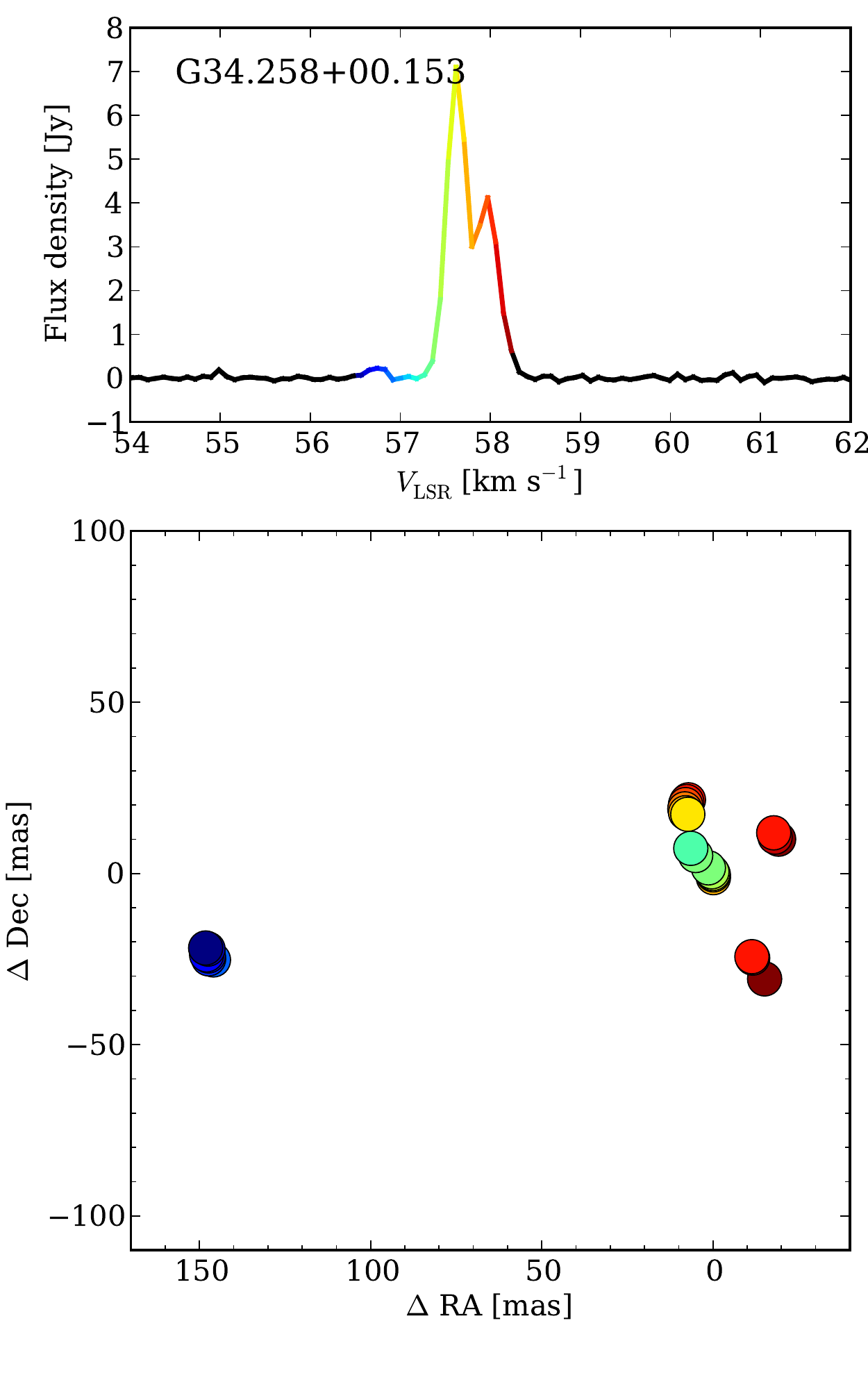}
\includegraphics[scale=0.6]{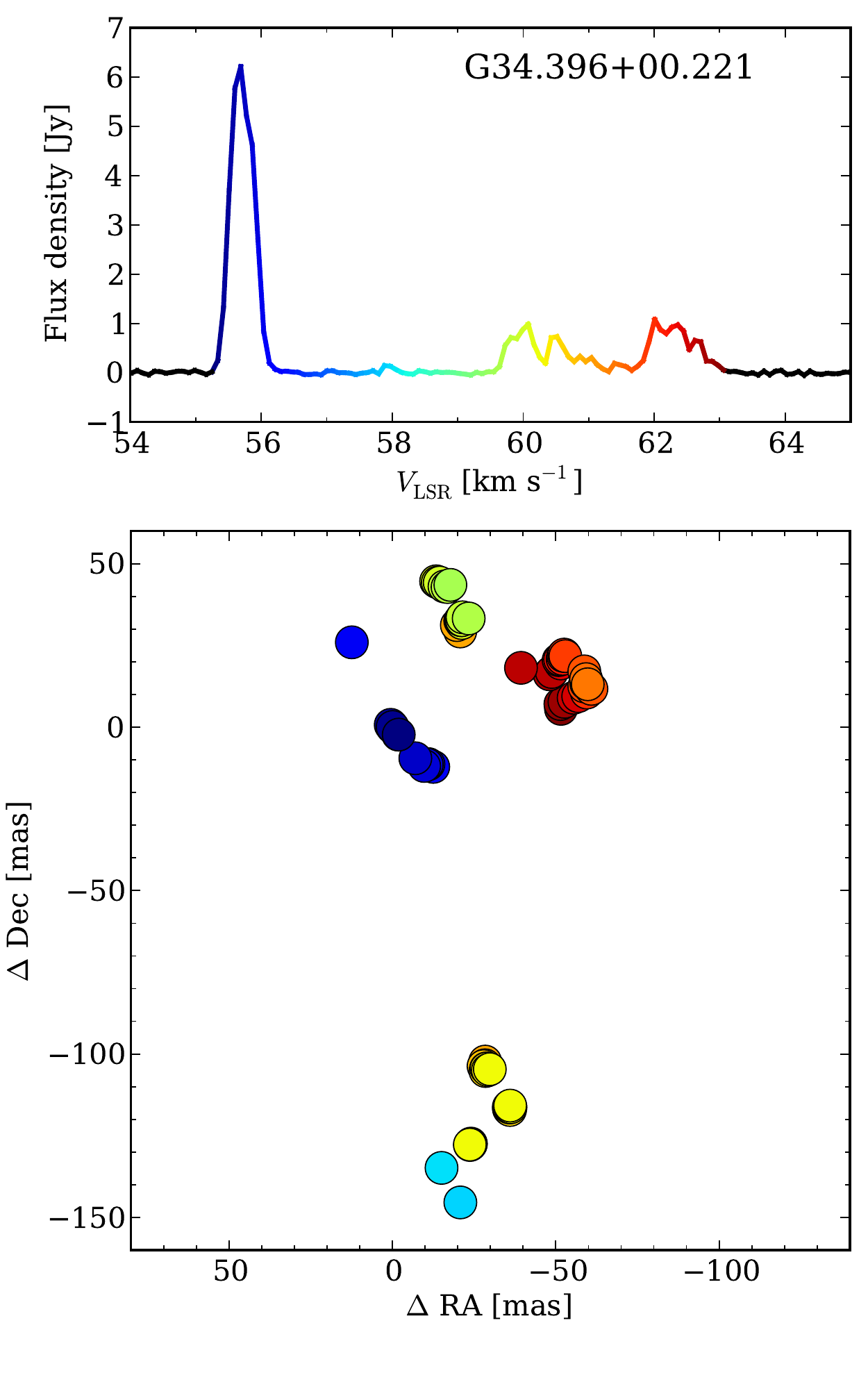}
\includegraphics[scale=0.6]{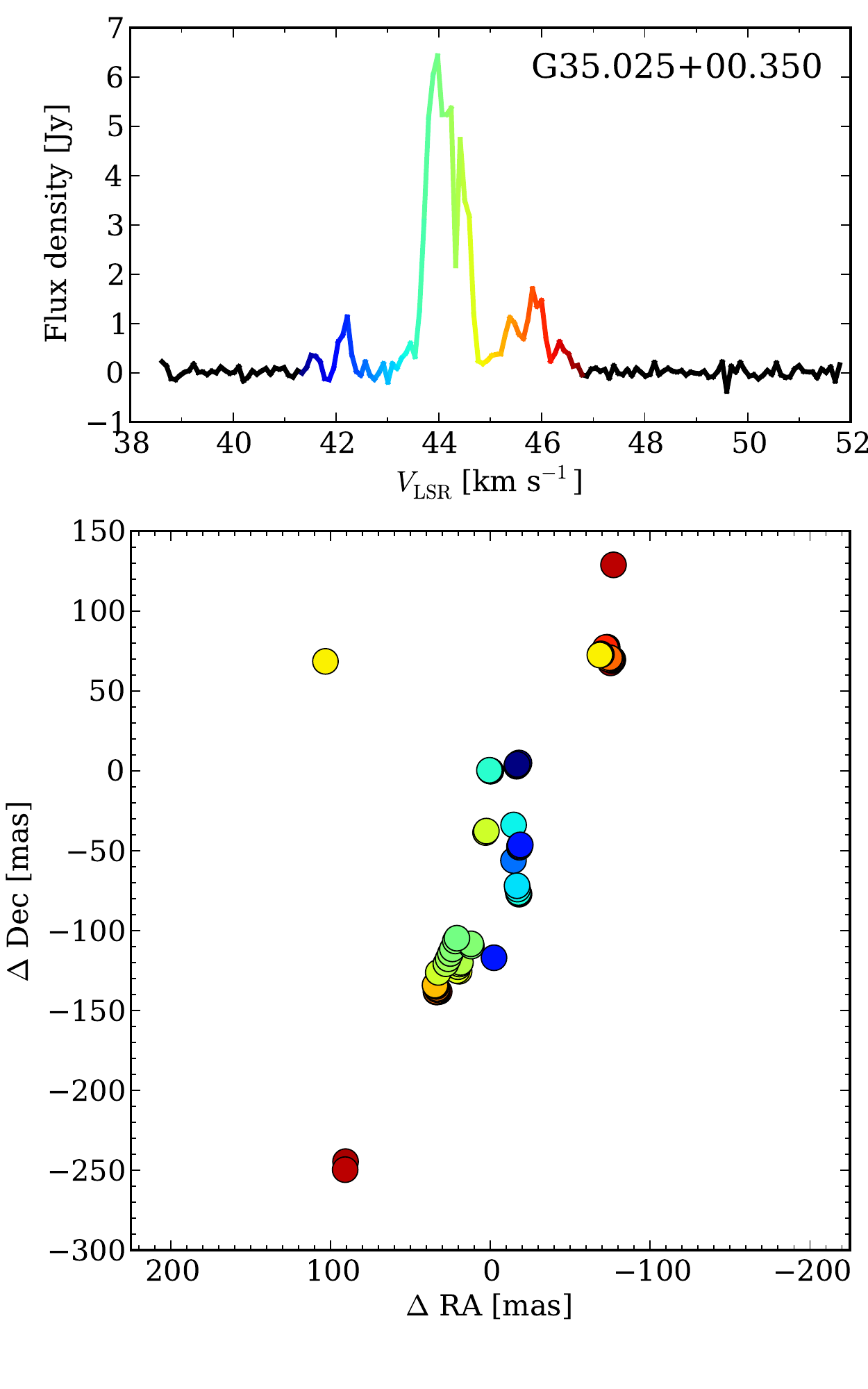}
\caption{continued.}
\label{distribapp}
\end{figure*}

\begin{figure*}
\centering
\includegraphics[scale=0.8]{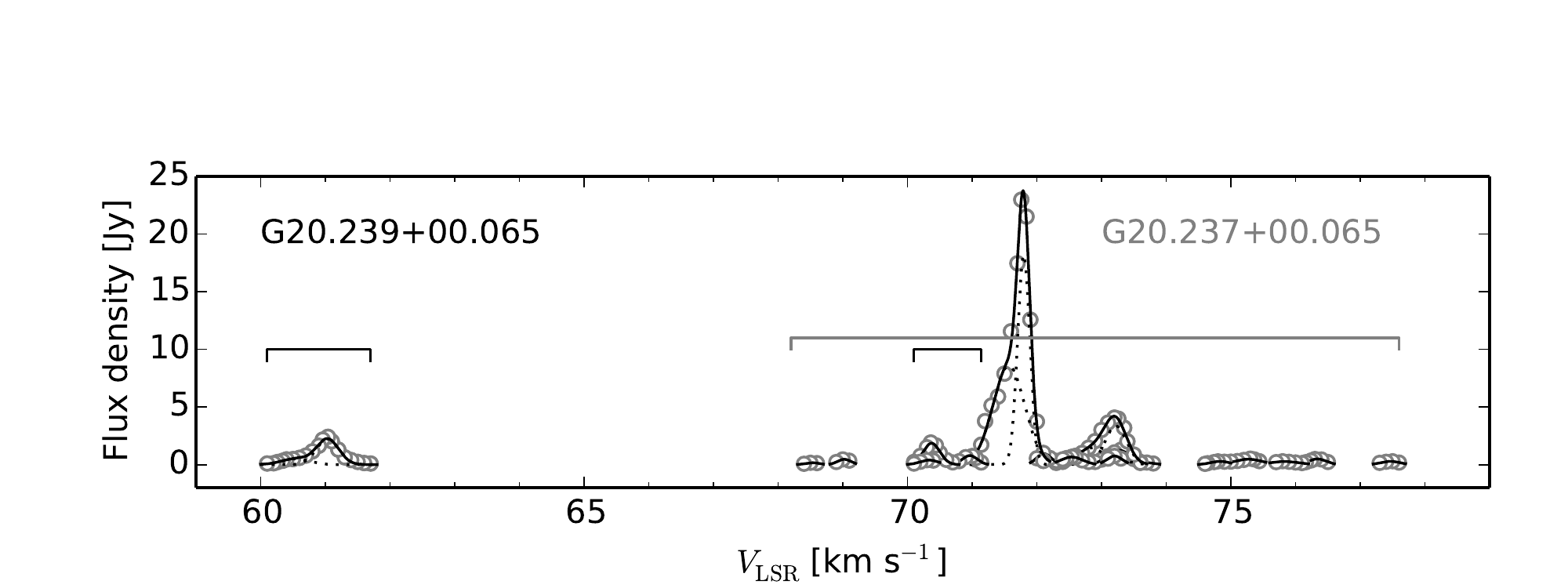}
\includegraphics[scale=0.8]{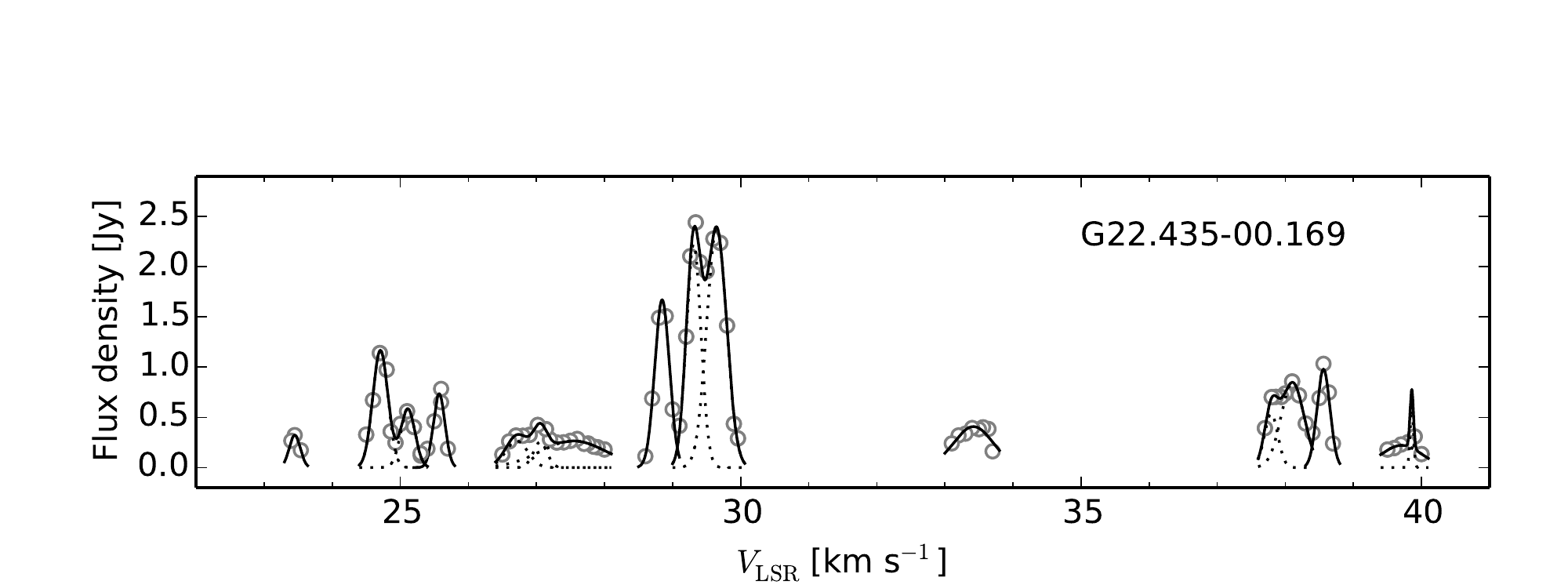}
\includegraphics[scale=0.8]{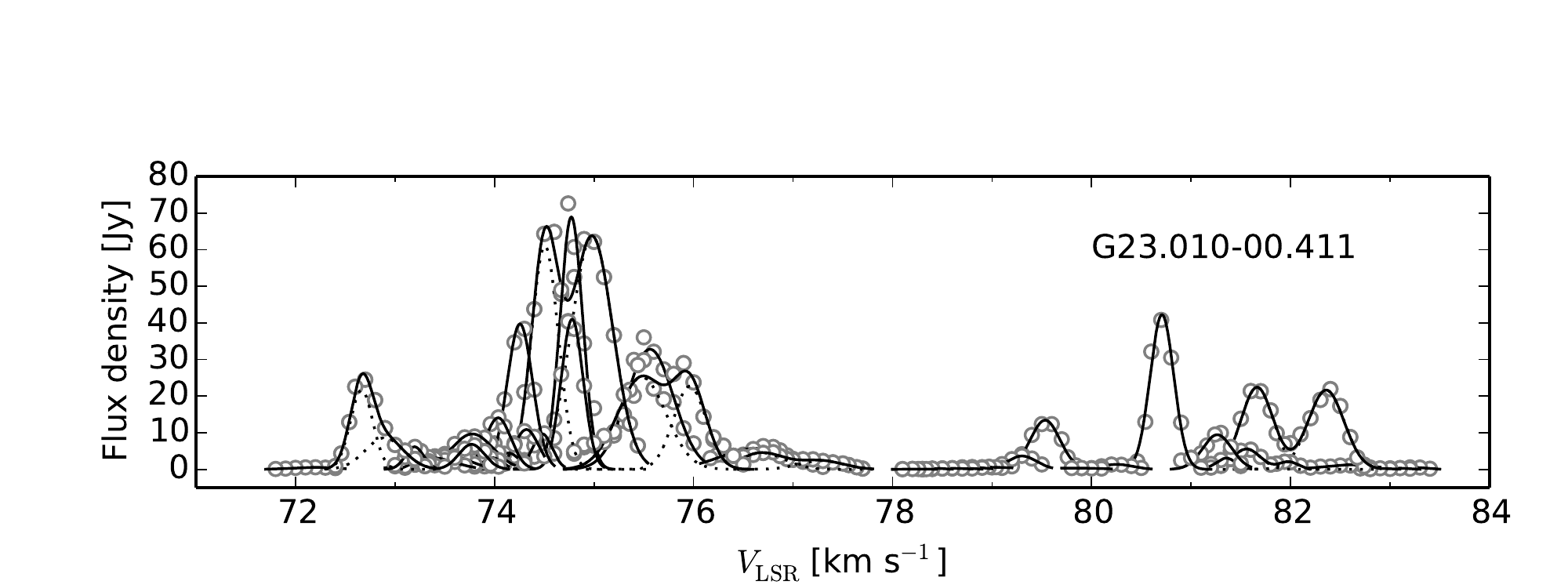}
\includegraphics[scale=0.8]{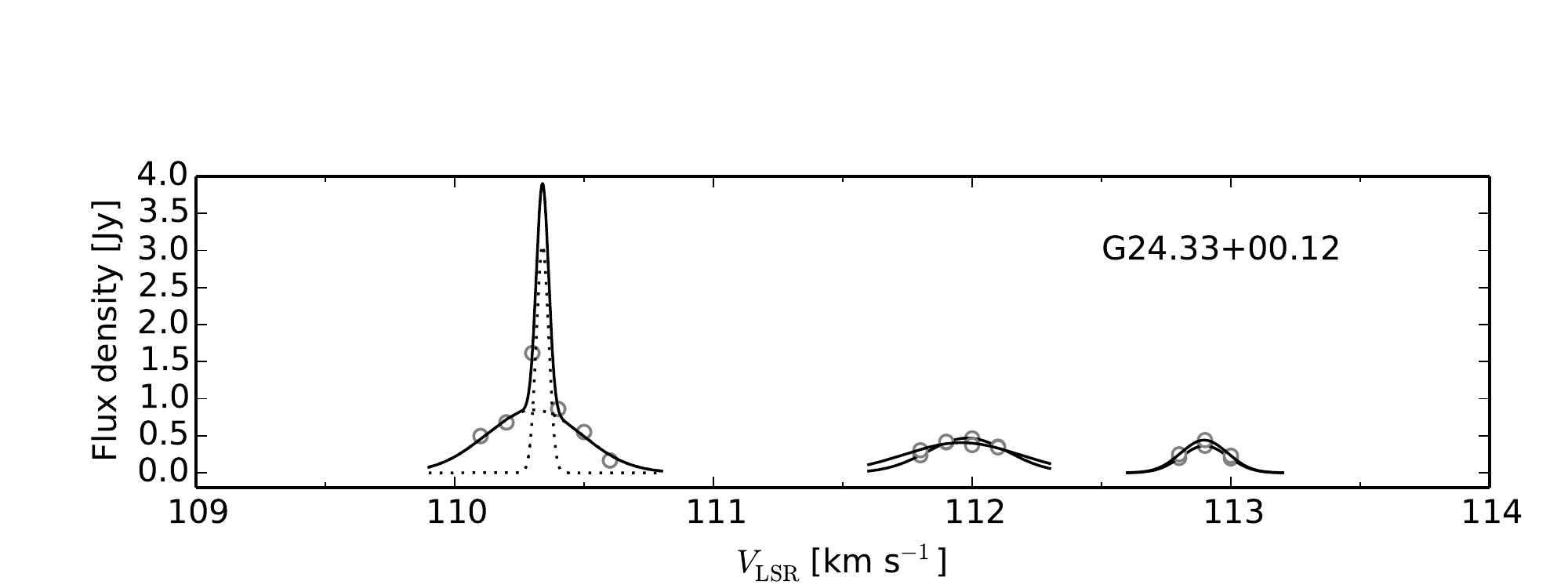}
\caption{Spectra of individual 6.7~GHz maser clouds with Gaussian velocity
profiles. Each circle traces the emission level of a single maser spot as
presented in Figs.~\ref{distrib} and \ref{distribapp}. The
black line represents the fitting of a Gaussian function (or functions) as summarised 
in Table~\ref{clouds}.}
\addtocounter{figure}{-1}
     \label{gauss}
   \end{figure*}

  \begin{figure*}
    \centering
     \includegraphics[scale=0.8]{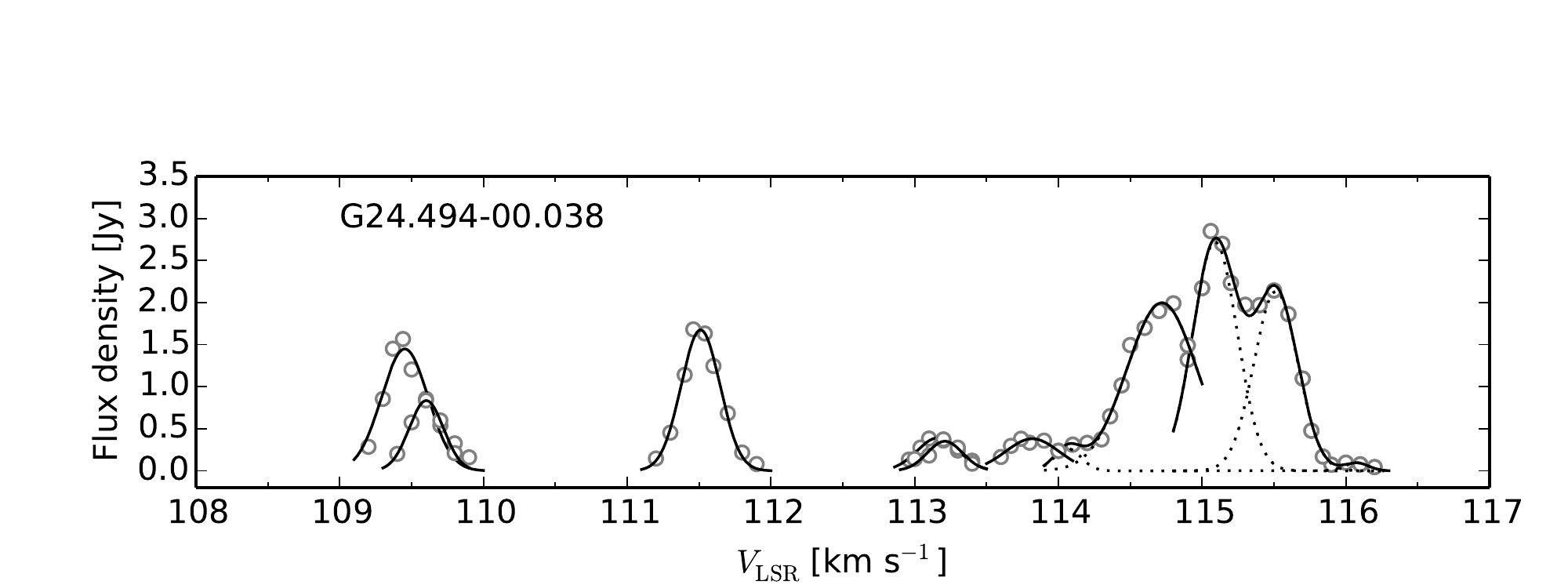}
     \includegraphics[scale=0.8]{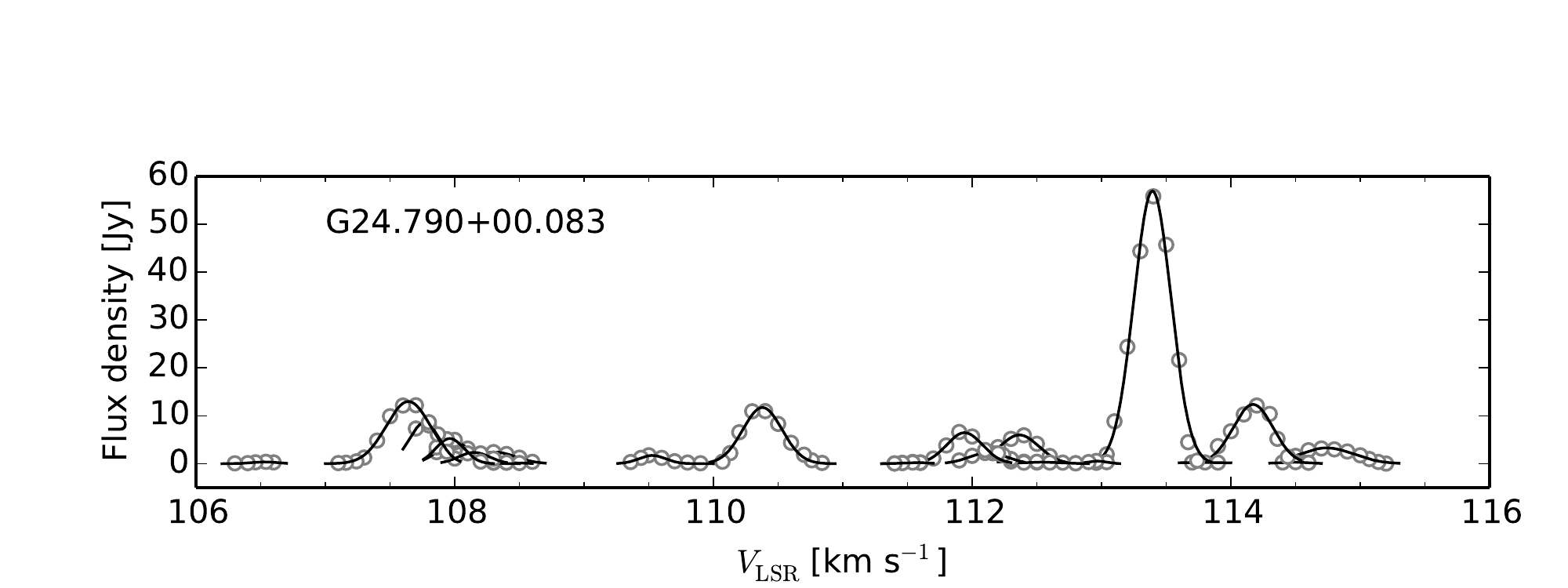}
     \includegraphics[scale=0.8]{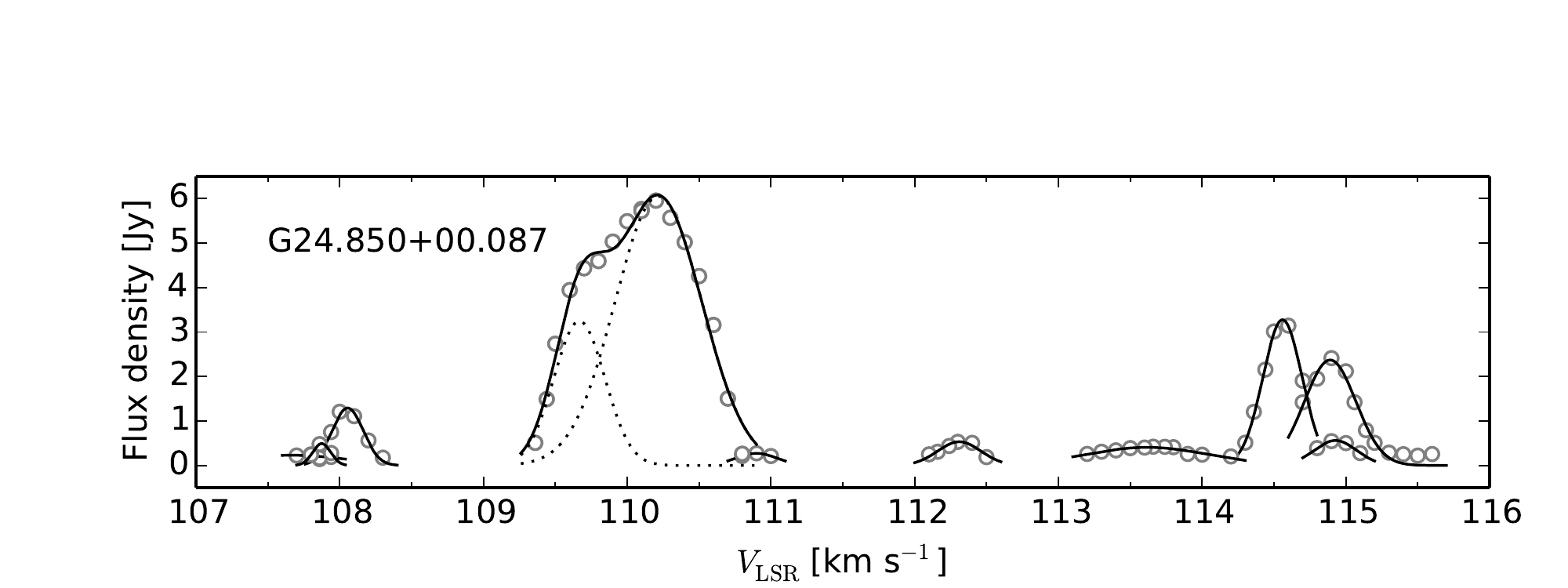}
     \includegraphics[scale=0.8]{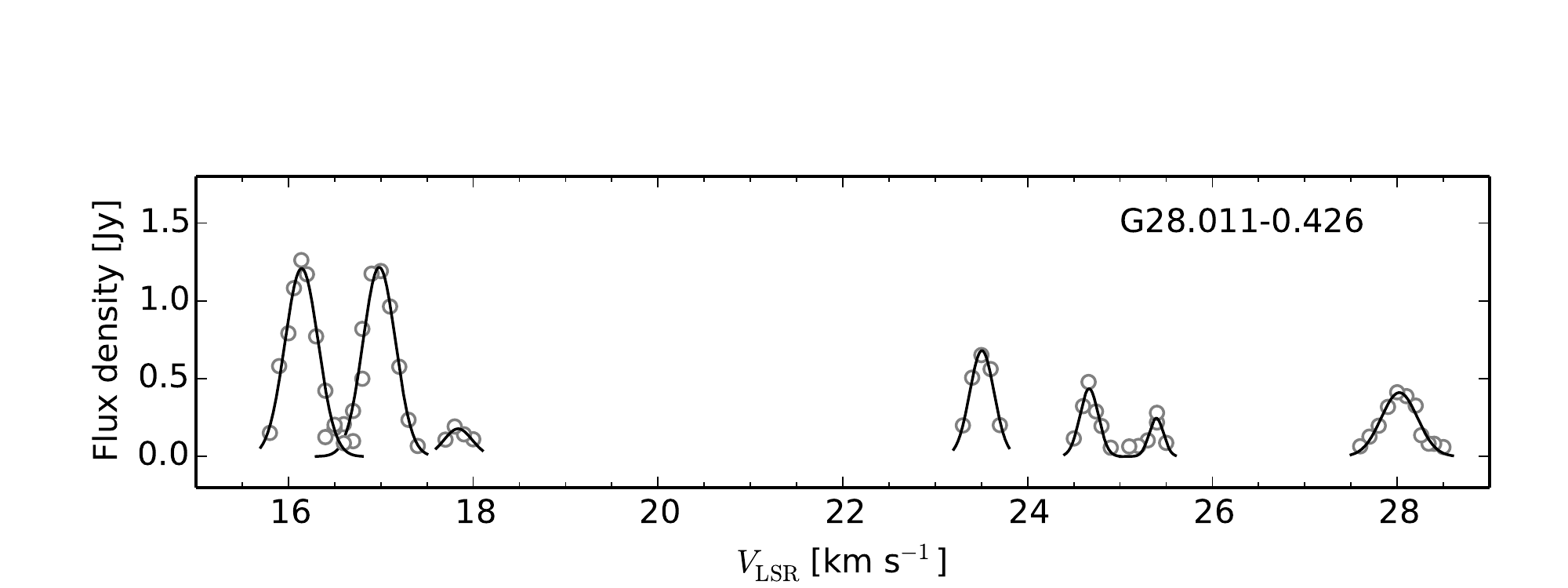}
     \caption{continued.}
\addtocounter{figure}{-1}
   \end{figure*}

  \begin{figure*}
    \centering
     \includegraphics[scale=0.8]{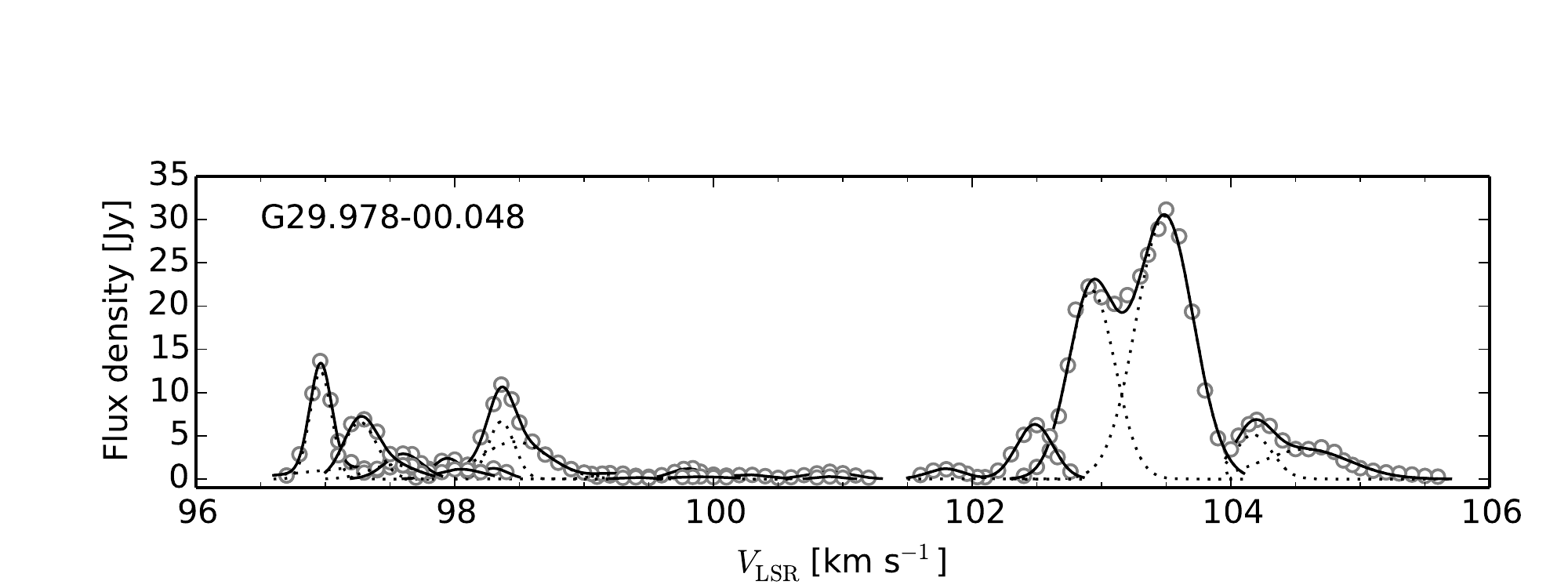}
     \includegraphics[scale=0.8]{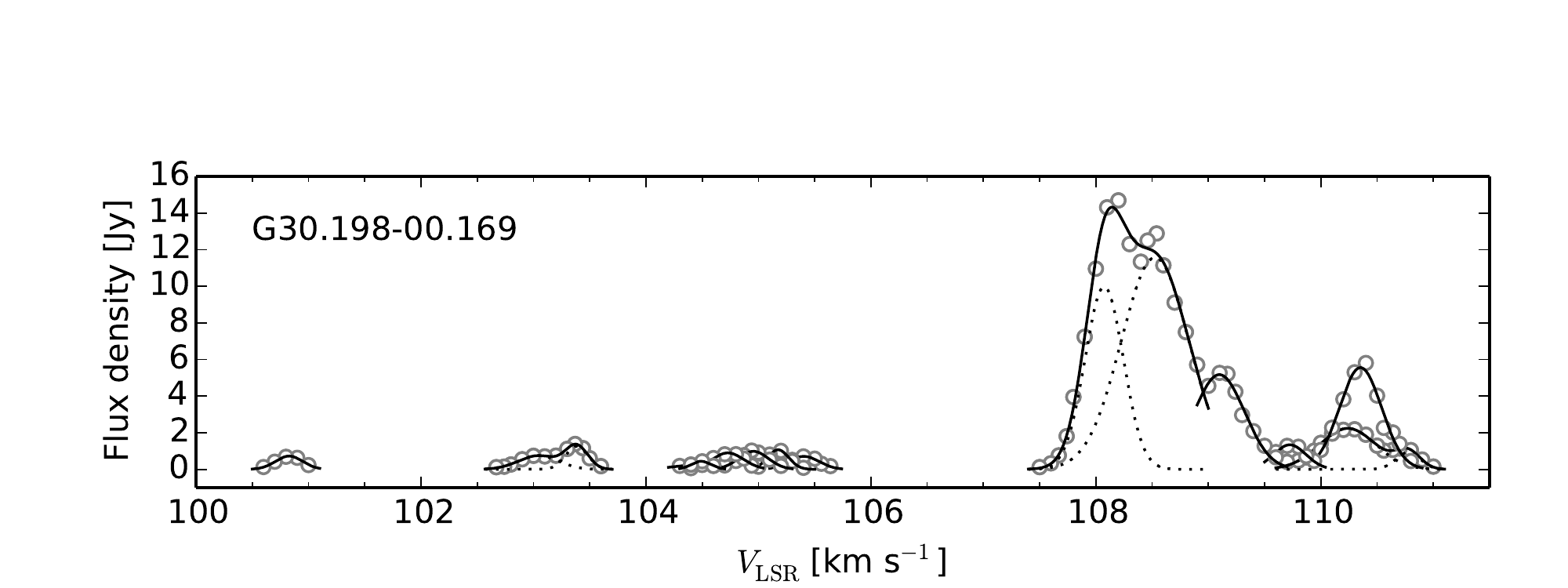}
     \includegraphics[scale=0.8]{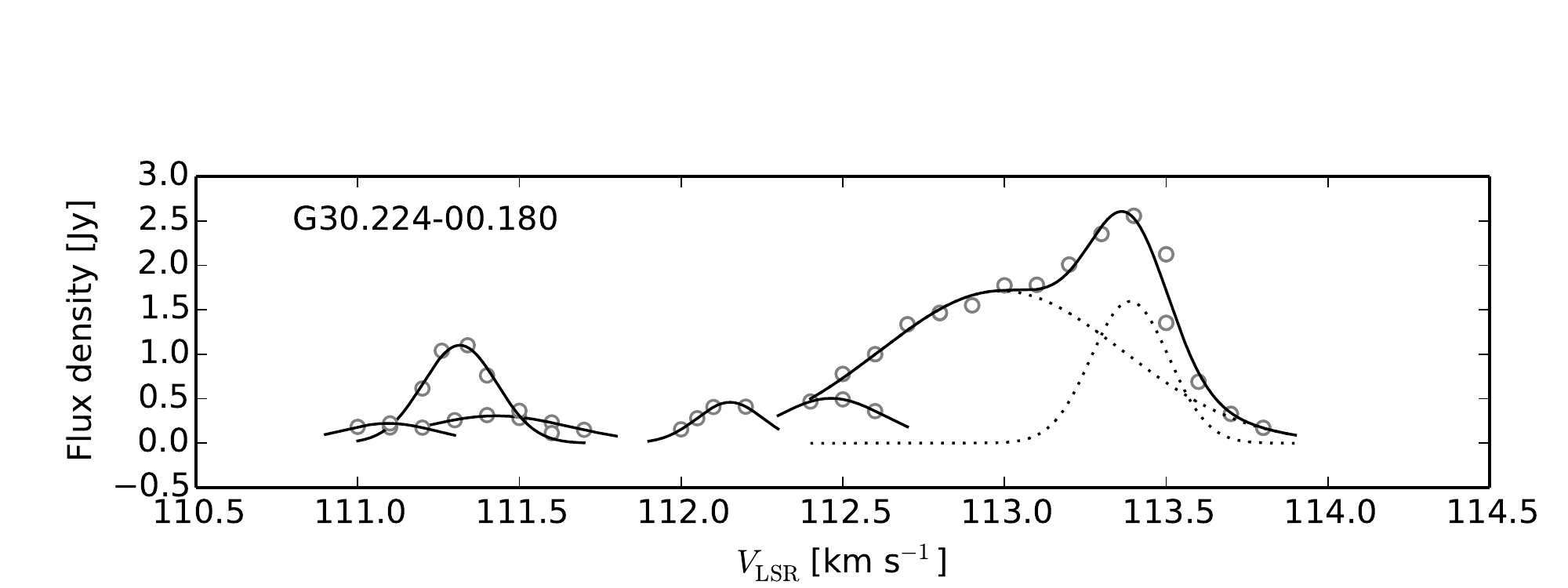}
     \includegraphics[scale=0.8]{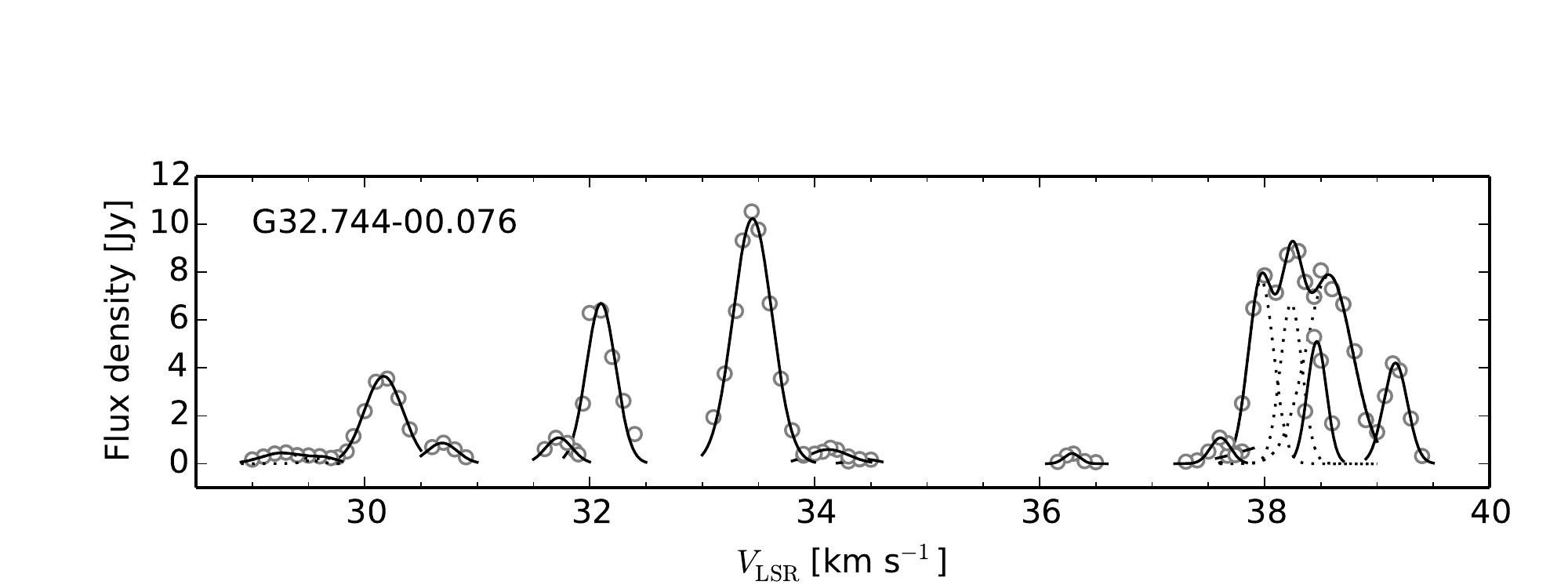}
     \caption{continued.}
\addtocounter{figure}{-1}
   \end{figure*}

  \begin{figure*}
    \centering
     \includegraphics[scale=0.8]{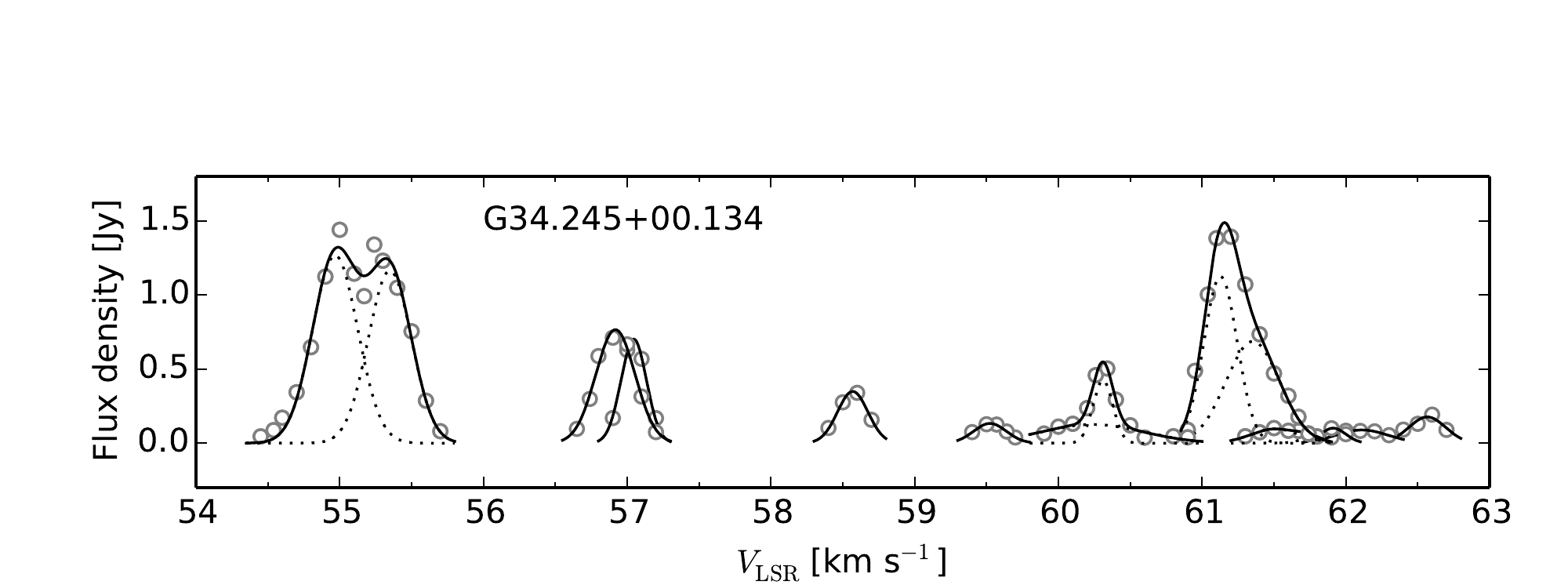}
     \includegraphics[scale=0.8]{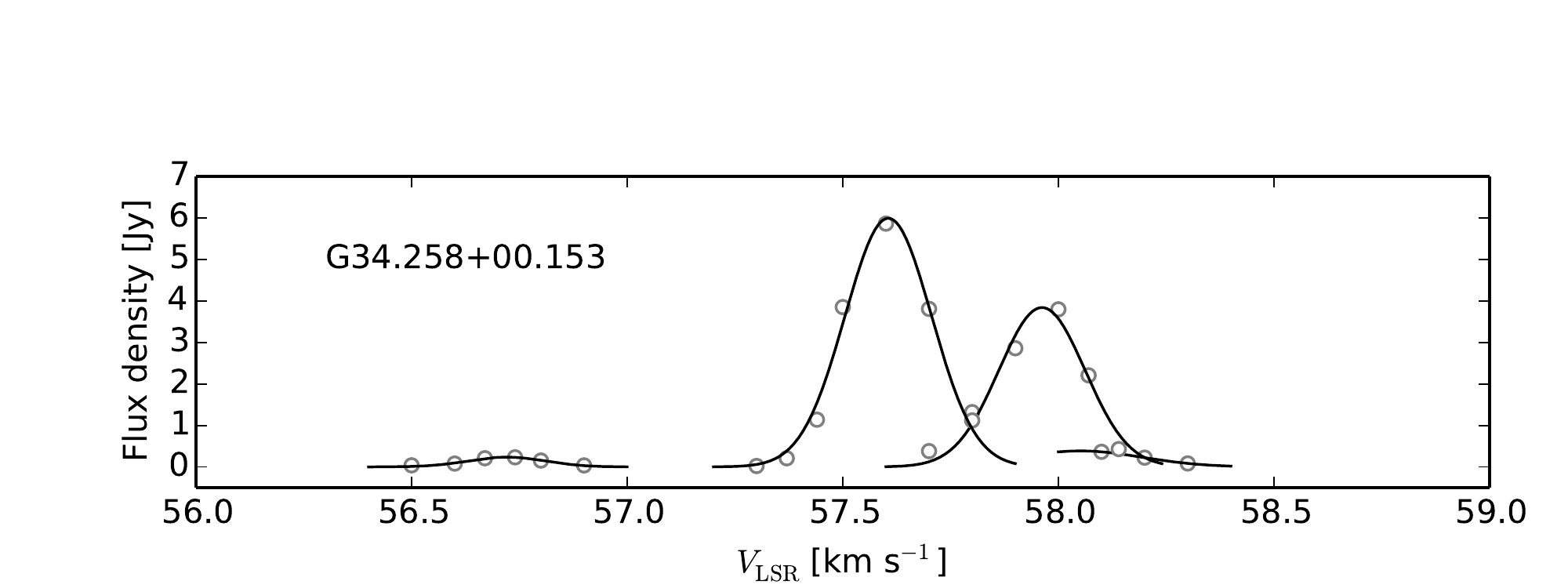}
     \includegraphics[scale=0.8]{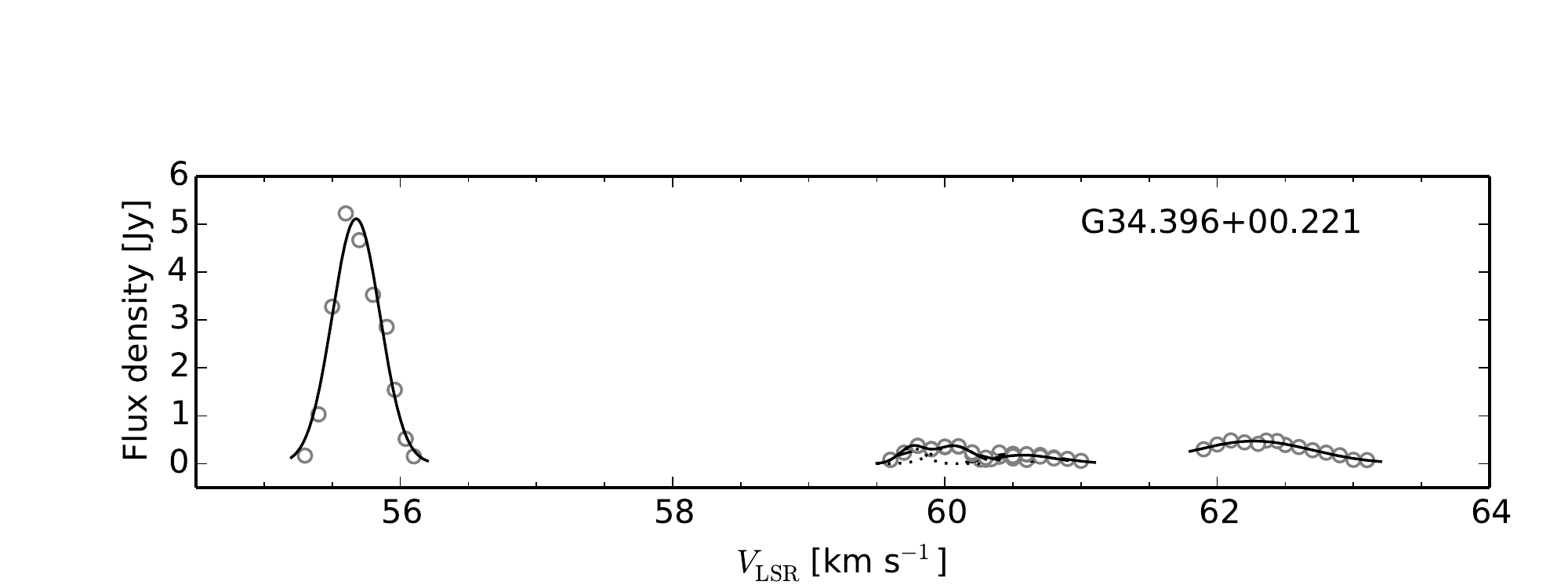}
     \includegraphics[scale=0.8]{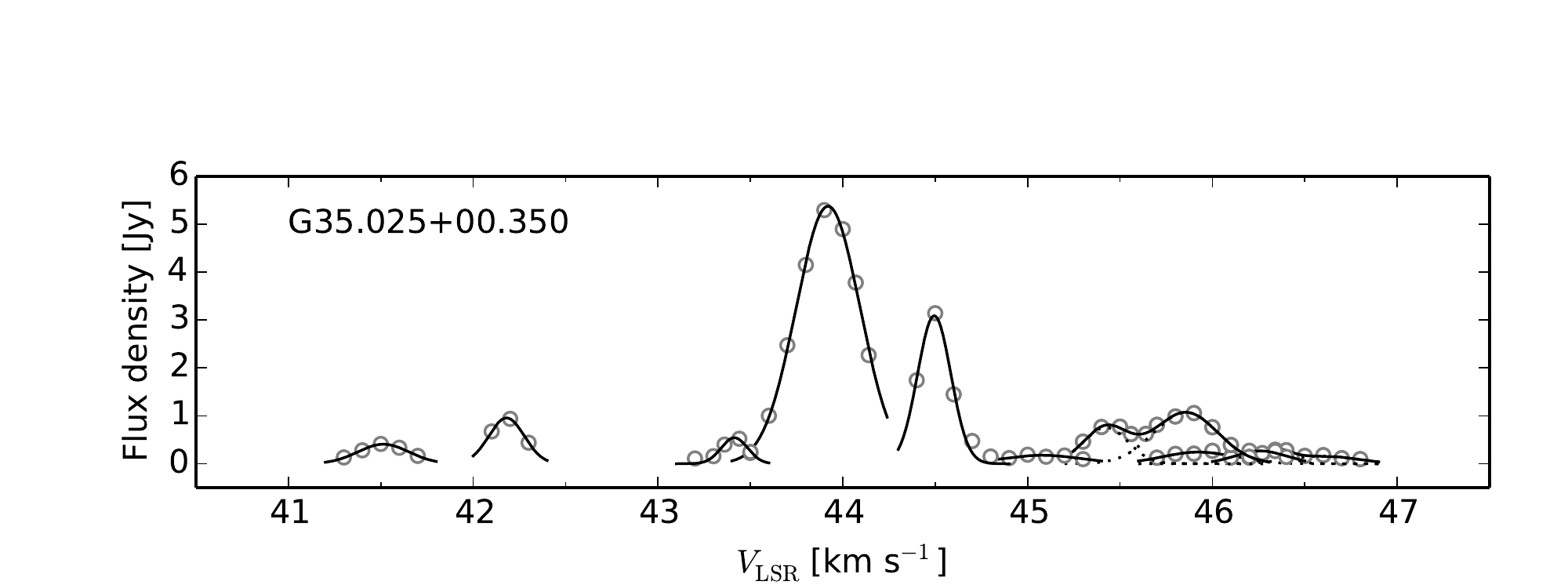}
     \caption{continued.}
   \end{figure*}

\begin{figure*}
\includegraphics[height=8cm,scale=0.30]{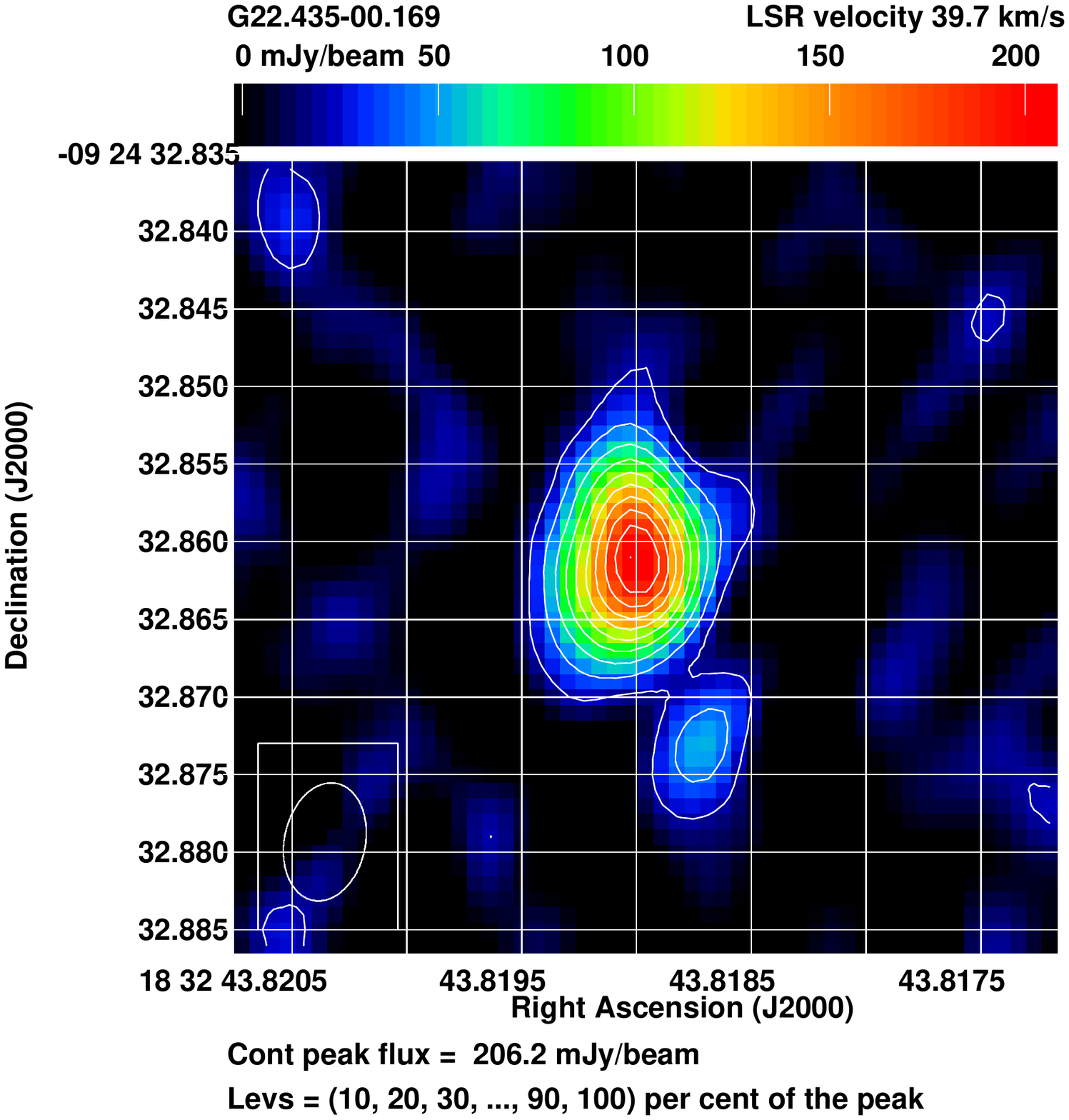}
\includegraphics[height=8cm,scale=0.30]{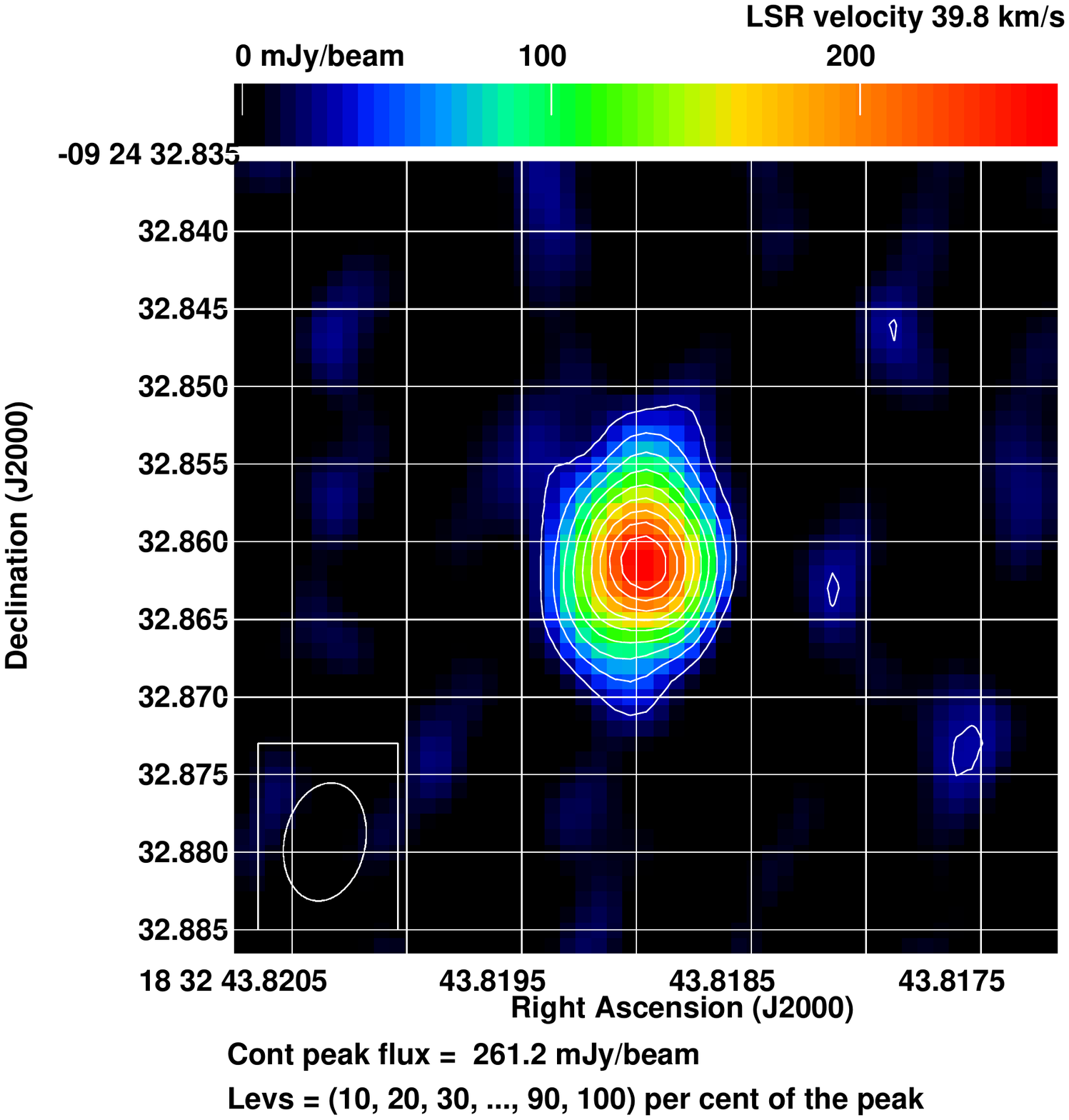}
\includegraphics[height=8cm,scale=0.30]{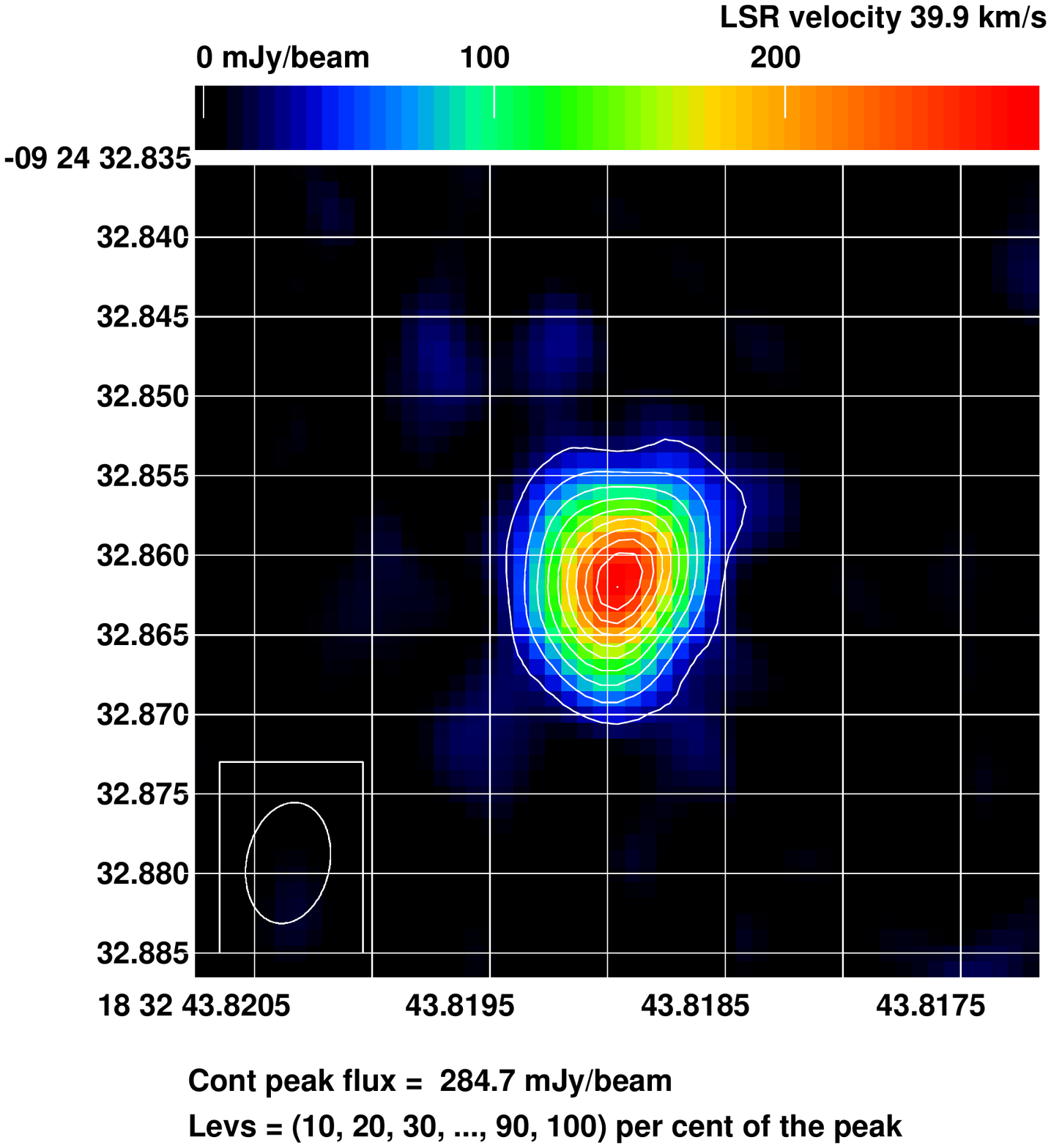}
\includegraphics[height=8cm,scale=0.30]{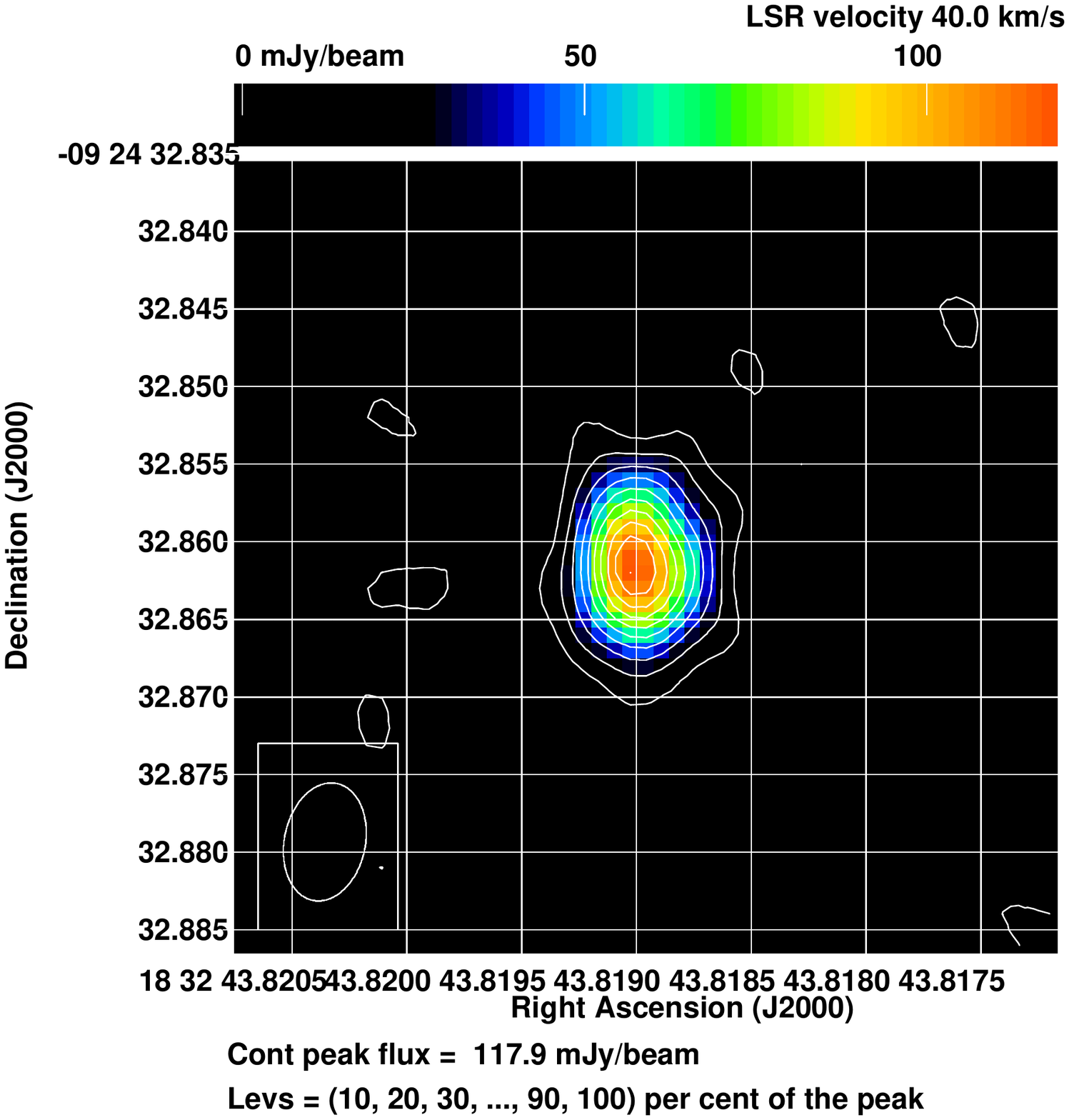}
\includegraphics[height=8cm,scale=0.30]{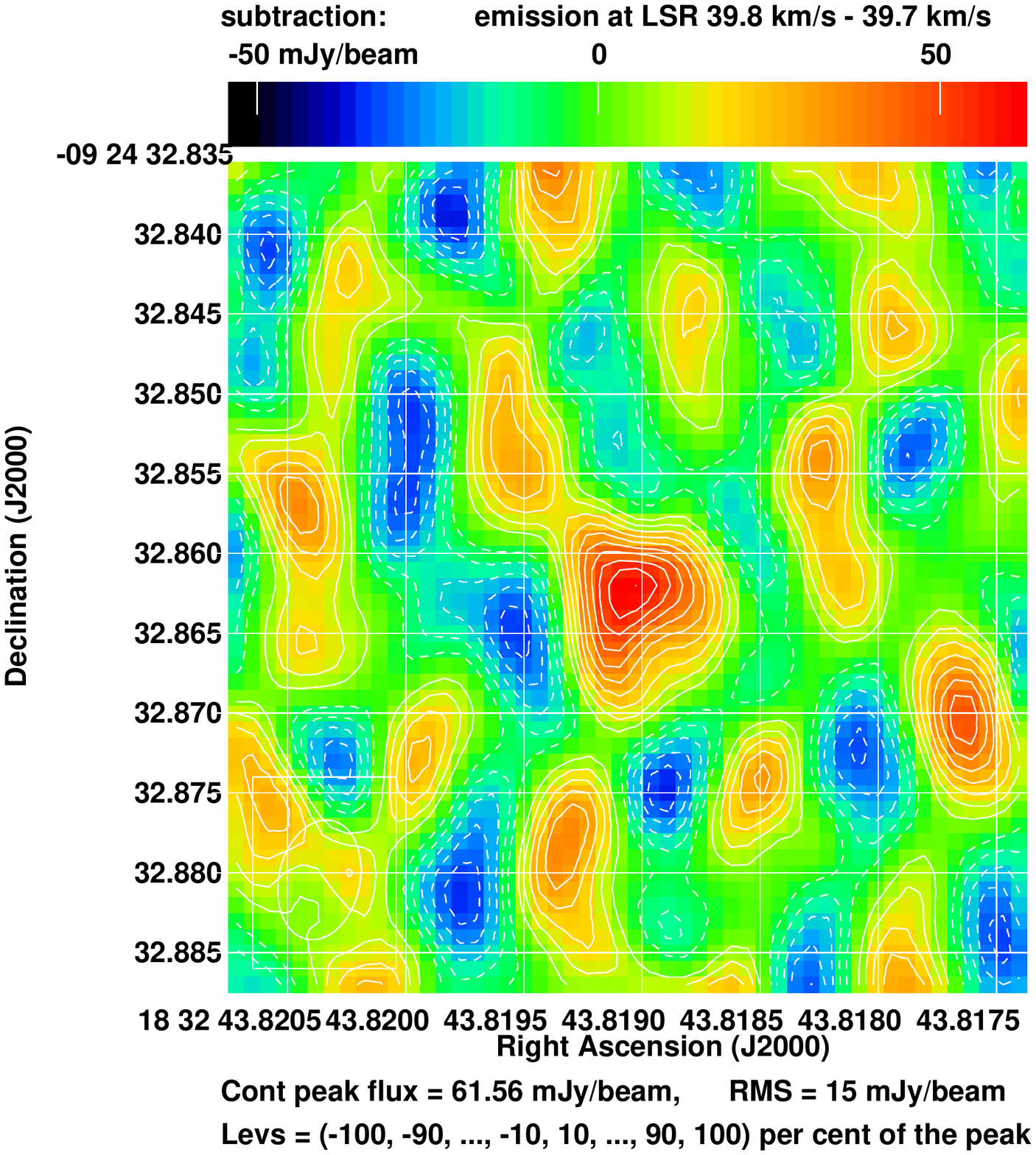}
\includegraphics[height=8cm,scale=0.30]{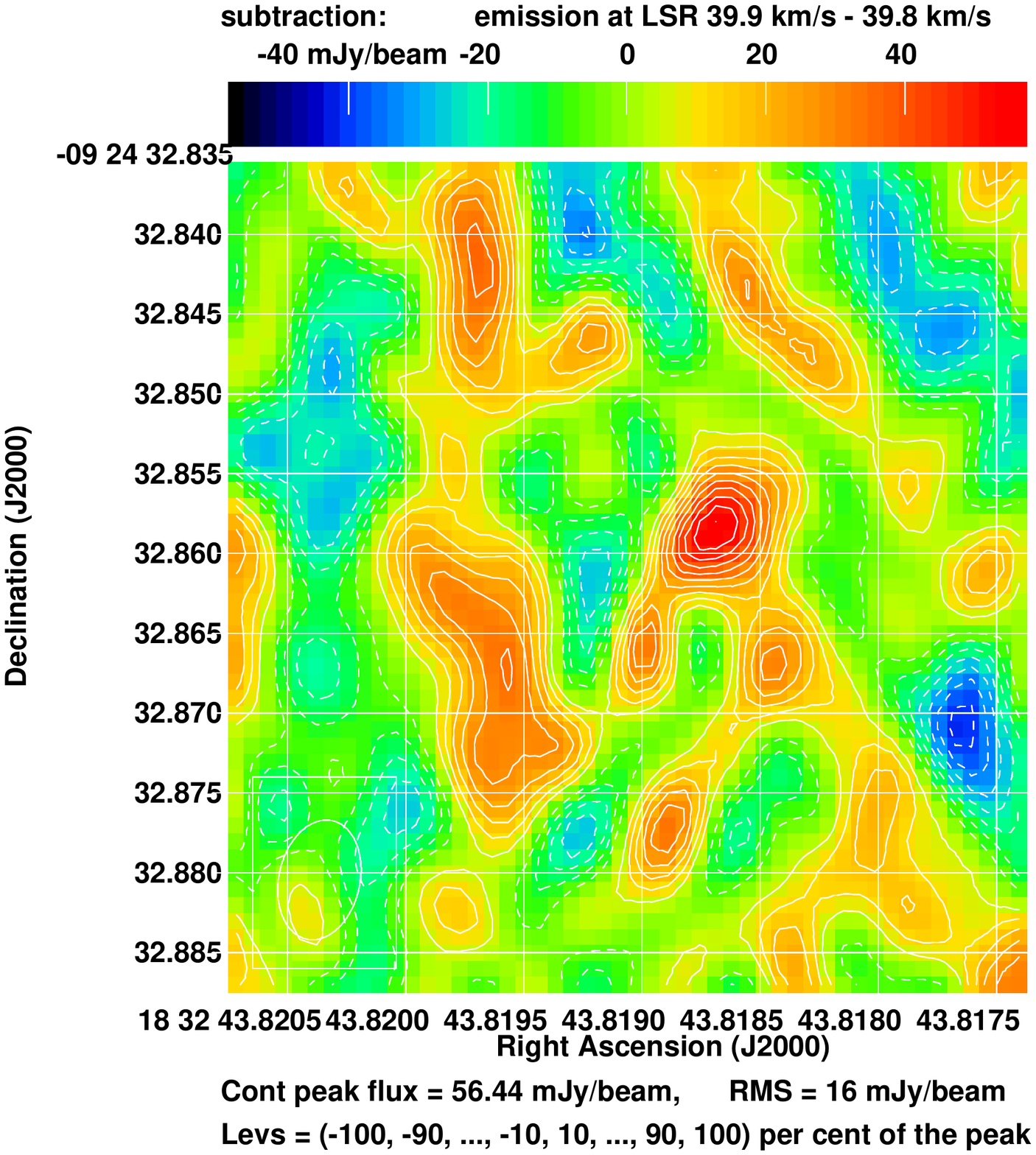}
\includegraphics[height=8cm,scale=0.30]{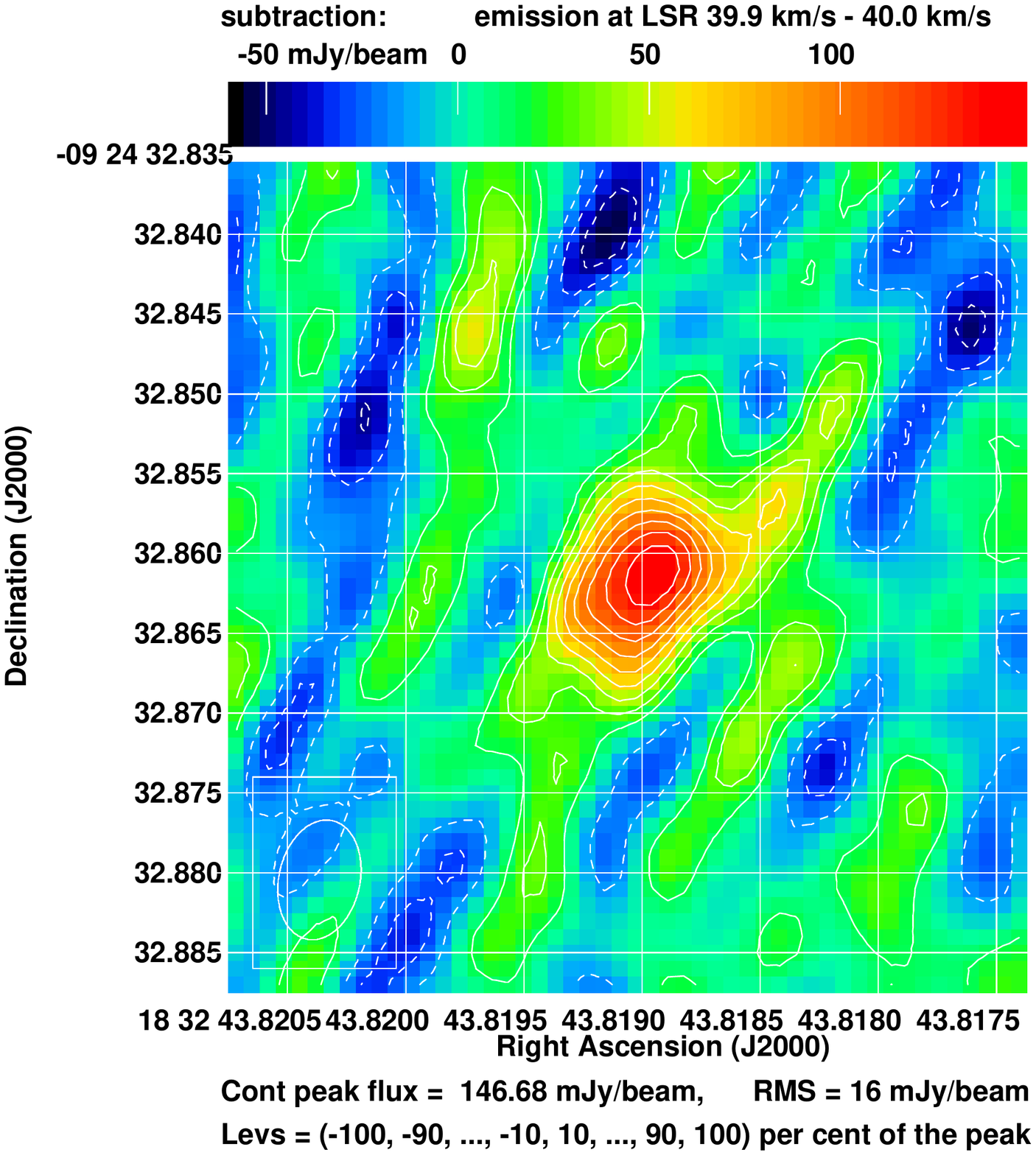}
\includegraphics[height=7cm,scale=0.35]{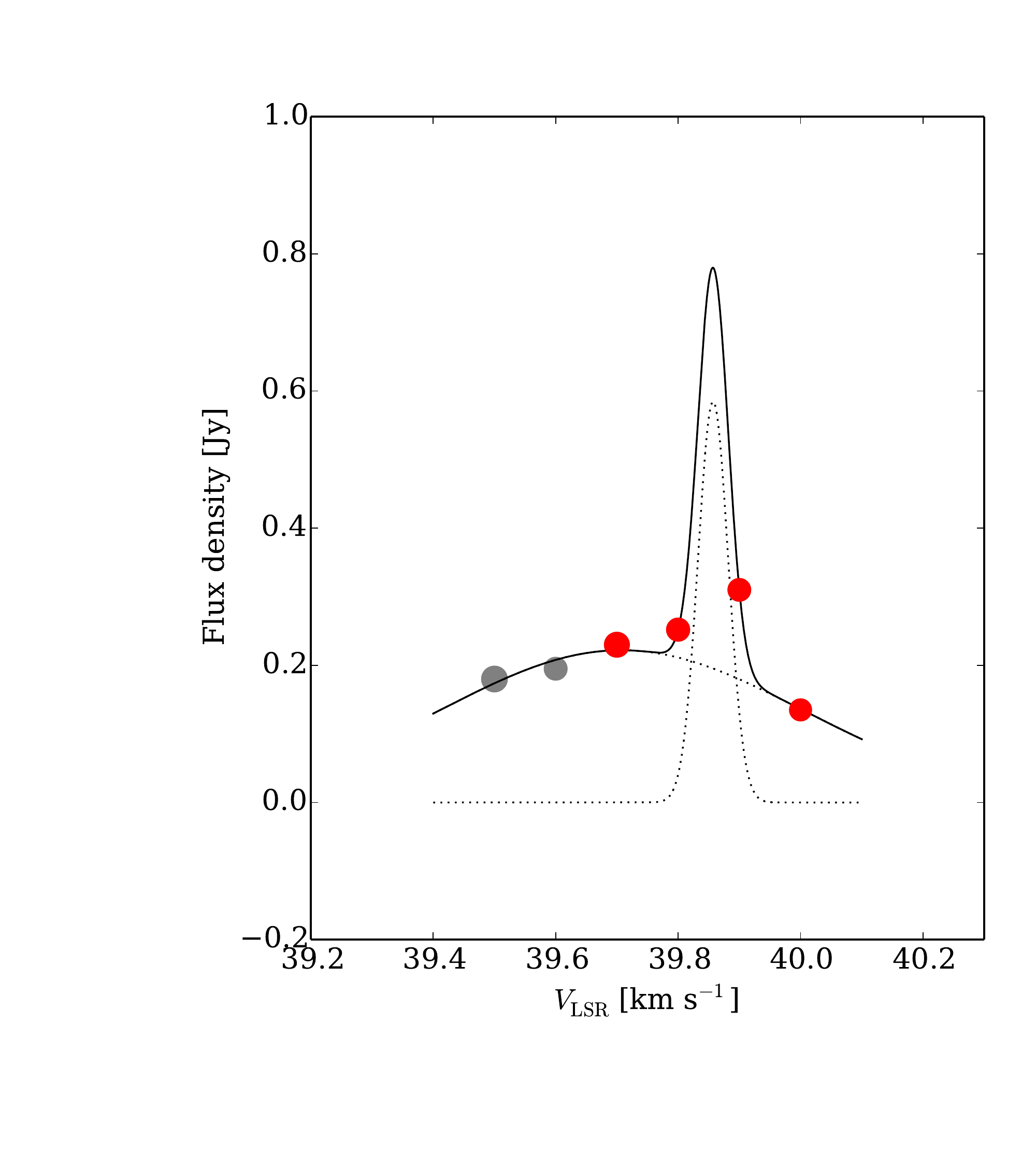}
\caption{Four of the brightest maser spots from Cloud {\it 1} of G22.435$-$00.169. They are indicated by red dots in the spectrum presented in the bottom panel. Their LSR velocities are given above each image. Three further images in the middle and bottom panels present the subtraction of emission as indicated above each image. 
The synthesized beam is presented in the left bottom corner of each image and the noise level is also given. The dot diameters in the spectrum are proportional to the deconvolved size of the maser spot.}
\label{cloud1}
\end{figure*}

\begin{figure*}
\includegraphics[scale=0.30]{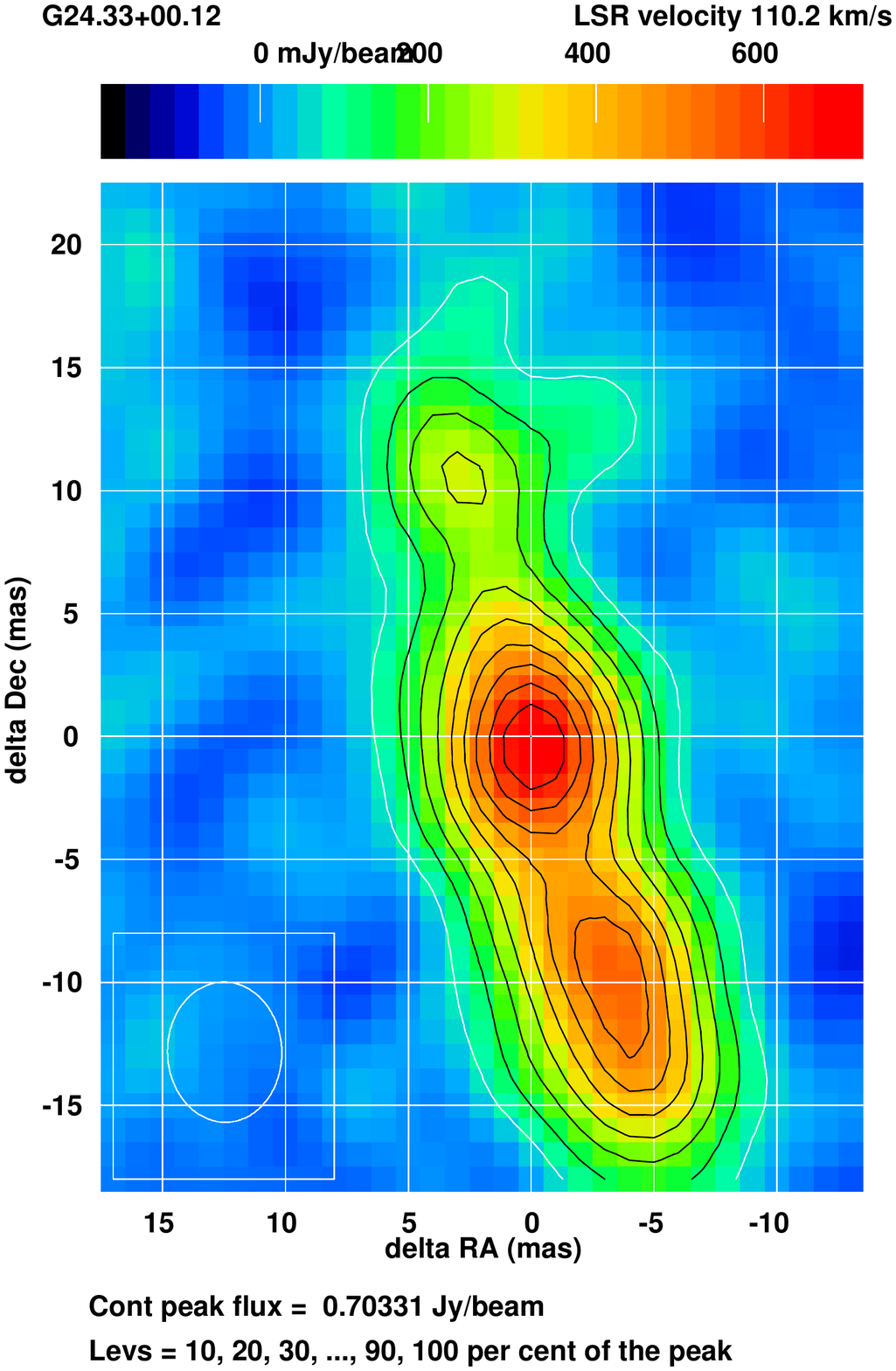}
\includegraphics[scale=0.30]{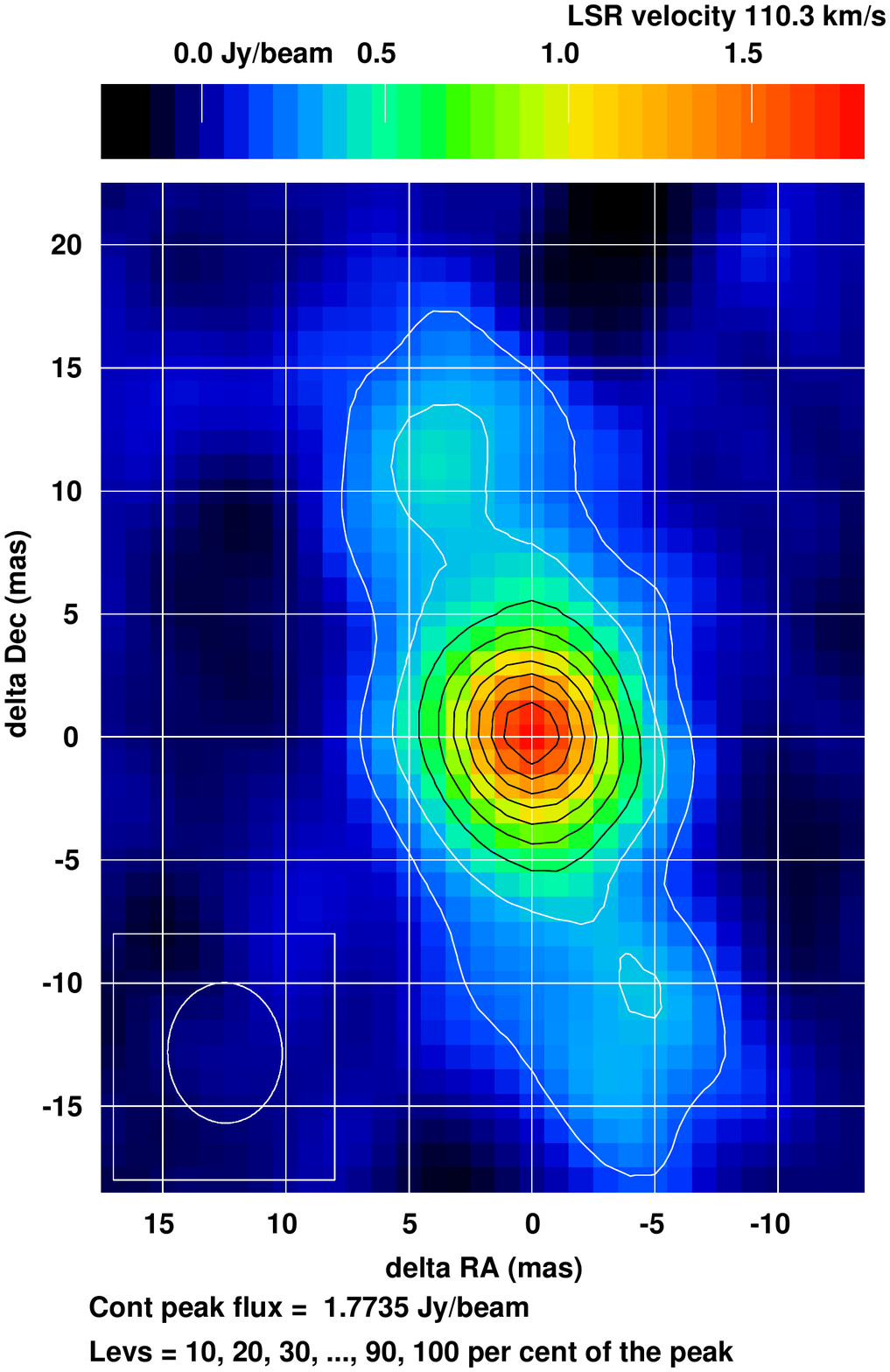}
\includegraphics[scale=0.30]{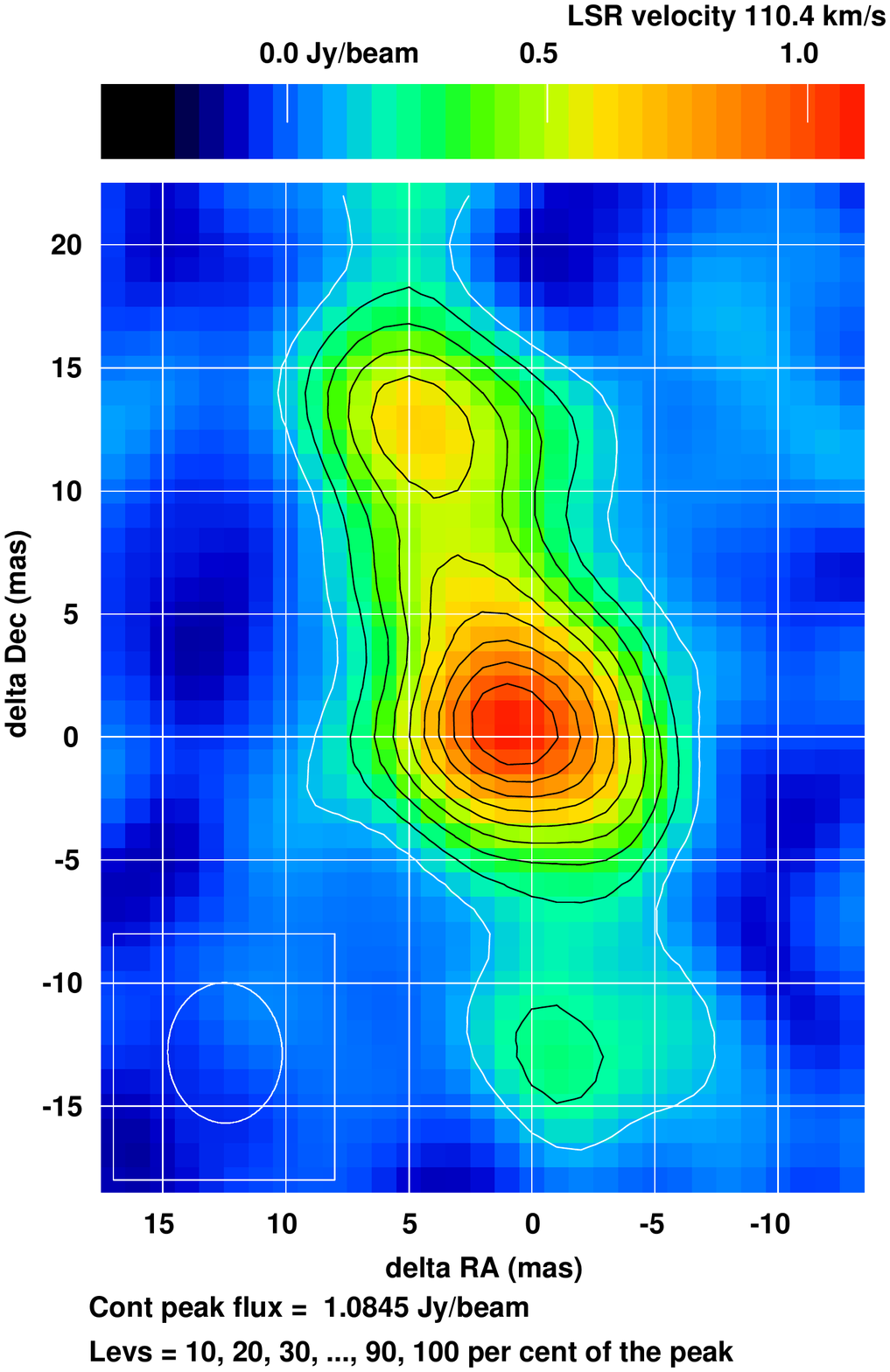}
\includegraphics[scale=0.30]{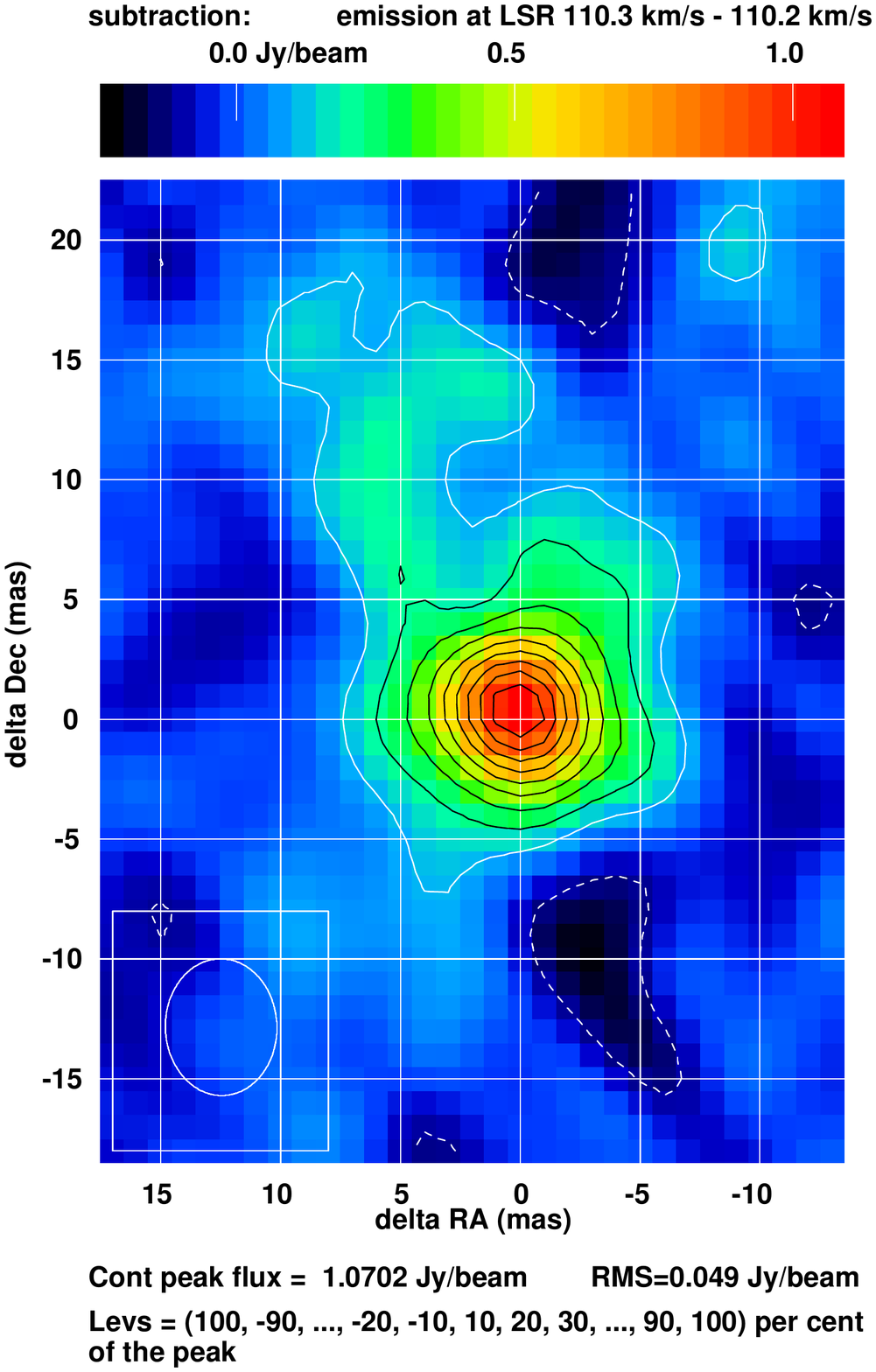}
\includegraphics[scale=0.30]{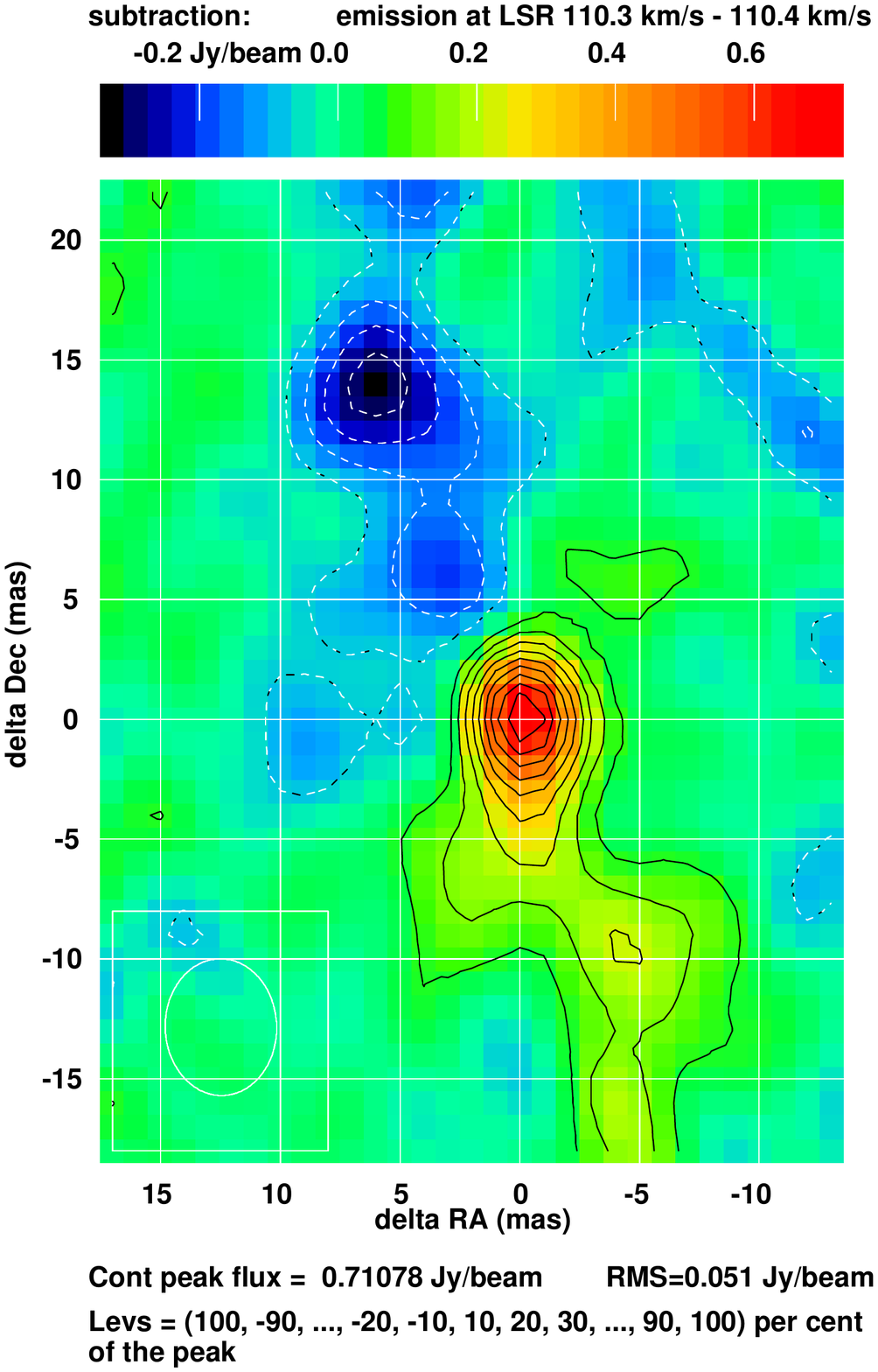}
\includegraphics[scale=0.35]{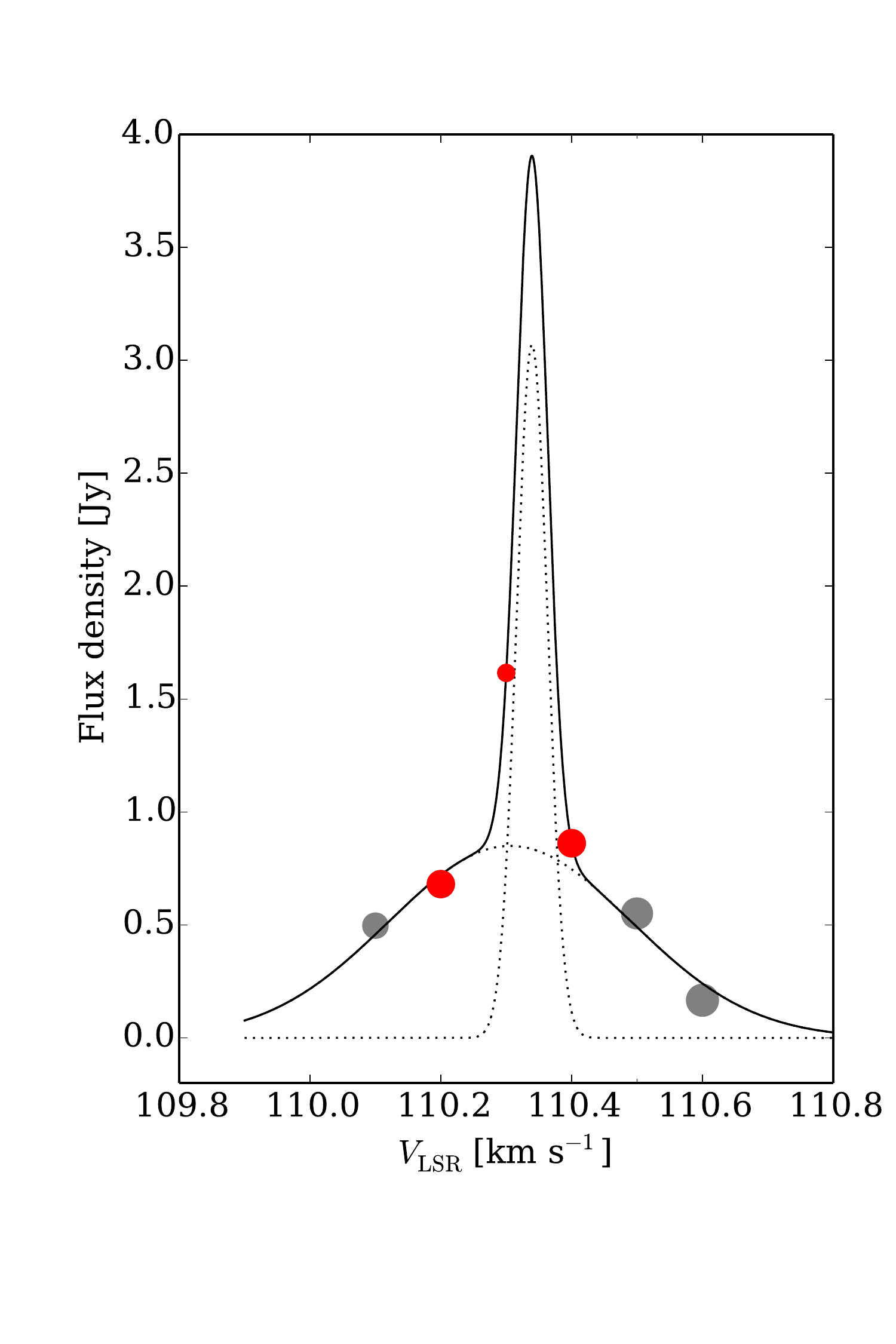}
\includegraphics[scale=0.35]{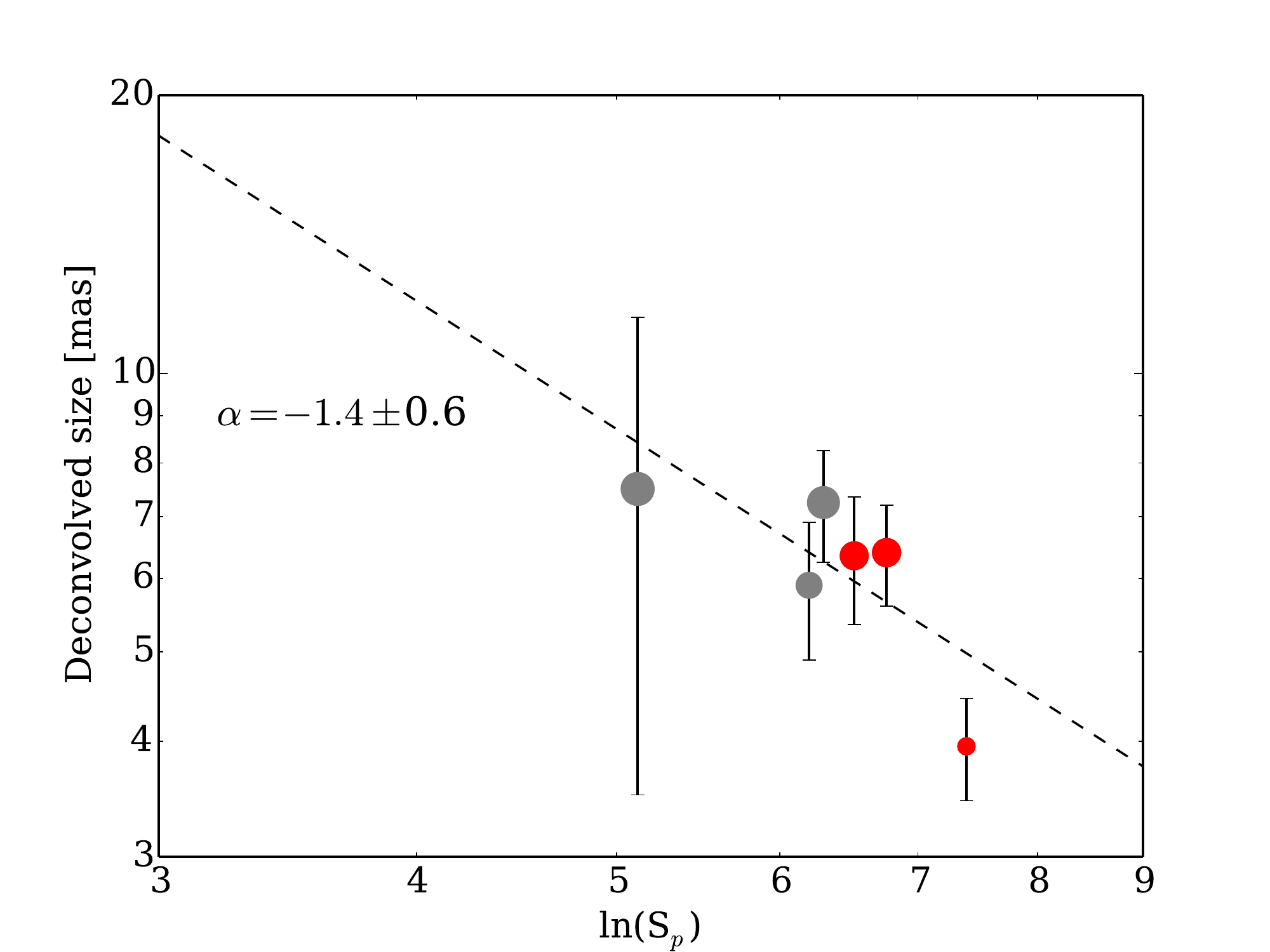}
\caption{Three of the brightest maser spots from Cloud {\it 5} of
G24.33$+$00.12. They are indicated by red dots in the spectrum presented in the
middle panel. Their LSR velocities are given above each image. The images in the middle panel present the subtraction of the emission
as indicated above each image. The synthesized beam is presented in the left bottom corners
of each image and the noise level is also given. The bottom panel presents the deconvolved size of a maser spot vs. natural logarithm of its peak flux density. The dot diameters in the spectrum and in the last dependence are proportional to the deconvolved size of the maser spot.}
\label{cloud2}
\end{figure*}

\end{appendix}

\begin{appendix}
\onecolumn
\section{Tables}
\begin{longtable}{c r r c c c c c ll}
\caption{\label{clouds} Parameters of 6.7~GHz methanol maser clouds with Gaussian velocity profiles.
The coordinates are relative to the brightest spot of each source 
listed in Table~\ref{results}. }\\     
\hline
Cloud & $\Delta$RA & $\Delta$Dec & V$_{\rm p}$ & V$_{\rm fit}$ & FWHM & S$_{\rm p}$ & S$_{\rm fit}$ & 
L$_{\rm proj}^a$ & V$_{\rm grad}^a$\\
& (mas) & (mas) & (km~s$^{-1}$) & (km~s$^{-1}$) & (km~s$^{-1}$) & (Jy~beam$^{-1}$) & (Jy~beam$^{-1}$) &
(mas) & (km~s$^{-1}$~mas$^{-1}$)\\
&&&&&&&&(AU)&(km~s$^{-1}$~au$^{-1}$)\\
\hline                    
\endfirsthead
\caption{Continued.}\\
\hline
Cloud & $\Delta$RA & $\Delta$Dec & V$_{\rm p}$ & V$_{\rm fit}$ & FWHM & S$_{\rm p}$ & S$_{\rm fit}$ & L$_{\rm
proj}^a$ & V$_{\rm grad}^a$\\
& (mas) & (mas) & (km~s$^{-1}$) & (km~s$^{-1}$) & (km~s$^{-1}$) & (Jy~beam$^{-1}$) & (Jy~beam$^{-1}$) & 
(mas) & (km~s$^{-1}$~mas$^{-1}$)\\
&&&&&&&&(AU)&(km~s$^{-1}$~au$^{-1}$)\\
\hline
\endhead
\hline
\endfoot
\hline
\endlastfoot
\multicolumn{10}{l}{\bf G20.237$+$00.065}\\
{\it 1} & -126.734 & 18.555 & 77.5 & 77.47 & 0.33 & 0.293 & 0.288 & 0.7(2.9) & 0.46(0.103) \\
{\it 2} & -59.880 & 45.139 & 76.3 & 76.34 & 0.31 & 0.487 & 0.498 & 1.0(4.4) & $-$\\
{\it 3} & -60.398 & 55.921 & 75.8 & 75.83 & 0.51 & 0.295 & 0.272 & 3.1(13.6) & 0.13(0.030) \\
{\it 4} & -30.345 & 9.301 & 75.3 & 75.27 & 0.50 & 0.522 & 0.475 & 0.9(3.8) & 0.51(0.116) \\
{\it 5} & -28.809 & 10.073 & 74.8 & 74.83 & 0.33 & 0.287 & 0.287 & 2.0(8.9) & $-$\\
{\it 6} & -66.879 & 96.597 & 73.6 & 73.56 & 0.41 & 0.242 & 0.247 & 2.0(9.0) & 0.10(0.022) \\
{\it 7} & -13.168 & -23.973 & 73.3 & 73.27 & 0.42 & 1.292 & 1.251 & 3.3(14.6) & 0.21(0.048) \\
{\it 8} & 35.952 & 60.392 & 73.2 & 73.22 & 0.39 & 4.078 & 3.356 & 5.9(25.8) & $-$\\
 &  & & & 72.91 & 0.64 & & 1.510 & &\\
{\it 9} & 10.576 & -7.160 & 73.2 & 73.20 & 0.27 & 0.755 & 0.755 & 0.8(3.3) & 0.26(0.060) \\
{\it 10} & -12.755 & -21.490 & 72.5 & 72.54 & 0.39 & 0.685 & 0.663 & 0.4(1.9) & $-$\\
{\it 11} & -7.290 & -19.179 & 72.1 & 72.11 & 0.27 & 1.024 & 0.950 & 4.0(17.7) & 0.07(0.017) \\
{\it 12} & 0.000 & 0.000 & 71.8 & 71.57 & 0.57 & 22.980 & 8.580 & 1.8(7.8) & 0.58(0.133)\\
& & & & 71.80 & 0.22 & & 18.189 & &\\
{\it 13} & 7.835 & -17.220 & 69.0 & 69.02 & 0.24 & 0.456 & 0.466 & 2.7(11.8) & 0.07(0.017) \\
{\it 14} & 6.614 & -21.423 & 68.5 & 68.53 & 0.24 & 0.160 & 0.167 & 1.2(5.2) & 0.17(0.039) \\
\multicolumn{10}{l}{\bf G20.239$+$00.065}\\
{\it 1} & -16.171 & 1.880 & 71.0 & 70.96 & 0.26 & 0.750 & 0.793 & 2.6(10.0) & 0.13(0.034) \\
{\it 2} & -21.830 & -0.856 & 70.4 & 70.37 & 0.30 & 1.925 & 1.863 & 2.5(9.7) & $-$\\
{\it 3} & 132.127 & -15.336 & 70.3 & 70.33 & 0.32 & 0.381 & 0.395 & 0.7(2.7) & 0.43(0.109) \\
{\it 4} & 0.000 & 0.000 & 61.0 & 61.03 & 0.41 & 2.414 & 2.231 & 1.8(7.2) & $-$\\
& & & & 60.54 & 0.51 & & 0.514 & & \\
\multicolumn{10}{l}{\bf G22.435$-$00.169}\\
{\it 1} & 4.274 & -77.749 & 39.9 & 39.86 & 0.06 & 0.310 & 0.584 & 1.2(2.7) & $-$\\
& & & & 39.71 & 0.70 & & 0.222 & &\\
{\it 2} & 1.846 & -69.665 & 38.6 & 38.56 & 0.22 & 1.034 & 0.981 & 4.9(10.7) & 0.06(0.028) \\
{\it 3} & -2.002 & -63.241 & 38.1 & 38.11 & 0.40 & 0.858 & 0.847 & 1.6(3.5) & $-$\\
& & & & 37.79 & 0.23 & & 0.538 & &\\
{\it 4} & 57.697 & -14.160 & 33.6 & 33.42 & 0.68 & 0.402 & 0.408 & 1.9(4.2) & $-$\\
{\it 5} & 0.000 & 0.000 & 29.3 & 29.31 & 0.26 & 2.442 & 2.226 & 10.9(23.9) & $-$\\
& & & & 29.65 & 0.34 & & 2.382 & & \\
{\it 6} & -70.486 & -173.053 & 28.9 & 28.85 & 0.25 & 1.509 & 1.673 & 2.3(5.0) & 0.18(0.080) \\
{\it 7} & -14.825 & -30.702 & 27.0 & 26.71 & 0.38 & 0.425 & 0.266 & 2.5(5.5) & $-$\\
& & & & 27.05 & 0.23 & & 0.256 & & \\
& & & & 27.54 & 1.14 & & 0.266 & & \\
{\it 8} & 89.834 & -168.271 & 25.6 & 25.57 & 0.20 & 0.784 & 0.737 & 1.7(3.6) & 0.24(0.110) \\
{\it 9} & 84.015 & -158.917 & 24.7 & 24.71 & 0.26 & 1.141 & 1.169 & 9.6(21.0) & 0.08(0.038)\\
& & & & 25.11 & 0.25 & & 0.586 & &\\
{\it 10} & 92.640 & -129.221 & 23.4 & 23.45 & 0.19 & 0.324 & 0.324 & 1.6(3.5) & 0.09(0.039) \\
\multicolumn{10}{l}{\bf G23.010$-$00.411}\\
{\it 1} & -194.798 & -213.716 & 83.2 & 83.21 & 0.33 & 0.550 & 0.562 & 0.4(1.9) & $-$\\
{\it 2} & 58.664 & -313.609 & 83.1 & 83.07 & 0.27 & 0.336 & 0.347 & 0.5(2.1) & 0.44(0.096)\\
{\it 3} & -191.247 & -220.462 & 82.9 & 82.81 & 0.40 & 0.303 & 0.306 & 0.7(3.1) & 0.44(0.096)\\
{\it 4} & -199.511 & -270.652 & 82.7 & 82.59 & 0.60 & 1.574 & 1.252 & 2.1(9.7) & $-$\\
{\it 5} & -145.399 & -311.958 & 82.4 & 82.36 & 0.42 & 21.919 & 21.659 & 3.5(16.1) & 0.43(0.093)\\
& & & & 81.66 & 0.38 & & 22.422 & & \\
{\it 6} & -194.522 & -207.374 & 81.6 & 81.98 & 0.22 & 5.371 & 2.081 & 1.8(8.0) & $-$\\
& & & & 81.56 & 0.33 & & 5.575 & & \\
{\it 7} & 193.502 & -466.083 & 81.4 & 81.35 & 0.23 & 2.845 & 3.078 & 3.2(14.6) & 0.09(0.021)\\
{\it 8} & -182.558 & -249.745 & 81.3 & 81.26 & 0.35 & 9.972 & 9.464 & 2.7(12.6) & 0.22(0.048)\\
{\it 9} & -193.125 & -266.819 & 80.7 & 80.71 & 0.28 & 40.806 & 42.257 & 1.2(5.6) & 0.53(0.115)\\
{\it 10} & 168.821 & -446.987 & 80.2 & 80.25 & 0.37 & 1.329 & 1.377 & 4.1(19.0) & 0.12(0.026)\\
{\it 11} & 4.018 & -415.900 & 79.9 & 79.83 & 0.64 & 0.378 & 0.366 & 5.7(26.2) & 0.05(0.011)\\
{\it 12} & -180.122 & -263.407 & 79.5 & 79.53 & 0.35 & 12.367 & 13.438 & 1.7(8.0) & $-$\\
{\it 13} & 100.850 & -366.987 & 79.3 & 79.30 & 0.34 & 3.695 & 3.662 & 2.0(9.0) & 0.25(0.055)\\
{\it 14} & -67.630 & -365.271 & 79.0 & 78.92 & 0.74 & 0.728 & 0.614 & 17.1(78.7) & 0.06(0.013)\\
{\it 15} & -291.282 & -16.504 & 78.8 & 78.78 & 0.32 & 0.642 & 0.663 & 0.5(2.3) & 0.60(0.130)\\
{\it 16} & -125.788 & -12.313 & 78.7 & 78.69 & 0.27 & 0.372 & 0.373 & 0.4(1.6) & 0.56(0.123)\\
{\it 17} & 219.223 & -371.722 & 78.6 & 78.59 & 0.36 & 0.301 & 0.284 & 5.4(24.6) & 0.07(0.016)\\
{\it 18} & 221.708 & -363.425 & 78.4 & 78.44 & 0.20 & 0.246 & 0.282 & 0.6(3.0) & 0.31(0.068)\\
{\it 19} & 22.345 & -337.273 & 78.2 & 78.21 & 0.30 & 0.122 & 0.121 & 1.1(5.0) & 0.22(0.048)\\
{\it 20} & 50.430 & -81.453 & 76.8 & 76.68 & 0.54 & 4.669 & 4.545 & 3.9(18.1) & 0.30(0.066)\\
& & & & 77.28 & 0.51 & & 2.369 & & \\
{\it 21} & 46.076 & -81.377 & 76.7 & 76.63 & 0.84 & 6.375 & 5.773 & 2.2(9.9) & $-$\\
{\it 22} & 31.486 & -81.789 & 75.5 & 75.56 & 0.55 & 36.047 & 32.809 & 3.9(18.0) & $-$\\
{\it 23} & 34.037 & -77.577 & 75.5 & 75.96 & 0.39 & 29.831 & 22.713 & 8.3(37.9) & 0.21(0.045)\\
& & & & 75.48 & 0.57 & & 25.191 & & \\
{\it 24} & 0.000 & 0.000 & 74.7 & 74.77 & 0.27 & 72.617 & 68.955 & 0.5(2.4) & 0.96(0.210)\\
{\it 25} & 23.300 & -69.860 & 74.7 & 74.78 & 0.26 & 40.359 & 41.013 & 12.7(58.5) & 0.05(0.012)\\
{\it 26} & 42.041 & -48.323 & 74.6 & 74.98 & 0.49 & 64.844 & 63.688 & 12.7(58.1) & 0.12(0.026)\\
& & & & 74.51 & 0.32 & & 61.091 & & \\
{\it 27} & 68.423 & -33.641 & 74.5 & 74.48 & 0.23 & 9.851 & 9.727 & 1.4(6.2) & 0.22(0.048)\\
{\it 28} & 202.398 & 114.089 & 74.3 & 74.26 & 0.30 & 38.381 & 39.732 & 3.5(16.0) & 0.14(0.031)\\
{\it 29} & 76.724 & -36.759 & 74.3 & 74.32 & 0.29 & 10.449 & 10.886 & 2.7(12.6) & $-$\\
{\it 30} & 100.084 & -35.954 & 73.8 & 73.77 & 0.30 & 6.590 & 6.848 & 4.3(19.6) & 0.13(0.028)\\
{\it 30} & 49.372 & -40.654 & 74.1 & 74.14 & 0.24 & 4.296 & 4.497 & 2.7(12.6) & 0.11(0.024)\\
{\it 31} & 12.488 & -6.688 & 74.0 & 74.04 & 0.30 & 14.210 & 14.086 & 0.7(3.1) & $-$\\
{\it 32} & 92.294 & -22.367 & 74.0 & 73.98 & 0.23 & 3.299 & 3.192 & 4.2(19.2) & 0.06(0.012)\\
{\it 33} & 37.229 & -38.862 & 73.8 & 73.78 & 0.48 & 8.999 & 9.690 & 2.4(11.0) & 0.29(0.064)\\
{\it 34} & 21.959 & -32.309 & 73.4 & 73.52 & 0.52 & 3.573 & 3.910 & 4.2(19.4) & 0.17(0.036)\\
{\it 35} & 76.315 & -87.397 & 73.3 & 73.36 & 0.54 & 3.357 & 3.203 & 0.9(4.0) & $-$\\
{\it 36} & 171.269 & 13.669 & 73.3 & 73.27 & 0.25 & 2.676 & 2.665 & 0.7(3.3) & $-$\\
{\it 37} & -71.484 & 30.307 & 73.2 & 73.20 & 0.22 & 6.161 & 6.171 & 1.2(5.5) & 0.28(0.061)\\
{\it 38} & 127.879 & -33.335 & 72.7 & 72.66 & 0.26 & 24.507 & 21.717 & 5.3(24.2) & 0.17(0.037)\\
& & & & 72.89 & 0.45 & & 8.629 & & \\
{\it 39} & 44.291 & -96.383 & 72.2 & 72.18 & 0.50 & 0.578 & 0.575 & 1.1(5.0) & $-$\\
\multicolumn{10}{l}{\bf G24.33$+$00.12}\\
{\it 1} & -234.435 & 104.783 & 112.9 & 112.90 & 0.21 & 0.365 & 0.365 & 1.0(9.9) & $-$\\
{\it 2} & -231.190 & 114.450 & 112.9 & 112.90 & 0.21 & 0.442 & 0.442 & 2.7(25.7) & $-$\\
{\it 3} & -20.363 & -49.060 & 112.0 & 111.98 & 0.37 & 0.465 & 0.470 & 1.3(12.4) & $-$\\
{\it 4} & -23.818 & -59.485 & 111.9 & 111.96 & 0.52 & 0.420 & 0.408 & 0.8(7.8) & $-$\\
{\it 5} & 0.000 & 0.000 & 110.3 & 110.34 & 0.06 & 1.616 & 3.069 & 3.5(32.9) & 0.14(0.015)\\
& & & & 110.31 & 0.44 & & 0.850 & & \\
\multicolumn{10}{l}{\bf G24.494$-$00.038}\\
{\it 1} & 0.000 & 0.000 & 115.1 & 115.51 & 0.36 & 2.850 & 2.133 & 3.6(21.1) & $-$\\
& & & & 115.09 & 0.36 & & 2.721 & & \\
{\it 2} & 0.842 & 9.978 & 114.8 & 114.72 & 0.57 & 1.991 & 1.997 & 5.5(31.9) & 0.16(0.028)\\
& & & & 114.06 & 0.22 & & 0.265 & &\\
{\it 3} & -7.780 & 20.377 & 113.7 & 113.82 & 0.44 & 0.382 & 0.382 & 1.6(9.4) & 0.25(0.042)\\
{\it 4} & 226.976 & 184.659 & 113.1 & 113.16 & 0.34 & 0.384 & 0.392 & 1.5(8.8) & 0.29(0.050)\\
{\it 5} & -137.792 & -36.379 & 113.2 & 113.21 & 0.29 & 0.374 & 0.352 & 2.0(11.9) & 0.20(0.034)\\
{\it 6} & 245.766 & 180.996 & 111.5 & 111.51 & 0.32 & 1.682 & 1.676 & 0.5(3.1) & $-$\\
{\it 7} & 168.173 & -47.488 & 109.4 & 109.45 & 0.38 & 1.567 & 1.449 & 2.5(14.5) & 0.28(0.048)\\
{\it 8} & 191.525 & -137.901 & 109.6 & 109.60 & 0.28 & 0.837 & 0.836 & 0.7(3.8) & $-$\\
\multicolumn{10}{l}{\bf G24.790$+$00.083}\\
{\it 1} & -254.737 & -117.751 & 116.6 & 116.49 & 0.49 & 0.142 & 0.134 & 2.8(21.9) & $-$\\
{\it 2} & -141.611 & 77.971 & 114.7 & 114.74 & 0.49 & 3.191 & 3.310 & 0.7(5.1) & $-$\\
{\it 3} & -38.114 & -28.927 & 114.5 & 114.48 & 0.29 & 0.331 & 0.338 & 1.3(10.2) & 0.15(0.020)\\
{\it 4} & -76.344 & 99.104 & 114.2 & 114.17 & 0.36 & 12.151 & 12.390 & 0.6(4.5) & 0.92(0.120)\\
{\it 5} & -36.854 & -27.814 & 113.8 & 113.78 & 0.55 & 0.276 & 0.277 & 0.6(4.6) & 0.33(0.043)\\
{\it 6} & 0.000 & 0.000 & 113.4 & 113.39 & 0.34 & 55.832 & 56.937 & 1.0(7.3) & 0.82(0.106)\\
{\it 7} & -1403.008 & 1082.429 & 113.0 & 112.97 & 0.18 & 0.585 & 0.562 & 2.6(19.9) & 0.09(0.012)\\
{\it 8} & -1389.840 & 1059.146 & 112.5 & 112.52 & 0.36 & 0.403 & 0.350 & 8.2(62.8) & 0.05(0.006)\\
{\it 9} & -1385.532 & 1035.473 & 112.4 & 112.36 & 0.35 & 5.931 & 6.011 & 2.2(16.9) & $-$\\
{\it 10} & -1378.794 & 1056.553 & 112.3 & 112.31 & 0.40 & 0.482 & 0.483 & 1.7(12.7) & 0.12(0.016)\\
{\it 11} & -1154.240 & 998.717 & 112.1 & 112.12 & 0.34 & 2.253 & 2.316 & 2.1(16.5) & 0.19(0.024)\\
{\it 12} & -1390.563 & 1036.280 & 111.9 & 111.94 & 0.33 & 6.627 & 6.488 & 4.3(32.9) & $-$\\
{\it 13} & -1386.732 & 1015.998 & 111.5 & 111.54 & 0.18 & 0.376 & 0.370 & 2.2(16.8) & 0.09(0.012)\\
{\it 14} & -1357.891 & 983.497 & 111.5 & 111.56 & 0.23 & 0.254 & 0.264 & 5.2(40.2) & $-$\\
{\it 15} &  -1323.012 & 695.680 & 110.4 & 110.38 & 0.36 & 10.985 & 11.739 & 2.2(17.3) & 0.34(0.044)\\
{\it 16} & -1314.433 & 702.126 & 109.5 & 109.53 & 0.24 & 1.767 & 1.736 & 3.1(24.1) & 0.17(0.022)\\
{\it 17} & -438.847 & -403.487 & 108.4 & 108.36 & 0.35 & 0.230 & 0.239 & 3.9(30.1) & 0.05(0.007)\\
{\it 18} & -426.166 & -404.449 & 108.3 & 108.31 & 0.38 & 2.431 & 2.431 & 0.6(4.9) & 0.62(0.081)\\
{\it 19} & -376.519 & -400.742 & 108.0 & 108.00 & 0.30 & 4.453 & 4.473 & 15.3(117.7) & 0.04(0.005)\\
{\it 20} & -399.395 & -404.246 & 108.1 & 108.15 & 0.28 & 2.173 & 2.369 & 1.6(12.2) & 0.19(0.025)\\
{\it 21} & -408.564 & -407.577 & 107.9 & 107.96 & 0.26 & 5.130 & 5.275 & 1.2(9.3) & 0.28(0.036)\\
{\it 22} & -432.878 & -405.869 & 107.8 & 107.77 & 0.27 & 8.003 & 8.592 & 2.3(17.6) & 0.10(0.014)\\
{\it 23} & -442.739 & -405.074 & 107.7 & 107.64 & 0.40 & 12.215 & 12.998 & 4.5(34.5) & 0.16(0.020)\\
{\it 24} & -1447.730 & 554.386 & 106.5 & 106.53 & 0.23 & 0.394 & 0.382 & 1.8(14.1) & 0.16(0.021)\\
\multicolumn{10}{l}{\bf G24.850$+$00.087}\\
{\it 1} & -109.496 & -35.007 & 114.9 & 114.89 & 0.42 & 2.413 & 2.372 & 2.7(21.3) & $-$\\
{\it 2} & -109.305 & -47.761 & 114.9 & 114.93 & 0.35 & 0.547 & 0.562 & 0.2(1.4) & $-$\\
{\it 3} & -106.147 & -28.145 & 114.6 & 114.56 & 0.32 & 3.144 & 3.277 & 1.8(14.2) & $-$\\
{\it 4} & -42.638 & -28.227 & 113.7 & 113.62 & 1.00 & 0.422 & 0.408 & 6.6(52.9) & $-$\\
{\it 5} & -49.674 & -15.365 & 112.3 & 112.31 & 0.35 & 0.531 & 0.534 & 1.5(11.6) & $-$\\
{\it 6} & -50.345 & -14.013 & 110.9 & 110.90 & 0.33 & 0.273 & 0.273 & 2.0(15.8) & $-$\\
{\it 7} & 0.000 & 0.000 & 110.2 & 110.22 & 0.71 & 5.955 & 6.054 & 4.1(33.0) & 0.35(0.044)\\
& & & & 109.67 & 0.42 & & 3.247 & & \\
{\it 8} & -49.529 & 36.453 & 108.0 & 108.06 & 0.25 & 1.207 & 1.292 & 0.9(7.3) & $-$\\
{\it 9} & -24.542 & 33.906 & 107.8 & 107.67 & 0.90 & 0.249 & 0.231 & 1.5(12.0) & 0.16(0.020)\\
{\it 10} & -36.860 & 29.541 & 107.9 & 107.87 & 0.15 & 0.480 & 0.493 & 0.6(4.8) & 0.23(0.029)\\
\multicolumn{10}{l}{\bf G28.011$-$00.426}\\
{\it 1} & 92.574 & 3.003 & 28.0 & 28.02 & 0.46 & 0.413 & 0.411 & 2.6(2.9) & $-$\\
{\it 2} & 86.825 & 0.042 & 25.4 & 25.39 & 0.18 & 0.281 & 0.247 & 6.7(7.4) & 0.06(0.054)\\
{\it 3} & 102.492 & -8.423 & 24.7 & 24.67 & 0.24 & 0.479 & 0.436 & 1.8(1.9) & $-$\\
{\it 4} & 291.51 & -13.000 & 23.5 & 23.51 & 0.31 & 0.652 & 0.681 & 0.9(1.0) & $-$\\
{\it 5} & 61.879 & 59.222 & 17.8 & 17.84 & 0.36 & 0.193 & 0.179 & 1.4(1.5) & $-$\\
{\it 6} & -13.026 & -17.706 & 17.0 & 16.98 & 0.41 & 1.192 & 1.216 & 4.9(5.4) & $-$\\
{\it 7} & 0.000 & -0.000 & 16.1 & 16.14 & 0.42 & 1.262 & 1.208 & 7.9(8.7) & 0.11(0.104)\\
\multicolumn{10}{l}{\bf G29.978$-$00.048}\\
{\it 1} & -4.834 & -0.926 & 104.2 & 104.17 & 0.34 & 6.845 & 5.133 & 3.3(29.2) & 0.46(0.053)\\
& & & & 104.58 & 0.80 & & 3.447 & &\\
{\it 2} & 0.000 & 0.000 & 103.5 & 103.49 & 0.52 & 31.162 & 30.431 & 3.5(30.4) & $-$\\
& & & & 102.92 & 0.43 & & 21.829 & & \\
{\it 3} & 90.625 & -42.016 & 102.5 & 102.48 & 0.33 & 6.235 & 6.322 & 6.0(52.7) & 0.19(0.022)\\
& & & & 101.79 & && 1.198 & &\\
{\it 4} & 34.319 & 74.130 & 100.9 & 100.90 & 0.41 & 0.858 & 0.825 & 0.4(3.5) & $-$\\
{\it 5} & -8.366 & 55.249 & 100.9 & 100.90 & 0.22 & 0.285 & 0.285 & 0.5(4.3) & 0.41(0.046)\\
{\it 6} & 16.719 & 54.398 & 99.9 & 99.89 & 0.53 & 0.296 & 0.264 & 0.7(6.0) & 0.50(0.056)\\
{\it 7} & 45.163 & 75.117 & 99.8 & 99.80 & 0.32 & 1.253 & 1.189 & 2.2(19.6) & $-$\\
& & & & 100.26 & 0.35 & & 0.493 & & \\
{\it 8} & 38.405 & 72.979 & 99.4 & 99.39 & 0.28 & 0.188 & 0.189 & 0.8(7.4) & 0.24(0.027)\\
{\it 9} & 20.622 & 57.507 & 99.3 & 99.30 & 0.46 & 0.460 & 0.455 & 1.4(12.2) & 0.29(0.033)\\
{\it 10} & -39.646 & 93.832 & 98.4 & 98.36 & 0.24 & 10.922 & 6.650 & 6.4(56.3) & 0.22(0.025)\\
& & & & 98.46 & 0.62 & & 4.320 & & \\
& & & & 99.24 & 0.35 & & 0.620 & & \\
{\it 11} & -18.450 & 99.110 & 98.3 & 98.30 & 0.26 & 1.242 & 1.242 & 0.2(2.2) & 0.81(0.093)\\
{\it 12} & 18.050 & 54.211 & 98.0 & 98.05 & 0.40 & 1.110 & 1.139 & 0.8(6.9) & $-$\\
{\it 13} & 36.858 & 75.497 & 98.0 & 97.94 & 0.27 & 2.256 & 2.398 & 0.3(2.5) & 1.06(0.120)\\
{\it 14} & 50.310 & 84.710 & 97.6 & 97.60 & 0.34 & 2.957 & 2.918 & 0.5(4.6) & 0.95(0.108)\\
{\it 15} & 58.784 & 72.340 & 97.3 & 97.27 & 0.30 & 6.905 & 6.595 & 3.1(27.4) & 0.21(0.023)\\
& & & & 97.52 & & & 1.655 & &\\
{\it 16} & 40.918 & 76.075 & 97.0 & 96.96 & 0.20 & 13.643 & 12.418 & 2.3(20.5) & 0.45(0.051)\\
& & & & 97.39 & & & 1.399 & &\\
\multicolumn{10}{l}{\bf G30.198$-$00.169}\\
{\it 1} & 2.829 & 10.127 & 110.4 & 110.35 & 0.44 & 5.812 & 5.559 & 0.3(1.8) & 2.56(0.434)\\
{\it 2} & 4.614 & -6.567 & 110.3 & 110.77 & 0.25 & 2.189 & 0.926 & 3.5(20.6) & 0.37(0.063)\\
& & & & 110.24 & 0.58 & & 2.241 & & \\
{\it 3} & 1.343 & 5.443 & 109.7 & 109.72 & 0.34 & 1.287 & 1.344 & 2.9(16.8) & 0.12(0.020)\\
{\it 4} & 1.063 & -1.381 & 109.1 & 109.10 & 0.53 & 5.271 & 5.176 & 4.4(26.0) & 0.14(0.023)\\
{\it 5} & 0.000 & 0.000 & 108.2 & 108.07 & 0.40 & 14.690 & 9.977 & 4.0(23.5) & 0.35(0.060)\\
& & & & 108.53 & 0.71 & & 11.575 & & \\
{\it 6} & -90.183 & -41.008 & 105.4 & 105.40 & 0.32 & 0.710 & 0.713 & 8.9(52.4) & 0.05(0.008)\\
{\it 7} & -166.889 & -62.443 & 105.2 & 105.17 & 0.22 & 1.015 & 1.073 & 1.8(10.7) & 0.22(0.037)\\
{\it 8} & -87.819 & -55.249 & 104.9 & 104.96 & 0.31 & 1.028 & 0.988 & 0.8(4.5) & 0.65(0.111)\\
{\it 9} & 5.698 & -146.731 & 104.8 & 104.73 & 0.33 & 0.830 & 0.902 & 0.5(2.8) & 1.13(0.192)\\
{\it 10} & -95.825 & -56.526 & 104.5 & 104.49 & 0.20 & 0.442 & 0.446 & 0.6(3.4) & 0.35(0.059)\\
{\it 11} & -175.386 & -23.978 & 104.4 & 104.45 & 0.45 & 0.264 & 0.262 & 2.9(17.4) & 0.10(0.017)\\
{\it 12} & -114.333 & -71.815 & 103.4 & 103.38 & 0.23 & 1.387 & 1.262 & 7.9(46.7) & $-$\\
& & & & 103.04 & 0.41 & & 0.752 & & \\
{\it 13} & -107.274 & -87.316 & 100.8 & 100.83 & 0.28 & 0.689 & 0.728 & 0.7(4.4) & 0.54(0.091)\\
\multicolumn{10}{l}{\bf G30.224$-$00.180}\\
{\it 1} & 0.000 & 0.000 & 113.4 & 113.39 & 0.28 & 2.557 & 1.595 & 3.8(24.9) & 0.34(0.052)\\
& & & & 112.99 & 0.88 & & 1.712 & & \\
{\it 2} & -19.347 & 14.031 & 112.5 & 112.46 & 0.39 & 0.493 & 0.505 & 1.2(8.0) & $-$\\
{\it 3} & 7.905 & 13.522 & 112.2 & 112.15 & 0.24 & 0.411 & 0.460 & 0.6(3.9) & 0.33(0.051)\\
{\it 4} & 9.897 & 9.227 & 111.4 & 111.43 & 0.54 & 0.316 & 0.308 & 1.4(9.0) & 0.29(0.044)\\
{\it 5} & 5.545 & 2.937 & 111.3 & 111.32 & 0.27 & 1.101 & 1.102 & 0.8(5.0) & $-$\\
{\it 6} & 18.440 & 0.031 & 111.1 & 111.09 & 0.36 & 0.223 & 0.223 & 0.4(2.7) & 0.48(0.073)\\
\multicolumn{10}{l}{\bf G32.744$-$00.076}\\
{\it 1} & -285.122 & 1075.249 & 39.1 & 39.16 & 0.25 & 4.192 & 4.211 & 0.9(10.5) & 0.45(0.038)\\
{\it 2} & -205.913 & 1064.347 & 38.3 & 37.97 & 0.27 & 8.882 & 7.659 & 9.6(113.8) & 0.12(0.011)\\
& & & & 38.24 & 0.23 & & 6.681 & & \\
& & & & 38.57 & 0.49 & & 7.878 & & \\
{\it 3} & -225.144 & 1068.596 & 38.4 & 38.46 & 0.20 & 5.298 & 5.112 & 0.3(3.0) & 0.95(0.080)\\
{\it 4} & -131.195 & 1053.808 & 37.8 & 38.18 & 0.88 & 0.512 & 0.870 & 0.2(2.3) & 0.68(0.057)\\
{\it 5} & -140.279 & 1057.763 & 37.6 & 37.61 & 0.21 & 1.092 & 1.086 & 4.0(47.0) & 0.11(0.009)\\
{\it 6} & -133.037 & 1057.820 & 36.3 & 36.29 & 0.16 & 0.422 & 0.430 & 0.5(6.5) & $-$\\
{\it 7} & -0.639 & -0.582 & 34.4 & 34.44 & 0.29 & 0.183 & 0.193 & 0.4(4.9) & 0.48(0.041)\\
{\it 8} & -137.005 & 72.654 & 34.1 & 34.12 & 0.42 & 0.668 & 0.595 & 2.9(34.4) & 0.17(0.015)\\
{\it 9} & 0.000 & 0.000 & 33.4 & 33.45 & 0.41 & 10.535 & 10.240 & 1.9(22.2) & 0.42(0.036)\\
{\it 10} & -26.065 & -20.175 & 32.1 & 32.10 & 0.31 & 6.395 & 6.699 & 1.3(14.9) & 0.42(0.036)\\
{\it 11} & -15.255 & -16.117 & 31.7 & 31.73 & 0.28 & 1.082 & 1.086 & 1.1(13.1) & 0.27(0.023)\\
{\it 12} & -41.825 & 245.805 & 30.7 & 30.69 & 0.32 & 0.878 & 0.868 & 0.1(1.0) & $-$\\
{\it 13} & -120.351 & 93.973 & 30.2 & 30.17 & 0.40 & 3.556 & 3.654 & 0.5(6.3) & 1.19(0.101)\\
{\it 14} & -115.868 & 91.832 & 29.3 & 29.26 & 0.43 & 0.468 & 0.453 & 2.7(31.6) & 0.26(0.022)\\
& & & & 29.63 & 0.29 & & 0.247 & & \\
\multicolumn{10}{l}{\bf G34.245$+$00.134}\\
{\it 1} & 46.804 & -0.858 & 62.6 & 62.57 & 0.30 & 0.193 & 0.177 & 0.6(2.0) & $-$\\
{\it 2} & 42.323 & -0.566 & 62.0 & 62.11 & 0.43 & 0.084 & 0.090 & 0.8(2.5) & 0.52(0.158)\\
{\it 3} & 58.040 & -9.770 & 61.9 & 61.91 & 0.21 & 0.101 & 0.102 & 0.9(2.9) & 0.23(0.068)\\
{\it 4} & 25.169 & 31.365 & 61.5 & 61.74 & 0.18 & 0.102 & 0.031 & 4.8(15.8) & 0.10(0.032)\\
& & & & 61.51 & 0.39 & & 0.097 & & \\
{\it 5} & 36.771 & 4.503 & 61.2 & 61.13 & 0.27 & 1.393 & 1.125 & 2.1(7.0) & $-$\\
& & & & 61.35 & 0.44 & & 0.687 & & \\
{\it 6} & 9.798 & 41.269 & 60.3 & 60.23 & 0.83 & 0.504 & 0.127 & 12.0(39.4) & 0.08(0.025)\\
& & & & 60.31 & 0.16 & & 0.425 & &\\
{\it 7} & 19.577 & 12.389 & 59.5 & 59.52 & 0.27 & 0.128 & 0.134 & 0.9(2.8) & $-$\\
{\it 8} & 2.494 & 3.544 & 58.6 & 58.57 & 0.25 & 0.340 & 0.350 & 3.0(9.9) & 0.10(0.030)\\
{\it 9} & -63.596 & 70.975 & 57.0 & 57.05 & 0.21 & 0.628 & 0.703 & 0.5(1.7) & 0.59(0.180)\\
{\it 10} & -36.674 & 6.257 & 56.9 & 56.92 & 0.32 & 0.712 & 0.766 & 1.5(5.0) & 0.36(0.109)\\
{\it 11} & 0.000 & 0.000 & 55.0 & 55.35 & 0.35 & 1.440 & 1.164 & 7.4(24.4) & 0.17(0.051)\\
& & & & 54.97 & 0.37 & & 1.268 & & \\
\multicolumn{10}{l}{\bf G34.258$+$00.153}\\
{\it 1} & -11.380 & -24.303 & 58.1 & 58.05 & 0.34 & 0.365 & 0.387 & 6.5(22.1) & 0.03(0.009)\\
{\it 2} & 8.011 & 20.121 & 58.0 & 57.96 & 0.24 & 3.797 & 3.841 & 4.2(14.4) & 0.10(0.031)\\
{\it 3} & 0.000 & 0.000 & 57.6 & 57.61 & 0.24 & 5.862 & 5.991 & 8.6(29.2) & 0.06(0.017)\\
{\it 4} & 147.389 & -24.092 & 56.7 & 56.72 & 0.21 & 0.232 & 0.237 & 3.5(11.9) & 0.11(0.034)\\
\multicolumn{10}{l}{\bf G34.396$+$00.222}\\
{\it 1} & -52.259 & 21.820 & 62.1 & 62.27 & 1.01 & 0.486 & 0.474 & 16.6(54.8) & $-$\\
{\it 2} & -35.981 & -117.128 & 60.6 & 60.58 & 0.54 & 0.199 & 0.181 & 1.3(4.4) & 0.40(0.122)\\
{\it 3} & -23.973 & -127.623 & 60.4 & 60.43 & 0.35 & 0.147 & 0.137 & 0.4(1.4) & $-$\\
{\it 4} & -29.204 & -104.930 & 60.4 & 60.54 & 0.65 & 0.234 & 0.214 & 2.9(9.5) & 0.24(0.074)\\
{\it 5} & -20.989 & 32.721 & 60.1 & 59.96 & 0.60 & 0.358 & 0.354 & 1.8(5.8) & 0.34(0.104)\\
{\it 6} & -16.871 & 42.875 & 59.8 & 59.76 & 0.22 & 0.375 & 0.337 & 4.5(14.9) & 0.13(0.040)\\
& & & & 60.06 & 0.34 & & 0.378 & & \\
{\it 7} & 0.000 & 0.000 & 55.6 & 55.68 & 0.41 & 5.227 & 5.115 & 12.9(42.7) & 0.06(0.019)\\
\multicolumn{10}{l}{\bf G35.025$+$00.350}\\
{\it 1} & -76.439 & 69.014 & 46.3 & 46.60 & 0.42 & 0.291 & 0.161 & 3.3(7.6) & 0.34(0.145)\\
& & & & 46.35 & 0.18 & & 0.221 & & \\
& & & & 45.92 & 0.45 & & 0.245 & & \\
{\it 2} & -71.930 & 76.612 & 46.2 & 46.26 & 0.32 & 0.272 & 0.269 & 0.7(1.7) & 0.42(0.179)\\
{\it 3} & 33.848 & -138.399 & 45.9 & 45.86 & 0.41 & 1.058 & 1.074 & 4.8(11.1) & 0.19(0.081)\\
& & & & 45.42 & 0.28 & & 0.760 & & \\
{\it 4} & -68.945 & 72.754 & 45.0 & 45.08 & 0.50 & 0.189 & 0.177 & 0.7(1.7) & $-$\\
{\it 5} & 19.442 & -120.938 & 44.5 & 44.49 & 0.21 & 3.143 & 3.090 & 4.8(11.1) & 0.08(0.036)\\
{\it 6} & 0.000 & 0.000 & 43.9 & 43.92 & 0.41 & 5.296 & 5.377 & 0.5(1.2) & 1.23(0.529)\\
{\it 7} & -17.610 & -76.766 & 43.4 & 43.41 & 0.17 & 0.522 & 0.544 & 5.3(12.3) & 0.06(0.024)\\
{\it 8} & -18.277 & -47.123 & 42.2 & 42.18 & 0.23 & 0.937 & 0.957 & 1.4(3.3) & 0.14(0.060)\\
{\it 9} & -16.817 & 3.389 & 41.5 & 41.51 & 0.33 & 0.413 & 0.408 & 2.0(4.7) & $-$\\
\hline
\end{longtable}

\end{appendix}

\end{document}